%% file: 1loopbb.tex
\overfullrule=0pt
\input harvmacM



\def\a{{\alpha}}
\def\l{{\lambda}}

\def\b{{\beta}}
\def\g{{\gamma}}

\def\d{{\delta}}

\def\s{{\sigma}}
\def\si{{\sigma}}

\def\half{{1\over 2}}
\def\p{{\partial}}

\def\t{{\theta}}

\def\bar{\overline}
\def\({\left(}
\def\){\right)}
\def\cF{{\cal F}}

\def\frac#1#2{{#1 \over #2}}
\def\ap{\alpha'}


\input label.defs


\input epsf
\def\figin{\epsfcheck\figin}\def\figins{\epsfcheck\figins}
\def\epsfcheck{\ifx\epsfbox\UnDeFiNeD
\message{(NO epsf.tex, FIGURES WILL BE IGNORED)}
\gdef\figin##1{\vskip2in}\gdef\figins##1{\hskip.5in}
\else\message{(FIGURES WILL BE INCLUDED)}%
\gdef\figin##1{##1}\gdef\figins##1{##1}\fi}
\def\DefWarn#1{}
\def\figinsert{\goodbreak\topinsert}
\def\ifig#1#2#3{\DefWarn#1\xdef#1{Fig.~\the\figno}
\writedef{#1\leftbracket fig.\noexpand~\the\figno}%
\figinsert\figin{\centerline{#3}}
\medskip
\smallskip
\leftskip=20pt \rightskip=20pt \baselineskip12pt\noindent
{{\bf Fig.~\the\figno}\ \ninepoint #2}
\medskip
\global\advance\figno by1\par\endinsert}


\def\DefWarn#1{}
\def\tikzcaption#1#2{\DefWarn#1\xdef#1{Fig.~\the\figno}
\writedef{#1\leftbracket Fig.\noexpand~\the\figno}%
{
\smallskip
\leftskip=20pt \rightskip=20pt \baselineskip12pt\noindent
{{\bf Fig.~\the\figno}\ \ninepoint #2}
\bigskip
\global\advance\figno by1 \par}}

\def\ntoalpha#1{%
\ifcase#1%
@%
\or A\or B\or C\or D\or E\or F\or G\or H\or I
\fi
}

\global\newcount\appno \global\appno=1
\def\applab#1{\xdef #1{\ntoalpha\appno}\writedef{#1\leftbracket#1}\wrlabeL{#1=#1}
\global\advance\appno by1}


\Title{\vbox{\vskip -2cm \rightline{AEI--2012--032}\rightline{DAMTP--2012-26}}}{
	       	\vbox{\vskip-1.5cm
		\centerline{ The Structure of n-Point One-Loop}\vskip6pt
		\centerline{Open Superstring Amplitudes}
		}
	}
\vskip-12pt

\centerline{Carlos R. Mafra$^{a,}$\foot{e-mail: c.r.mafra@damtp.cam.ac.uk} and
Oliver Schlotterer$^{b,}$\foot{e-mail: olivers@aei.mpg.de}}

\bigskip
\centerline{\it $^a$ DAMTP, University of Cambridge}
\centerline{\it Wilberforce Road, Cambridge, CB3 0WA, UK}
\medskip
\centerline{\it $^b$ Max--Planck--Institut f\"ur Gravitationsphysik} 
\centerline{\it Albert--Einstein--Institut, 14476 Potsdam, Germany}

\bigskip\medskip\medskip
\noindent
In this article we investigate one-loop amplitudes in maximally supersymmetric superstring theory.
The non-anomalous part of the worldsheet integrand is presented for any number of massless
open-string states. The polarization dependence is organized into the same
BRST-invariant kinematic combinations which also govern the leading string correction to
tree-level amplitudes. The dimensions of the bases for both the kinematics and the associated worldsheet
integrals is found to be the unsigned Stirling number $S_3^{n-1}$ of first kind. We explain why
the same combinatorial structures govern on the one hand finite one-loop amplitudes of equal helicity
states in pure Yang--Mills theory and on the other hand the color tensors at order $\ap^2$ of the
color-dressed tree amplitude.

\Date{March 2012}
\goodbreak

\lref\GoncharovJF{
  A.~B.~Goncharov, M.~Spradlin, C.~Vergu and A.~Volovich,
  ``Classical Polylogarithms for Amplitudes and Wilson Loops,''
Phys.\ Rev.\ Lett.\  {\bf 105}, 151605 (2010).
[arXiv:1006.5703 [hep-th]].
}

\lref\ArkaniHamedKV{
  N.~Arkani-Hamed, J.~L.~Bourjaily, F.~Cachazo, S.~Caron-Huot and J.~Trnka,
  ``The All-Loop Integrand For Scattering Amplitudes in Planar N=4 SYM,''
JHEP {\bf 1101}, 041 (2011).
[arXiv:1008.2958 [hep-th]].
}

\lref\CaronHuotKK{
  S.~Caron-Huot and S.~He,
  ``Jumpstarting the All-Loop S-Matrix of Planar N=4 Super Yang-Mills,''
[arXiv:1112.1060 [hep-th]].
}

\lref\psf{
  N.~Berkovits,
  ``Super-Poincare covariant quantization of the superstring,''
  JHEP {\bf 0004}, 018 (2000)
  [arXiv:hep-th/0001035].
}
\lref\MPS{
  N.~Berkovits,
  ``Multiloop amplitudes and vanishing theorems using the pure spinor formalism for the superstring,''
JHEP {\bf 0409}, 047 (2004).
[hep-th/0406055].
}
\lref\MSSI{
  C.R.~Mafra, O.~Schlotterer and S.~Stieberger,
  ``Complete N-Point Superstring Disk Amplitude I. Pure Spinor Computation,''
[arXiv:1106.2645 [hep-th]].
}
\lref\MSSII{
  C.R.~Mafra, O.~Schlotterer and S.~Stieberger,
  ``Complete N-Point Superstring Disk Amplitude II. Amplitude and Hypergeometric Function Structure,''
[arXiv:1106.2646 [hep-th]].
}
\lref\MSST{
  C.R.~Mafra, O.~Schlotterer, S.~Stieberger and D.~Tsimpis,
  ``A recursive method for SYM n-point tree amplitudes,''
Phys.\ Rev.\ D {\bf 83}, 126012 (2011).
[arXiv:1012.3981 [hep-th]].
}
\lref\fiveptone{
  C.R.~Mafra and C.~Stahn,
  ``The One-loop Open Superstring Massless Five-point Amplitude with the Non-Minimal Pure Spinor Formalism,''
JHEP {\bf 0903}, 126 (2009).
[arXiv:0902.1539 [hep-th]].
}
\lref\verlindes{
  E.P.~Verlinde and H.L.~Verlinde,
  ``Chiral Bosonization, Determinants and the String Partition Function,''
Nucl.\ Phys.\ B {\bf 288}, 357 (1987)..
}
\lref\GS{
  M.B.~Green and J.H.~Schwarz,
  ``Covariant Description of Superstrings,''
Phys.\ Lett.\ B {\bf 136}, 367 (1984),
  ``Properties of the Covariant Formulation of Superstring Theories,''
Nucl.\ Phys.\ B {\bf 243}, 285 (1984).
}
\lref\GSWII{
  M.B.~Green, J.H.~Schwarz and E.~Witten,
  ``Superstring Theory. Vol. 2: Loop Amplitudes, Anomalies And Phenomenology,''
Cambridge, UK: Univ.~Pr.~(1987) 596 P. (Cambridge Monographs On Mathematical Physics).
}
\lref\PSS{
  C.R.~Mafra,
  ``PSS: A FORM Program to Evaluate Pure Spinor Superspace Expressions,''
[arXiv:1007.4999 [hep-th]].
}
\lref\FORM{
  J.A.M.~Vermaseren,
  ``New features of FORM,''
  arXiv:math-ph/0010025.
\semi
  M.~Tentyukov and J.A.M.~Vermaseren,
  ``The multithreaded version of FORM,''
  arXiv:hep-ph/0702279.
}

\lref\towardsFT{
  C.R.~Mafra,
  ``Towards Field Theory Amplitudes From the Cohomology of Pure Spinor Superspace,''
JHEP {\bf 1011}, 096 (2010).
[arXiv:1007.3639 [hep-th]].
}
\lref\MSSTsix{
  C.R.~Mafra, O.~Schlotterer, S.~Stieberger and D.~Tsimpis,
  ``Six Open String Disk Amplitude in Pure Spinor Superspace,''
Nucl.\ Phys.\ B {\bf 846}, 359 (2011).
[arXiv:1011.0994 [hep-th]].
}
\lref\wittentwistor{
  E.~Witten,
  ``Twistor-Like Transform In Ten-Dimensions,''
  Nucl.\ Phys.\  B {\bf 266}, 245 (1986).
}
\lref\dhokerRev{
  E.~D'Hoker, D.H.~Phong,
  ``The Geometry of String Perturbation Theory,''
Rev.\ Mod.\ Phys.\  {\bf 60}, 917 (1988).
}
\lref\GSanom{
  M.B.~Green and J.H.~Schwarz,
  ``Anomaly Cancellation in Supersymmetric D=10 Gauge Theory and Superstring Theory,''
Phys.\ Lett.\ B {\bf 149}, 117 (1984),
``The Hexagon Gauge Anomaly in Type I Superstring Theory,''
Nucl.\ Phys.\ B {\bf 255}, 93 (1985).
}

\lref\StiebergerHQ{
  S.~Stieberger,
  ``Open \& Closed vs. Pure Open String Disk Amplitudes,''
[arXiv:0907.2211 [hep-th]].
}

\lref\BjerrumBohrRD{
  N.~E.~J.~Bjerrum-Bohr, P.~H.~Damgaard and P.~Vanhove,
  ``Minimal Basis for Gauge Theory Amplitudes,''
Phys.\ Rev.\ Lett.\  {\bf 103}, 161602 (2009).
[arXiv:0907.1425 [hep-th]].
}

\lref\symBSS{
  L.~Brink, J.H.~Schwarz and J.~Scherk,
  ``Supersymmetric Yang-Mills Theories,''
Nucl.\ Phys.\ B {\bf 121}, 77 (1977)..
}
\lref\BGS{
  M.B.~Green, J.H.~Schwarz and L.~Brink,
  ``N=4 Yang-Mills and N=8 Supergravity as Limits of String Theories,''
Nucl.\ Phys.\ B {\bf 198}, 474 (1982)..
}
\lref\BCJ{
  Z.~Bern, J.J.M.~Carrasco and H.~Johansson,
  ``New Relations for Gauge-Theory Amplitudes,''
Phys.\ Rev.\ D {\bf 78}, 085011 (2008).
[arXiv:0805.3993 [hep-ph]].
}
\lref\BCJps{
  C.R.~Mafra, O.~Schlotterer and S.~Stieberger,
  ``Explicit BCJ Numerators from Pure Spinors,''
JHEP {\bf 1107}, 092 (2011).
[arXiv:1104.5224 [hep-th]].
}
\lref\dixonduca{
  V.~Del Duca, L.J.~Dixon and F.~Maltoni,
  ``New color decompositions for gauge amplitudes at tree and loop level,''
Nucl.\ Phys.\ B {\bf 571}, 51 (2000).
[hep-ph/9910563].
}
\lref\odaY{
  I.~Oda and M.~Tonin,
  ``Y-formalism in pure spinor quantization of superstrings,''
Nucl.\ Phys.\ B {\bf 727}, 176 (2005).
[hep-th/0505277].
}
\lref\copenhagen{
  N.~E.~J.~Bjerrum-Bohr, P.~H.~Damgaard, H.~Johansson and T.~Sondergaard,
  ``Monodromy--like Relations for Finite Loop Amplitudes,''
JHEP {\bf 1105}, 039 (2011).
[arXiv:1103.6190 [hep-th]].
}
\lref\Vcolor{
  T.~van Ritbergen, A.~N.~Schellekens and J.~A.~M.~Vermaseren,
  ``Group theory factors for Feynman diagrams,''
Int.\ J.\ Mod.\ Phys.\ A {\bf 14}, 41 (1999).
[hep-ph/9802376].
}
\lref\bilal{
  A.~Bilal,
  ``Higher derivative corrections to the nonAbelian Born-Infeld action,''
Nucl.\ Phys.\ B {\bf 618}, 21 (2001).
[hep-th/0106062].
}
\lref\LiE{
M.A.A. van Leeuwen, A.M. Cohen and B. Lisser, ``LiE, A Package for Lie Group Computations'',
Computer Algebra Nederland, Amsterdam, ISBN 90-74116-02-7, 1992
}
\lref\Hunter{
  D.~Lust, O.~Schlotterer, S.~Stieberger and T.~R.~Taylor,
  ``The LHC String Hunter's Companion (II): Five-Particle Amplitudes and Universal Properties,''
Nucl.\ Phys.\ B {\bf 828}, 139 (2010).
[arXiv:0908.0409 [hep-th]].
}
\lref\HoweMF{
  P.S.~Howe,
  ``Pure spinors lines in superspace and ten-dimensional supersymmetric theories,''
Phys.\ Lett.\  {\bf B258}, 141-144 (1991).
}
\lref\expPSS{
  N.~Berkovits,
  ``Explaining Pure Spinor Superspace,''
  [hep-th/0612021].
}
\lref\cohoSO{
  N.~Berkovits,
  ``Cohomology in the pure spinor formalism for the superstring,''
JHEP {\bf 0009}, 046 (2000).
[hep-th/0006003].
}
\lref\breno{
  N.~Berkovits, B.C.~Vallilo,
  ``Consistency of superPoincare covariant superstring tree amplitudes,''
JHEP {\bf 0007}, 015 (2000).
[hep-th/0004171].
}
\lref\twoloop{
  N.~Berkovits,
  ``Super-Poincare covariant two-loop superstring amplitudes,''
JHEP {\bf 0601}, 005 (2006).
[hep-th/0503197].
}
\lref\twolooptwo{
  N.~Berkovits, C.R.~Mafra,
  ``Equivalence of two-loop superstring amplitudes in the pure spinor and RNS formalisms,''
Phys.\ Rev.\ Lett.\  {\bf 96}, 011602 (2006).
[hep-th/0509234].
}
\lref\KK{
  R.~Kleiss, H.~Kuijf,
  ``Multi - Gluon Cross-sections And Five Jet Production At Hadron Colliders,''
Nucl.\ Phys.\  {\bf B312}, 616 (1989).
}
\lref\BG{
  F.A.~Berends, W.T.~Giele,
  ``Recursive Calculations for Processes with n Gluons,''
Nucl.\ Phys.\  {\bf B306}, 759 (1988).
}
\lref\tsimpis{
  G.~Policastro and D.~Tsimpis,
``$R^4$, purified,''
  Class.\ Quant.\ Grav.\  {\bf 23}, 4753 (2006)
  [arXiv:hep-th/0603165].
}
\lref\thetaSYM{
  	J.P.~Harnad and S.~Shnider,
	``Constraints And Field Equations For Ten-Dimensional Superyang-Mills
  	Theory,''
  	Commun.\ Math.\ Phys.\  {\bf 106}, 183 (1986)
\semi
	P.A.~Grassi and L.~Tamassia,
        ``Vertex operators for closed superstrings,''
        JHEP {\bf 0407}, 071 (2004)
        [arXiv:hep-th/0405072].
}
\lref\ictp{
  N.~Berkovits,
  ``ICTP lectures on covariant quantization of the superstring,''
[hep-th/0209059].
}
\lref\humberto{
  H.~Gomez,
  ``One-loop Superstring Amplitude From Integrals on Pure Spinors Space,''
JHEP {\bf 0912}, 034 (2009).
[arXiv:0910.3405 [hep-th]].
}
\lref\humbertoT{
  H.~Gomez and C.R.~Mafra,
  ``The Overall Coefficient of the Two-loop Superstring Amplitude Using Pure
  Spinors,''
  JHEP {\bf 1005}, 017 (2010)
  [arXiv:1003.0678 [hep-th]].
}
\lref\aisaka{
  Y.~Aisaka and N.~Berkovits,
  ``Pure Spinor Vertex Operators in Siegel Gauge and Loop Amplitude Regularization,''
JHEP {\bf 0907}, 062 (2009).
[arXiv:0903.3443 [hep-th]].
}
\lref\schabinger{
  R.~M.~Schabinger,
  ``One-loop N = 4 super Yang-Mills scattering amplitudes in d dimensions, relation to open strings and polygonal Wilson loops,''
J.\ Phys.\ A A {\bf 44}, 454007 (2011).
[arXiv:1104.3873 [hep-th]].
}
\lref\notriangle{
  Z.~Bern, L.~J.~Dixon, D.~C.~Dunbar and D.~A.~Kosower,
  ``One loop n point gauge theory amplitudes, unitarity and collinear limits,''
Nucl.\ Phys.\ B {\bf 425}, 217 (1994).
[hep-ph/9403226].
}

\lref\StiebergerBH{
  S.~Stieberger and T.~R.~Taylor,
  ``Amplitude for N-Gluon Superstring Scattering,''
Phys.\ Rev.\ Lett.\  {\bf 97}, 211601 (2006).
[hep-th/0607184].
}

\lref\stienotation{
  S.~Stieberger and T.~R.~Taylor,
  ``Multi-Gluon Scattering in Open Superstring Theory,''
Phys.\ Rev.\ D {\bf 74}, 126007 (2006).
[hep-th/0609175].
}

\lref\BernQK{
  Z.~Bern, G.~Chalmers, L.J.~Dixon and D.A.~Kosower,
  ``One loop N gluon amplitudes with maximal helicity violation via collinear limits,''
Phys.\ Rev.\ Lett.\  {\bf 72}, 2134 (1994).
[hep-ph/9312333].
}

\lref\ParkeGB{
  S.J.~Parke and T.R.~Taylor,
  ``An Amplitude for $n$ Gluon Scattering,''
Phys.\ Rev.\ Lett.\  {\bf 56}, 2459 (1986)..
}

\lref\MafraAR{
  C.R.~Mafra,
  ``Pure Spinor Superspace Identities for Massless Four-point Kinematic Factors,''
JHEP {\bf 0804}, 093 (2008).
[arXiv:0801.0580 [hep-th]].
}
\lref\BjornO{
  J.~Bjornsson and M.B.~Green,
  ``5 loops in 24/5 dimensions,''
JHEP {\bf 1008}, 132 (2010).
[arXiv:1004.2692 [hep-th]].
}
\lref\BjornD{
  J.~Bjornsson,
  ``Multi-loop amplitudes in maximally supersymmetric pure spinor field theory,''
JHEP {\bf 1101}, 002 (2011).
[arXiv:1009.5906 [hep-th]].
}
\lref\BerkovitsAW{
  N.~Berkovits, M.B.~Green, J.G.~Russo and P.~Vanhove,
  ``Non-renormalization conditions for four-gluon scattering in supersymmetric string and field theory,''
JHEP {\bf 0911}, 063 (2009).
[arXiv:0908.1923 [hep-th]].
}
\lref\NMPS{
  N.~Berkovits,
  ``Pure spinor formalism as an N=2 topological string,''
JHEP {\bf 0510}, 089 (2005).
[hep-th/0509120].
}
\lref\catalan{
  N.~Arkani-Hamed, F.~Cachazo, C.~Cheung and J.~Kaplan,
  ``The S-Matrix in Twistor Space,''
JHEP {\bf 1003}, 110 (2010).
[arXiv:0903.2110 [hep-th]].
}
\lref\oneloopM{
  C.R.~Mafra,
  ``Four-point one-loop amplitude computation in the pure spinor formalism,''
JHEP {\bf 0601}, 075 (2006).
[hep-th/0512052].
}
\lref\medinaD{
  L.A.~Barreiro and R.~Medina,
  ``5-field terms in the open superstring effective action,''
JHEP {\bf 0503}, 055 (2005).
[hep-th/0503182].
}

\lref\TsuchiyaVA{
  A.~Tsuchiya,
  ``More On One Loop Massless Amplitudes Of Superstring Theories,''
Phys.\ Rev.\ D {\bf 39}, 1626 (1989)..
}

\lref\AtickRS{
  J.~J.~Atick and A.~Sen,
  ``Covariant One Loop Fermion Emission Amplitudes In Closed String Theories,''
Nucl.\ Phys.\ B {\bf 293}, 317 (1987)..
}

\lref\StiebergerWK{
  S.~Stieberger and T.~R.~Taylor,
  ``NonAbelian Born-Infeld action and type 1. - heterotic duality 2: Nonrenormalization theorems,''
Nucl.\ Phys.\ B {\bf 648}, 3 (2003).
[hep-th/0209064].
}

\lref\ArkaniHamedGZ{
  N.~Arkani-Hamed, F.~Cachazo and J.~Kaplan,
  ``What is the Simplest Quantum Field Theory?,''
JHEP {\bf 1009}, 016 (2010).
[arXiv:0808.1446 [hep-th]].
}

\lref\BernUE{
  Z.~Bern, J.~J.~M.~Carrasco and H.~Johansson,
  ``Perturbative Quantum Gravity as a Double Copy of Gauge Theory,''
Phys.\ Rev.\ Lett.\  {\bf 105}, 061602 (2010).
[arXiv:1004.0476 [hep-th]].
}

\lref\BernUF{
  Z.~Bern, J.~J.~M.~Carrasco, L.~J.~Dixon, H.~Johansson and R.~Roiban,
  ``Simplifying Multiloop Integrands and Ultraviolet Divergences of Gauge Theory and Gravity Amplitudes,''
[arXiv:1201.5366 [hep-th]].
}

\lref\DrummondFD{
  J.~M.~Drummond, J.~M.~Henn and J.~Plefka,
  ``Yangian symmetry of scattering amplitudes in N=4 super Yang-Mills theory,''
JHEP {\bf 0905}, 046 (2009).
[arXiv:0902.2987 [hep-th]].
}

\lref\DrummondVQ{
  J.~M.~Drummond, J.~Henn, G.~P.~Korchemsky and E.~Sokatchev,
  ``Dual superconformal symmetry of scattering amplitudes in N=4 super-Yang-Mills theory,''
Nucl.\ Phys.\ B {\bf 828}, 317 (2010).
[arXiv:0807.1095 [hep-th]].
}

\lref\BoelsMN{
  R.~H.~Boels and R.~S.~Isermann,
  ``Yang-Mills amplitude relations at loop level from non-adjacent BCFW shifts,''
[arXiv:1110.4462 [hep-th]].
}

\lref\BerkovitsBK{
  N.~Berkovits and C.~R.~Mafra,
  ``Some Superstring Amplitude Computations with the Non-Minimal Pure Spinor Formalism,''
JHEP {\bf 0611}, 079 (2006).
[hep-th/0607187].
}
\lref\DixonXS{
  L.J.~Dixon,
  ``Scattering amplitudes: the most perfect microscopic structures in the universe,''
J.\ Phys.\ A A {\bf 44}, 454001 (2011).
[arXiv:1105.0771 [hep-th]].
}
\lref\GreenYA{
  M.~B.~Green and J.H.~Schwarz,
  ``Supersymmetrical Dual String Theory. 3. Loops and Renormalization,''
Nucl.\ Phys.\ B {\bf 198}, 441 (1982)..
}
\lref\dhokerS{
  E.~D'Hoker, M.~Gutperle and D.~H.~Phong,
  ``Two-loop superstrings and S-duality,''
  Nucl.\ Phys.\  B {\bf 722}, 81 (2005)
  [arXiv:hep-th/0503180].
}
\lref\dhokerVI{
  E.~D'Hoker and D.~H.~Phong,
  ``Two-Loop Superstrings VI: Non-Renormalization Theorems and the 4-Point
  Function,''
  Nucl.\ Phys.\  B {\bf 715}, 3 (2005)
  [arXiv:hep-th/0501197].
}

\lref\BernJA{
  Z.~Bern, L.~J.~Dixon, D.~C.~Dunbar and D.~A.~Kosower,
  ``One loop selfdual and N=4 superYang-Mills,''
Phys.\ Lett.\ B {\bf 394}, 105 (1997).
[hep-th/9611127].
}

\lref\BernTQ{
  Z.~Bern, J.~J.~M.~Carrasco, L.~J.~Dixon, H.~Johansson and R.~Roiban,
  ``The Complete Four-Loop Four-Point Amplitude in N=4 Super-Yang-Mills Theory,''
Phys.\ Rev.\ D {\bf 82}, 125040 (2010).
[arXiv:1008.3327 [hep-th]].
}

\lref\CarrascoMN{
  J.~J.~.Carrasco and H.~Johansson,
  ``Five-Point Amplitudes in N=4 Super-Yang-Mills Theory and N=8 Supergravity,''
Phys.\ Rev.\ D {\bf 85}, 025006 (2012).
[arXiv:1106.4711 [hep-th]].
}

\lref\NilssonCM{
  B.E.W.~Nilsson,
  ``Pure Spinors as Auxiliary Fields in the Ten-dimensional Supersymmetric {Yang-Mills} Theory,''
Class.\ Quant.\ Grav.\  {\bf 3}, L41 (1986)..
}

\lref\BernUX{
  Z.~Bern and D.~A.~Kosower,
  ``Color decomposition of one loop amplitudes in gauge theories,''
Nucl.\ Phys.\ B {\bf 362}, 389 (1991)..
}

\lref\BernAQ{
  Z.~Bern and D.~A.~Kosower,
  ``The Computation of loop amplitudes in gauge theories,''
Nucl.\ Phys.\ B {\bf 379}, 451 (1992)..
}

\hoffset=-.48truein
\hsize=7.0 true in
\listtoc
\writetoc
\break
\hoffset=0true in
\hsize=6.5 true in

\newsec{Introduction}

\noindent
In recent years the pure spinor formalism \psf\ allowed striking compactness
in the computations of scattering amplitudes both in string theory
\refs{\MPS\breno\twoloop\NMPS\twolooptwo\MafraAR\humberto\humbertoT\aisaka\BerkovitsAW{--}\MSSI}
and directly in its field-theory limit \refs{\MSSI\towardsFT\MSSTsix\BjornO\MSST{--}\BjornD}.
It has been known since the work
of Nilsson \NilssonCM\ and Howe \HoweMF\ and  that the use of a pure spinor simplifies the description of $N=1$
super-Yang--Mills (SYM). With
the advent of the pure spinor formalism this rewarding description was put into
the context of the full superstring theory with a underlying BRST symmetry and a new
kind of superspace \expPSS.

Using the ideas of \BCJ\ for the field theory amplitudes, it was suggested in \towardsFT\ and
proven in \refs{\MSSI,\MSST} that BRST invariance together with the propagator structure of cubic
diagrams are sufficient to determine tree-level amplitudes of $D=10$ SYM to any multiplicity.
The recursive BRST cohomology method obtained in \MSST\ leads
to compact and elegant
supersymmetric answers and makes use of so-called BRST building blocks which can be regarded
as superspace representatives of cubic diagrams.
The {\it field-theory} techniques of \MSST\ were
subsequentely exploited to also calculate the general color-ordered open {\it superstring} tree amplitudes
in \refs{\MSSI,\MSSII}. The punchline is that the $n$-point string amplitudes are written as
a sum of $(n-3)!$ field theory subamplitudes dressed by hypergeometric integrals \MSSII.

The problem of computing one-loop amplitudes in open superstring theory has been dealt with since
the 1980's, the first successful result at four-points being \GreenYA\ in the NS sector and \AtickRS\
in the R sector. In spite of the technical difficulties caused by the
spin structure sums required by the RNS model, \TsuchiyaVA\ provides progress towards higher
multiplicity up to seven-points. In the context of heterotic theories, five- and six gluon
amplitudes as well as their implications for effective actions were analyzed in \StiebergerWK.
Pure spinor techniques have been applied to one-loop scattering in
\refs{\MPS,\oneloopM,\fiveptone,\humberto}, superspace results up to five-points are available from these
references. As for two-loop amplitudes, after an amazing effort by D'Hoker and Phong the four-point
amplitude was computed within the RNS formalism in \dhokerVI\ (see also \dhokerS). Two-loop
calculations using the pure spinor formalism can be found in
\refs{\twoloop,\twolooptwo,\humbertoT}.

Can this BRST line of reasoning within the pure spinor formalism
be extended to loop amplitudes? With this intention in mind, in this paper we apply the
technique of BRST-covariant building blocks to address one-loop
amplitudes in superstring theory. For any number of massless
SYM states, we determine the BRST invariant part of their worldsheet integrand which
is unaffected by the hexagon anomaly \GSanom. The complete kinematic factor turns out to be organized in terms of color-ordered
tree-level amplitudes at order $\ap^2$ that are dressed with worldsheet functions in a minimal basis.
A beautiful harmony in the combinatorics of both ingredients arises. However, evaluating the
(worldsheet- and modular) integrals is left for future work, in particular the extraction of
field theory loop integrals as $\ap \rightarrow 0$ along the lines of \BGS.

Superstring theory has proven to be a fruitful laboratory to learn about hidden structures in the
S matrix of its low energy field theories. The open superstring did not only inspire the color
organization of gauge theory amplitudes but also provided an elegant proof for Bern--Carrasco--Johansson (BCJ) relations among
color-ordered tree amplitudes \refs{\StiebergerHQ,\BjerrumBohrRD}, based on monodromy properties
on the worldsheet. Another difficult field theory problem which found a string-inspired answer is
the explicit construction of local kinematic numerators for gauge theory tree amplitudes which
satisfy all the dual Jacobi identities, see \BCJps. After these tree-level examples of
cross-fertilization between superstring and field theory amplitudes, we hope that this work helps
to provide further guidelines to organize multileg one-loop amplitudes in maximally supersymmetric
SYM in both ten and four dimensions\foot{The four-dimensional $N=4$ SYM theory can be obtained by standard dimensional reduction
from its ten-dimensional version with $N=1$ supersymmetry \symBSS.}. Even though the low energy behaviour of the worldsheet
integrals is not addressed, our result for the kinematic factor heavily constrains the form of
these field theory amplitudes. In particular, the gauge invariant kinematic building blocks $C_{1,\ldots}$ to be defined later on appear
to be a promising starting point to construct kinematic numerators for higher multiplicity. They
could potentially generalize the crossing symmetric factor
$s_{12} s_{23} {\cal A}^{\rm YM}(1,2,3,4)$ omnipresent in multiloop four-point amplitudes of $N=4$
SYM (where ${\cal A}^{\rm YM}(1,2,\ldots, n)$ denotes the color-ordered $n$-point tree amplitude
in maximally supersymmetric Yang Mills theory).

This paper is organized as follows. In section two, we review the construction of the $n$-point
SYM tree amplitude from first principles. We start with the massless vertex operators in terms of
SYM superfields and sketch how their singularity structure give rise to BRST building blocks representing
cubic subdiagrams. As we will argue, BRST invariance forces them to pair up such that color-ordered SYM amplitudes
emerge. Section three sets the formal foundation for the computation of one-loop amplitudes using the
minimal pure spinor formalism. It motivates the construction of a
further family of BRST building blocks which is carried out in section four. The fourth section
follows a line of reasoning similar to the tree-level review -- the BRST variation of the
one-loop specific building blocks allows to a priori determine any BRST invariant to be expected in
a one-loop computation. Then in section five, these BRST invariants are derived from an explicit
conformal field theory (CFT) computation, in particular the associated worldsheet functions are determined. Section six
connects the BRST invariants with $\ap$ corrections to tree-level amplitudes and explains why their symmetry
properties agree with those of finite one-loop amplitudes in pure Yang-Mills theory.
Finally, in the last section, we point out that also the color factors present at the
$\ap^2$ order of tree amplitudes align into the same combinatorial patterns. This leads to a
duality between the worldsheet integrand of one-loop amplitudes and color-dressed tree amplitudes
at ${\cal O}(\ap^2)$.

To give a brief reference to the main results of this work -- the final form for the $n$-point
kinematic factor can be found in equation \nfinal\ whose notation is explained in subsection
\Stirnot. Subsection \AFvloopSec\ contains the general conversion rule \AFasC\ between the BRST
invariants $C_{1,\ldots}$ and color-stripped ${\cal O}(\ap^2)$ trees ${\cal A}^{F^4}$ as well as
low multiplicity examples thereof. According to subsection \auchnoch, the representation \nCf\ of
the color-dressed ${\cal O}(\ap^2)$ tree manifests a duality to the one-loop kinematic factor
\nfinal.

\newsec{Review of tree-level cohomology building blocks}
\seclab\revtree

\noindent
In this section, we shall review the construction of tree-level amplitudes in ten-dimensional
SYM, based on BRST building blocks in pure spinor superspace
\refs{\towardsFT,\MSSTsix,\MSST}. Although the problem at hand is of purely field theoretic nature, we shall use
the vertex operators and the BRST charge of the pure spinor superstring \psf\ as the starting point.
These ingredients suggest a pure spinor superspace representation for color-ordered tree subdiagrams with one
off-shell leg. BRST invariance and the pole structure in the kinematic invariants
\eqn\treeone{
s_{12\ldots p} \ \equiv \ \frac{1}{2} \, (k_1+k_2+\cdots + k_p)^2
}
turn out to be sufficient in
order to determine the tree-level SYM amplitude ${\cal A}_n^{\rm YM}$ with any number $n$ of
external legs \refs{\MSSI,\MSST}. The compactness of the final expression
\eqn\treetwo{
{\cal A}^{{\rm YM}}(1,2, \ldots,n) = \sum_{j=1}^{n-2} \, \langle  M_{12\ldots j} \, M_{j+1\ldots n-1} \, V^n \rangle
}
(see sections 2.1 and 2.3 for the definitions of $V^n$ and $M_{12\ldots j}$, respectively) suggests to apply a similar program at loops, we will
follow these lines in section \BRSTbb\ and introduce similar superspace variables.

At the level of the full-fledged superstring theory, the main virtue of the BRST building block
representation for ${\cal A}^{{\rm YM}}$ is the possibility to identify these SYM constituents within the CFT computation of the superstring disk amplitude. The supersymmetric $n$-point tree amplitude in superstring theory
was shown in \refs{\MSSI,\MSSII} to decompose into a sum of $(n-3)!$ color-ordered field theory
amplitudes, each one of them being weighted by a separate function of $\ap$. The main result of the current
work is a similar decomposition of one-loop supersymmetric amplitudes, based on a new
family of BRST building blocks.

\subsec{From vertex operators to OPE residues}
\subseclab\secSYMfields

\noindent
One of the major tasks in computing the open string tree-level amplitude is the evaluation of
the CFT correlation function
\eqn\treethree{
\langle V^1(z_1) \, V^{n-1}(z_{n-1}) \, V^n(z_{n}) \, U^2(z_2) \, \ldots \, U^{n-2}(z_{n-2}) \rangle
}
where $V^1$ and $U^2$ denote
the vertex operators for the gluon multiplet with conformal-weight zero and one, respectively. They are conformal
fields on the worldsheet parametrized by a complex coordinate $z$. The
8+8 physical degrees of freedom are described by the superfields\foot{Throughout this work, SO$(1,9)$ vector
indices are taken from the middle of Latin alphabet $m,n,p,\ldots = 0,1,\ldots,9$ whereas Weyl spinor
indices $\alpha,\beta,\ldots = 1,2,\ldots 16$ are taken from the beginning of the Greek alphabet.} $A_\a,A^m,W^\a$
and $\cF_{mn}$ of $D=10$ SYM \wittentwistor
\eqn\PSvertices{
V^1 = \l^\a A^1_\a, \qquad
U^i = \p\t^\a A^i_\a + \Pi^m A^i_m + d_\a W^\a_i + \half {\cal F}_{mn}^i N^{mn},
}
where $\lambda^\alpha$ denotes the pure spinor ghost subject to $(\lambda \gamma^m \lambda) =0$ \psf.
The remaining ingredients $\partial \theta^\alpha, \Pi^m, d_\alpha$ and $N^{mn}$ of \PSvertices\ are
conformal weight-one fields on the worldsheet. The ten-dimensional superfields $A_\a,A^m,W^\a$
and $\cF_{mn}$, depending on the bosonic and fermionic superspace variables $x^m$ and $\theta_\alpha$,
obey the following equations of motion \refs{\wittentwistor,\ictp},
\eqn\SYM{
\eqalign{
2 D_{(\a} A_{\b)} & = \g^m_{\a\b} A_m\cr
D_\a{\cal F}_{mn} & = 2k_{[m} (\g_{n]} W)_\a
}\qquad\eqalign{
D_\a A_m &= (\g_m W)_\a + k_m A_\a  \cr
D_\a W^{\b} &= {1\over 4}(\g^{mn})^{\phantom{m}\b}_\a{\cal F}_{mn}.
}}
As shown in \thetaSYM, their $\t$ expansions can be computed
in the gauge $\t^\a A_\a =0$ and read \tsimpis
\eqnn\expansions
$$\eqalignno{
A_{\a}(x,\t)& ={1\over 2}a_m(\g^m\t)_\a -{1\over 3}(\xi\g_m\t)(\g^m\t)_\a
-{1\over 32}F_{mn}(\g_p\t)_\a (\t\g^{mnp}\t) + \cdots \cr
A_{m}(x,\t) &= a_m - (\xi\g_m\t) - {1\over 8}(\t\g_m\g^{pq}\t)F_{pq}
         + {1\over 12}(\t\g_m\g^{pq}\t)(\p_p\xi\g_q\t) + \cdots & \expansions \cr
W^{\a}(x,\t) &= \xi^{\a}\kern-0.15cm - {1\over 4}(\g^{mn}\t)^{\a} F_{mn}
           \kern-0.05cm + {1\over 4}(\g^{mn}\t)^{\a}(\p_m\xi\g_n\t)
	   \kern-0.05cm + {1\over 48}(\g^{mn}\t)^{\a}(\t\g_n\g^{pq}\t)\p_m F_{pq} 
	   + \cdots \cr
\cF_{mn}(x,\t) &= F_{mn} - 2(\p_{[m}\xi\g_{n]}\t) + {1\over
4}(\t\g_{[m}\g^{pq}\t)\p_{n]}F_{pq}
+ {1\over 6}\p_{[m}(\t\g_{n]}^{\phantom{m}pq}\t)(\xi\g_q\t)\p_p + \cdots
}$$
where $a_m(x) = e_m {\rm e}^{ik\cdot x}$, $\xi^{\a}(x) =\chi^\a {\rm e}^{ik\cdot x}$ are the gluon
and gluino polarizations and $F_{mn} = 2\p_{[m} a_{n]}$ is the linearized field-strength.

The equations of motion \SYM\ imply that the vertex operators in \PSvertices\ obey $Q V^i =0$ and
$Q U^j = \partial V^j$. Since their ingredients $V^i$ and $\int U^j$ are BRST closed, superstring
amplitudes (and in particular their field theory limit) should inherit this property.

The correlator \treethree\ can be computed by integrating out the conformal worldsheet fields of unit
weight within the $U^j$ vertex operator. This amounts to summing over all worldsheet singularities in $z_i
\rightarrow z_j$ which the fields in question can produce. In any CFT, this information is carried
by operator product expansions (OPEs), the first
example being
\eqn\treefour{
V^1(z_1) \, U^2 (z_2) \ \rightarrow  \ { L_{21} \over z_{21} } .
}
This defines a composite superfield $L_{21}$ associated with the degrees of freedom of the
states with labels $1$ and $2$, respectively. By iterating this OPE fusion, we define a family of
superfields of arbitrary rank \MSST
\eqn\treefive{
L_{21}(z_1) \, U^3 (z_3) \  \rightarrow  \ { L_{2131} \over z_{31} }, \qquad
L_{2131\ldots l1}(z_1) \, U^m (z_m) \  \rightarrow  \ { L_{2131\ldots l1m1} \over z_{m1} }
}
which will be referred to as OPE residues\foot{It
turns out that even if OPE contractions are firstly carried out among $U^i(z_i) U^j (z_j)$ and
then merged with $V^1$, the result is still a combination of $L_{2131\ldots m1}$ permutations.
In other words, at tree-level the OPE $U^i(z_i) U^j (z_j)$ does not introduce any independent
composite superfields.}. After the fields with conformal weight one
have been integrated out using their OPEs, the zero modes of the pure spinor $\l^\a$ and $\t^\a$
are integrated using the $\langle (\l^3\t^5)\rangle = 1$ prescription reviewed in \MSSI.

\subsec{From OPE residues to BRST building blocks}
\subseclab\opetobb

\noindent
A major shortcoming of the OPE residues $L_{2131\ldots m1}$ is their lack of symmetry under
exchange of labels $1,2,3,\ldots,m$. However, the obstructions to well-defined symmetry properties
can be shown to conspire to BRST-exact terms.
As a simple example, the symmetric rank-two combination is
\eqn\treesix{
L_{21} \ +
\ L_{12} = -\, Q(A_1 \cdot A_2)
}
where $Q=\l^\a D_\a$ denotes the BRST operator of the pure spinor formalism \psf\ and $A^m_i$ is
the vectorial superfield of $D=10$ SYM. Using the BRST transformation properties of
$L_{2131 \ldots}$, these BRST-exact admixtures have been identified in \refs{\MSSI,\MSST} up to rank five,
and their removal leads to a redefinition of the OPE residues\foot{We define (anti-)symmetrization of $p$
indices to include ${1\over p!}$, e.g. $L_{[21]}= {1\over 2}(L_{21}-L_{12})$.}
\eqn\treeseven{
T_{12}  \ \equiv  \ L_{21} -  \frac{1}{2} \,(L_{21}  +  L_{12}) = L_{[21]},\qquad
T_{123\ldots m} \ \equiv  \ L_{2131\ldots m1}  - {\rm corrections} \ .
}
The outcome of \treeseven\ is an improved family of superfields
$T_{123\ldots m}$ which we call BRST building blocks. They are covariant under the action of the BRST charge,
e.g.
\eqnn\Told\
$$\eqalignno{
Q \, T_{1}  &=  0 \cr
Q \, T_{12}  &=  s_{12} \, T_1 \, T_2 \cr
Q \, T_{123} &= (s_{123}-s_{12}) \, T_{12} \, T_3 - s_{12}\, ( T_{23} \, T_1  +  T_{31} \, T_{2}) \cr
Q \, T_{1234}  &=  (s_{1234}-s_{123}) \, T_{123} \, T_4  +  (s_{123}-s_{12}) \, ( T_{12} \, T_{34} + T_{124} \, T_3 ) \cr
& \ \ \ \ \ \ \ \  + s_{12} \, (  T_{134} \, T_2  + T_{13} \, T_{24}  + T_{14} \, T_{23} +  T_1 \, T_{234})\cr
Q T_{12{\dots} k} &= \sum_{j=2}^{k} \sum_{\a \in P(\beta_j)}  (s_{12 \ldots j} - s_{12 \ldots j-1})
T_{12{\ldots} j-1,\{\a\}} T_{j, \{\beta_j \backslash \a \} } &\Told
}$$
where $V_1 \equiv T_1$. The set $\b_j = \{j+1, j+2,{\ldots},k\}$ encompasses the $k-j$ labels
to the right of $j$, and $P(\beta_j)$ denotes its power set.
In other words, $Q$ acting on a BRST building block of higher rank yields products of two lower
rank analogues together with a Mandelstam variable.

As discussed in \MSSI, at each rank the BRST building blocks obey one new symmetry in its labels while
still respecting all the lower-rank symmetries. For example,
since the rank-two building block satisfies $T_{(12)} = 0$ all higher-order building blocks
also obey $T_{(12)34 \ldots} = 0$. At rank-three there is one new symmetry $T_{[123]}=0$
which is respected by all higher-order ranks, $T_{[123]4 \ldots m}=0$ and so forth.
The generalization to rank $m\ge 3$ is given by \MSSI,
\eqn\TrankNu{
\eqalign{
m=2p+1 &: \quad T_{12\ldots p+1[p+2[\ldots [2p-1[2p,2p+1]] \ldots ]]} - 2 T_{2p+1\ldots p+2[p+1[\ldots [3[21]] \ldots ]]}  = 0 \cr
m=2p &: \quad T_{12\ldots p[p+1[\ldots [2p-2[2p-1,2p]] \ldots ]]}
+ T_{2p\ldots p+1[p[\ldots [3[21]] \ldots ]]}  = 0,
}}
and leaves $(m-1)!$ independent components at rank $m$. It turns out that the above symmetries
are shared by color factors of nonabelian gauge theories formed by contracting structure
constants $f^{ijk}$ of the gauge group. At lowest ranks, we have
\eqn\jacTex{
0 = f^{(12) 3} \leftrightarrow T_{(12)} = 0,\qquad
0 = f^{[12|a} f^{3] 4a} \leftrightarrow T_{[123]} = 0,
}
which states their total
antisymmetry and Lie algebraic Jacobi identities, and similarly
\eqn\nonumb{
0= f^{12a} f^{a[3|b} f^{b|4]c} + f^{34a} f^{a[1|b} f^{b|2]c} \leftrightarrow T_{12[34]} + T_{34[12]}=0 . 
}
In general, the symmetries of a rank
$m$ building block are the same as those of a string of structure constants with
$m+1$ labels,
\eqn\treeten{
f^{12 a_2} \, f^{a_2 3 a_3} \, f^{a_3 4 a_4} \ldots f^{a_{m-1} m a_m} \leftrightarrow\; T_{1234 \ldots m} \, ,
}
where the free color index $a_m$ reflects an off-shell leg $m+1$ in the associated cubic diagram.

Therefore the basis of rank $m$ building blocks being $(m-1)!$-dimensional
is equivalent to the well-known fact that the basis of contractions of
structure constants with $p$ free adjoint indices has dimension $(p-2)!$ after Jacobi identities.

This similarity of building blocks with color factors as well as their BRST variations
suggest a diagrammatic interpretation for $T_{123\ldots m}$ in terms of tree
subdiagrams with cubic vertices \BCJ\ as seen on \figone. Firstly, the color structure of this diagram is given by \treeten\ via Feynman
rules and secondly each propagator can be cancelled by one of the Mandelstam
variables in the BRST variation $Q T_{123\ldots m} \rightarrow s_{12},s_{123},s_{1234},\ldots
,s_{1234\ldots m}$. In other words, the role of the BRST operator is to cancel
propagators.
\ifig\figone{The correspondence of tree graphs with cubic vertices and BRST building blocks.}
{\epsfxsize=0.80\hsize\epsfbox{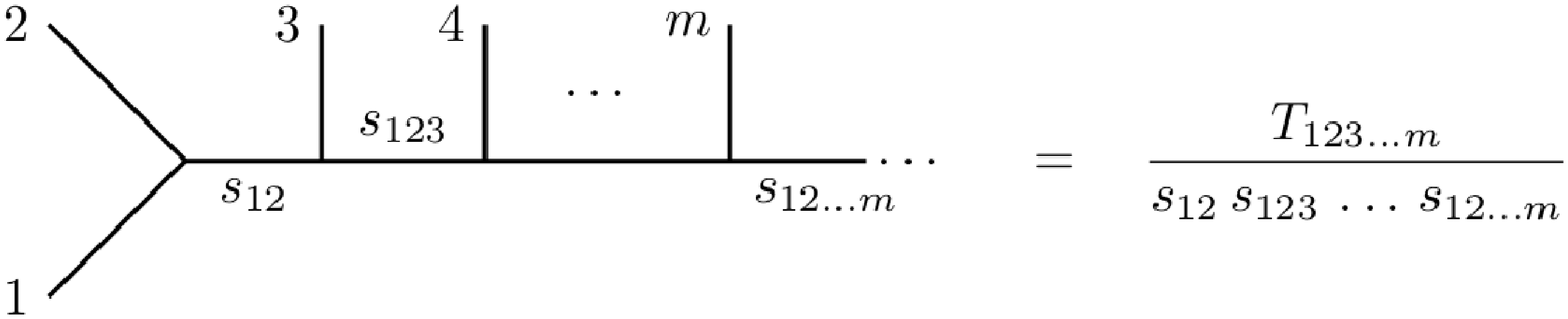}}

\subsec{From BRST building blocks to Berends--Giele currents}
\subseclab\brsttoBG

\noindent
Given the dictionary between cubic tree subdiagrams and BRST building blocks, the next challenge is to combine
different diagrams in order to arrive at BRST-invariant SYM amplitudes. The next
hierarchy level of building blocks consists of superspace representations $M_{123\ldots m}$ of so-called Berends--Giele currents \BG\ which can be thought of as color-ordered SYM tree amplitudes with one leg
off-shell. They encompass all the cubic diagrams present in the associated SYM tree and consist of kinematic
numerators $T_{123\ldots m}$ dressed by their propagators $(s_{12} s_{123} \ldots s_{12\ldots m})^{-1}$,
e.g.
\eqn\treeelevena{
M_{12} = { T_{12} \over s_{12}}, \quad \quad M_{123} = {T_{123} \over s_{12} s_{123}} + { T_{321} \over s_{23} s_{123}}
}
corresponding to the three- and four-point amplitudes with one leg off-shell. At rank four,
\eqn\treetwelve{
M_{1234} = {1 \over s_{1234}} \; \bigg(\, {T_{1234}\over s_{12}s_{123} }  +  {T_{3214}\over s_{23}s_{123} }
+ {T_{3421} \over s_{34}s_{234} } + {T_{3241} \over s_{23}s_{234} }  + {2T_{12[34]}\over
s_{12}s_{34} } \,\bigg)
}
collects the five cubic diagrams of a color-ordered five-point amplitude. The two diagrams present in $M_{123}$ are shown in \figtwo.

 \ifig\figtwo{(a) The cubic graphs with one leg off shell which compose the rank three Berends--Giele current
 $M_{123}$.
 (b) The factorization of the current $M_{12 \ldots m}$ under
 the action of the BRST charge. The right-hand side involves the sum over all partitions of $m$
 legs which is compatible with the color ordering set by $\{1,2, \ldots,m\}$.}
{\epsfxsize=0.90\hsize\epsfbox{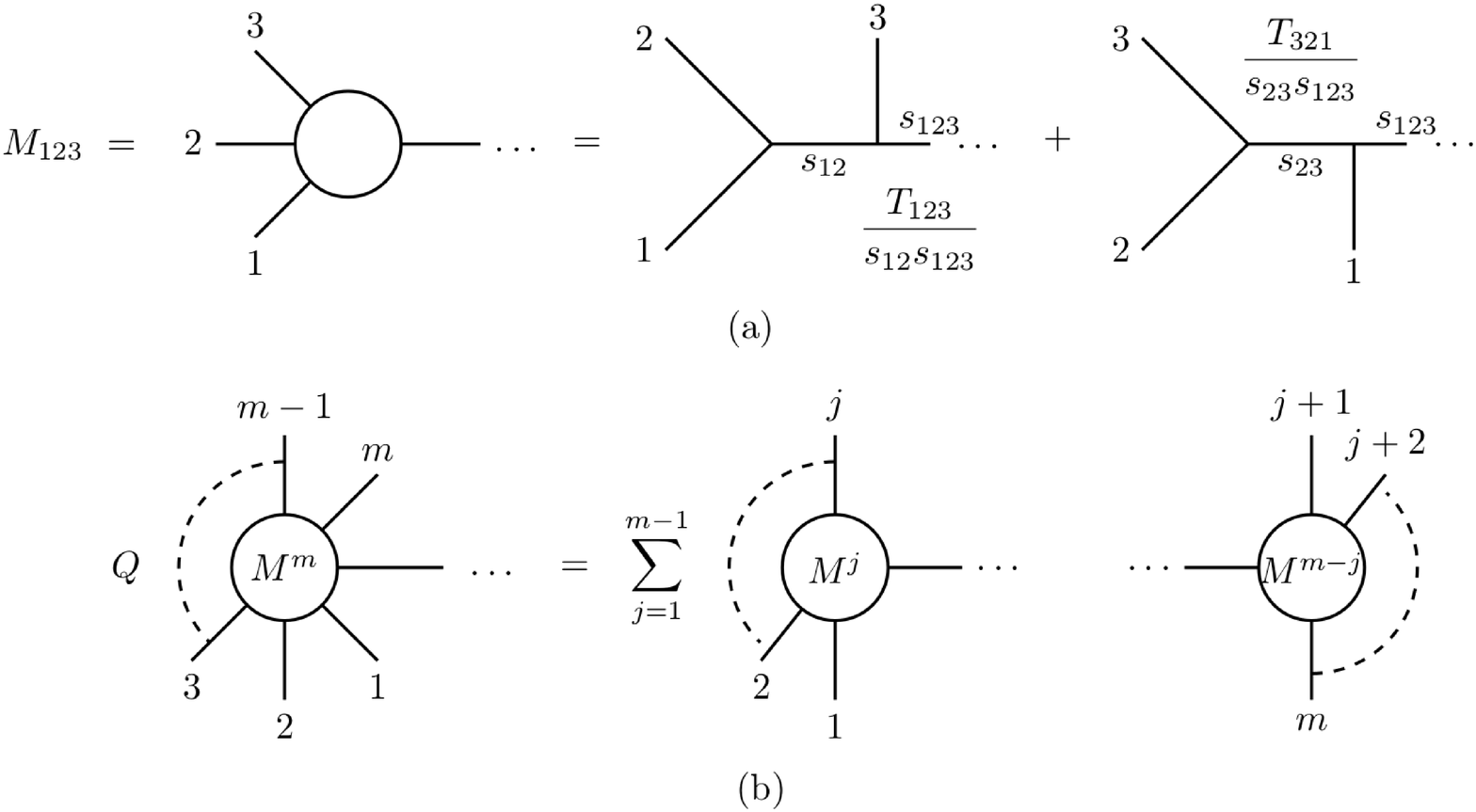}}

The necessity to combine BRST building blocks to full-fledged Berends--Giele currents can be seen from
their $Q$ variation: Their fine-tuned diagrammatic content makes sure that also the $M_{123\ldots
m}$ are covariant under the BRST charge, i.e. (where for rank one, $M_1 \equiv V_1$)
\eqnn\treethirteen
$$\eqalignno{
Q \, M_{1} & = 0 \cr
Q \, M_{12} & = M_1 \, M_2 \cr
Q \, M_{123} & = M_{12} \, M_3 + M_1 \, M_{23} \cr
Q \, M_{1234} & = M_{123} \, M_4 + M_{12} \, M_{34}  +  M_1 \, M_{234} \,. &\treethirteen\cr
}$$
In contrast to $Q T_{123\ldots m}$ as given by \Told, there are no explicit Mandelstam variables in \treethirteen\
because the rank $m$ current already incorporates $m-1$ simultaneous poles. The generalization of
\treethirteen\ to higher rank,
\eqn\treefourteen{
Q \, M_{12\ldots m} = \sum_{j=1}^{m-1} M_{12\ldots j} \, M_{j+1\ldots m}
}
involves all partitions of the $m$ on-shell legs on two
Berends--Giele currents which are compatible with the color ordering. The situation is depicted in
\figtwo b.

\subsec{The $D=10$ SYM amplitude as a pure spinor cohomology problem}
\subseclab\cohoProbSec

\noindent
Using the Berends--Giele currents reviewed in the previous subsection, a method to recursively compute
the ten-dimensional SYM tree-level scattering amplitudes was developed in \MSST. It was later shown
in \MSSI\ that the expressions found in \MSST\ also follow from the field theory limit of
tree-level superstring amplitudes computed with the pure spinor formalism.

The method relies on finding an expression in the cohomology of the pure spinor BRST charge, i.e.\ 
which is BRST-closed but non-exact,
$$
Q {\cal A}^{\rm YM}(1,2, \ldots,n) = 0 , \qquad {\cal A}^{\rm YM}(1,2, \ldots,n) \neq \langle Q {\cal X}_n \rangle .
$$
If we additionally require this cohomology element to reproduce the kinematic poles of a
color-ordered SYM subamplitude, the result is uniquely determined to be
\eqn\treeseventeen{
{\cal A}^{\rm YM}(1,2, \ldots,n) = \sum_{j=1}^{n-2} \, \langle  M_{12\ldots j} \, M_{j+1\ldots
n-1} \, V^n \rangle.
}
In order to show that the right-hand side is in the BRST cohomology first note that $QV_n = 0$, whereas
\eqn\treesixteen{
Q \, \sum_{j=1}^{n-2} M_{12\ldots j} \, M_{j+1\ldots n-1} = 0
}
follows from \treefourteen. And secondly, in the momentum phase space
of $n$ massless particles where the Mandelstam variable $s_{12\ldots n-1}$ vanishes,
$\sum_{j=1}^{n-1} M_{12\ldots j} \, M_{j+1\ldots n-1}$ can not be written as $QM_{12 \ldots n-1}$
since $M_{12 \ldots n-1}$ contains a divergent propagator $1/s_{12\ldots n-1}$. This rules out BRST-exactness of \treeseventeen.

The number of cubic diagrams in the color-ordered $n$-point tree amplitude
is given by the Catalan number $C_{n-2}$, see \catalan, which satisfies the recurrence relation $C_{p+1} =
\sum_{i=0}^p C_i C_{p-i}$ with $C_0 = 1$. By its diagrammatic construction, $M_{12 \ldots
j}$ gathers $C_{j-1}$ pole channels, so the number of poles in the expression \treeseventeen\ for
the $n$-point subamplitude is given by $\sum_{i=0}^{n-3} C_i C_{n-3-i}$, which is precisely
the recursive definition of $C_{n-2}$. The expression \treeseventeen\ therefore contains the same number of
cubic diagrams as the color-ordered $n$-point amplitude, and the fact that Berends--Giele currents have a
notion of color ordering guarantees that the pole channels in \treeseventeen\ are precisely those
of ${\cal A}^{\rm YM}(1,2, \ldots,n)$.
The factorization properties of the expression \treeseventeen\ are depicted in \figfour, and
the reader is referred to \MSSI\ for more details.
\ifig\figfour{Diagrammatic interpretation of the expression
$\sum_{j=1}^{n-2} \langle  M_{12\ldots j} M_{j+1\ldots n-1} V^n \rangle$ for the $n$-point SYM
tree amplitudes. The $j$ sum runs over all partitions of the first $n-1$ legs among two Berends--Giele currents.}
{\epsfxsize=0.85\hsize\epsfbox{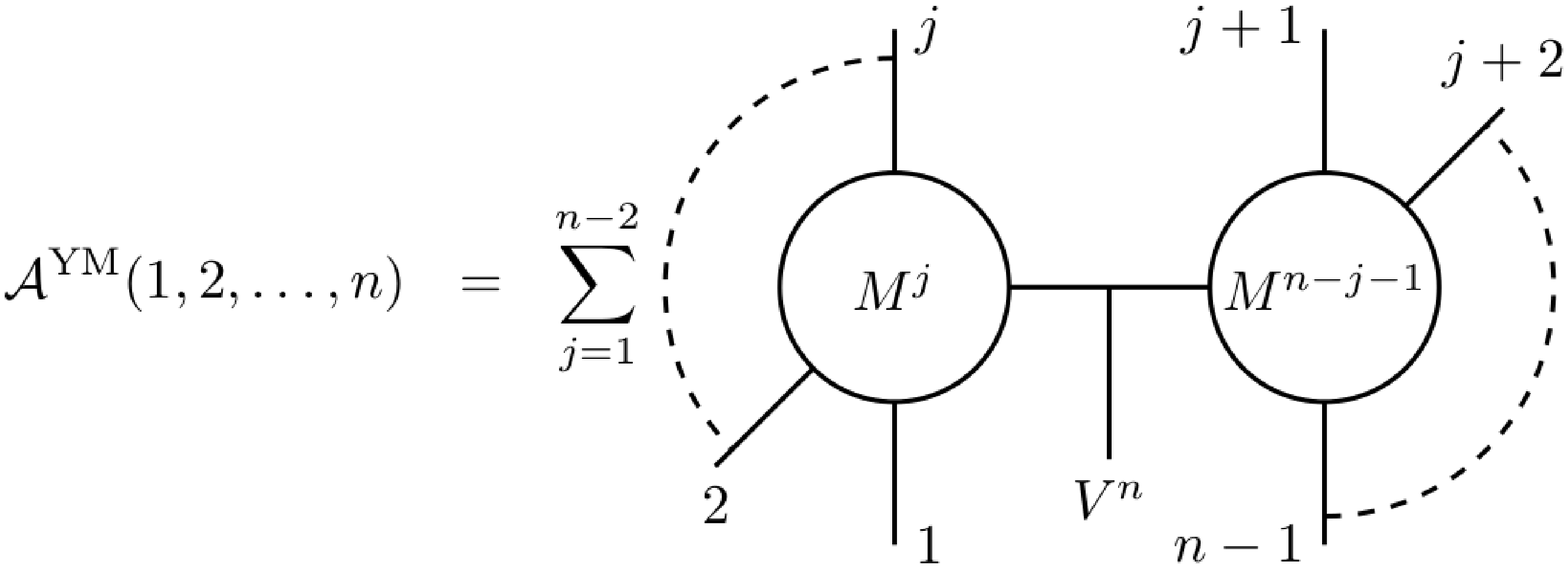}}

\newsec{One-loop amplitudes with the minimal pure spinor formalism}
\seclab\secOnePS
\noindent
This section sketches the prescription towards one-loop amplitudes within the minimal pure spinor
formalism. The main goal is to make the one-loop zero mode saturation rule \empir\ for the
correlator
$\langle V^1 \prod_{j=2}^n U^j \rangle$ plausible instead of giving an exhaustive review. The reader is
referred to \MPS\ for the details omitted in the following discussion.

The prescription to compute $n$-point one-loop amplitudes for open superstrings is \MPS\
\eqn\onepresc{
{\cal A}_n^{1- {\rm loop}} = \sum_{\rm top} C_{\rm top} \int_{D_{\rm top}} dt\, \langle (\mu, b)\prod_{P=2}^{10} Z_{B_P}
Z_J \prod_{I=1}^{11} Y_{C_I} \, V_1(z_1)\prod_{j=2}^n \int\, dz_j U^j(z_j)\rangle,
}
where $\mu$ is the Beltrami differential, $t$ is the Teichm\"uller parameter and
$b$ is the b-ghost whose contribution will be discussed below.
The sum runs over all one-loop open string worldsheet topologies, i.e.\ over planar and non-planar
cylinder diagrams as well as the Moebius strip, see \GSWII. The integration domain $D_{\rm top}$ for $t$ has to be adjusted accordingly, and the associated color factors
$C_{\rm top}$ are single- or double traces over Chan--Paton generators associated with the external states.
Both the Chan--Paton traces and the integration region for the $z_j$ must reflect the cyclic ordering
of the vertex operators on the boundaries of the genus one worldsheet. The main focus of this work is the simplification of the $t$ integrand in \onepresc, so we do not specify further details of $D_{\rm top}$ and $C_{\rm top}$ or comment on the interplay between the topologies.

In order to introduce
the remaining elements appearing in \onepresc, note that the computation of the CFT correlator at one-loop starts
by separating off the zero mode of the conformal weight one variables.
The role of the picture-changing operators $Z_B, Z_J$ and $Y_C$ is to ensure that the zero modes of bosonic and fermionic variables are absorbed correctly, see \MPS.
The angle brackets $\langle \ldots\rangle$ in \onepresc\ initially denote the path integral
over all the worldsheet variables in the pure spinor formalism. The non-zero modes are integrated out
using their OPEs as described below and we will
follow a procedure where the zero modes of $d_\a,N^{mn}$ and the ghost current $J$ are integrated out
first, leaving those of $\l^\a$ and $\t^\a$ for a last step in the computation, e.g.\ after
the superfield expansions of \expansions\ are substituted in the expressions of
various building blocks. And since general group theory arguments will be used to determine the integrals over zero modes of $d_\a,N^{mn}$ and $J$
the precise details of the zero-mode measures of \MPS\ will not be needed.

So unless
otherwise stated, every appearance of the pure spinor angle brackets $\langle \ldots \rangle$ in this paper
denotes the zero-mode integration of $\l^\a$ and $\t^\a$ only and will be taken as
the definition of pure spinor superspace \expPSS. This integration can be performed using
symmetry arguments alone and follows from the
tree-level prescription $\langle (\l^3\t^5)\rangle =1$ of \psf. Since this tedious process has been mostly
automated in \PSS\ we will restrict ourselves to presenting our one-loop results in compact
pure spinor superspace form as in the tree-level approach of \MSSI.
Furthermore, the correlation function of the matter variables
$x^m(z,\bar z)$ and $\Pi^m(z)$ is computed as in \refs{\verlindes,\dhokerRev}
and will receive no special treatment in the following.

The non-zero-modes
are integrated out using their OPEs \verlindes
\eqn\opesone{
d_\a(z_i)\t^\b(z_j) \rightarrow \eta_{ij} \d^\b_\a, \qquad \Pi_m(z_i) x^n(z_j,\bar z_j) \rightarrow -\eta_{ij}\delta_m^{n}.
}
Singularities in colliding worldsheet positions enter through the
function $\eta_{ij}$ which is defined on
a given Riemann surface as the derivative of the bosonic Green's function
$$
\eta_{ij}  :=  {\p\over \p z_i} \; \langle  x(z_i,\bar z_i) \, x (z_j,\bar z_j) \rangle \ .
$$
It behaves as $z_{ij}^{-1}$ as the positions approach each other but respects the periodicity
properties required by a higher-genus Riemann surface. The representation in terms of Jacobi theta
functions will not be needed in the following discussions, only its antisymmetry $\eta_{ij}= -
\eta_{ji}$ will play a fundamental role.

In the amplitude prescription \onepresc, the b-ghost is a composite operator whose form
is given schematically by \refs{\MPS,\odaY},
\eqnn\bghost
$$\eqalignno{
b & = (\Pi d + N\p\t + J\p\t)\,d\,\d(N) + (w\p\l + J\p N + N\p J + N\p N)\d(N) \cr
&+ (N\Pi + J\Pi + \p\Pi + d^2)(\Pi\d(N) + d^2\d'(N))\cr
&+ (Nd + Jd)(\p\t\d(N) + d\Pi\d'(N) + d^3\d''(N))\cr
&+ (N^2 + JN + J^2)(d\p\t\d'(N) + \Pi^2\d'(N) + \Pi d^2 \d''(N) + d^4 \d'''(N))
&\bghost
}$$
where $\d'(x) = {\p\over\p x}\d(x)$ is defined through integration by parts and the precise index contractions are being omitted.
It will
be argued in the appendix \bghostapp\ that the zero-mode contribution from the b-ghost is unique and given
by an expression of the form $d^4\d'(N)$. Furthermore, the result of the zero-mode integrations
in this case is fixed by group theory up to an overall constant, and this is the contribution which
will concern us in this paper.

We do not have a constructive proof that the b-ghost does not contribute via OPE contractions
(i.e. via nonzero modes), but an indirect argument based on total symmetry of the kinematic factor
will follow in subsection \auchnochmal.

In general, the evaluation of the one-loop amplitude \onepresc\ involves two separate challenges
summarized by the formula\foot{Since the Koba Nielsen factor
KN$=\left \langle \prod_{i=1}^n e^{i k_i \cdot x(z_i,\bar z_i) }\right \rangle$ due to the
plane wave correlator is a universal
prefactor, we define the kinematic factor $K_n$ not to contain KN. Nevertheless, its presence is
relevant for integration by parts relating different worldsheet functions, see subsection~\IBP.}

$$
{\cal A}_n^{1-{\rm loop}} = \sum_{\rm top} C_{\rm top} \int_{D_{\rm top}} dt \prod_{j=2}^n \int\, dz_j \left \langle \prod_{i=1}^n e^{i k_i \cdot x(z_i,\bar z_i) }\right \rangle
\times K_n \ .
$$
Firstly, the computation of the
kinematic factor $K_n$ in pure spinor superspace whose generic form is given by
\eqn\zeroint{
K_n = \eta^{n-4} \langle f(\l^\a,\t^\a;1,2,\dots,n)\rangle
}
(where $1,2,\ldots,n$ denote the physical degrees of freedom of the $n$ external states), and secondly, the evaluation of the integrals
over vertex operator positions on the boundary of the Riemann surface as well as the modular
parameter $t$. The form of the kinematic part is unique to the pure spinor formalism
and will be dealt with in the following sections. {Its BRST invariant
ingredients will be identified and related to the $\ap^2$ terms in the expansion of the corresponding
tree-level amplitudes. The expressions for the integrals over the Riemann surface are exactly like in
RNS or Green-Schwarz \GS\ formalisms and will not play a role in this article. Extracting 
information on the integrals -- in particular their field theory limits -- will be left for future work.

\subsec{The one-loop prescription for $d_\alpha, N^{mn}$ zero mode saturation}
\noindent
When the number of external states is four, the saturation of $d_\a$ zero modes in
\eqn\onefourpt{
{\cal A}_4^{1-{\rm loop}} = \sum_{\rm top} C_{\rm top}\int_{D_{\rm top}} dt \langle (\mu, b)\prod_{P=2}^{10} Z_{B_P} 
Z_J \prod_{I=1}^{11} Y_{C_I} \, V_1(z_1) \prod_{j=2}^4 \int dz_j U^j(z_j)\rangle,
}
is unique and determines the amplitude up to an overall coefficient \refs{\MPS,\oneloopM}. The picture
changing operators, the b-ghost and the external vertices provide ten, four and two $d_\a$ zero-modes,
respectively, thereby saturating all the sixteen zero modes of $d_\a$. Furthermore, as mentioned after \bghost,
the terms with four $d_\a$ zero modes from the b-ghost also contain factors which absorb extra
zero modes of $N^{mn}$, either $1,2$ or $3$. For the four-point amplitude the only possibility is the
absorption of one zero mode of $N^{mn}$ through an overall
factor of $\d'(N)$.
Summing it all up, the contribution from the external vertices is proportional to
\eqn\contex{
\half V_1 (dW^2)(dW^3) {\cal F}^4_{mn} N^{mn} + {\rm cyclic}(234)
}
and the remaining zero mode integration is given schematically by
\eqnn \fourzero
$$\eqalignno{
K_4 = \int &[{\cal D}\l][{\cal D}N]\, d^{16}\t\, d^{16}d\, (\l)^{10} \, (d)^{14} \,  (\t)^{11}  \, \d^{11}(\l)  \, \d^{10}(N) \, \d(J) \,  \d'(N)&\fourzero \cr &\times\, 
\half V_1 (dW^2)(dW^3) {\cal F}^4_{mn} N^{mn} + {\rm cyc}(234).
}
$$
As one can check in the expressions given in \MPS, the measure factor $[{\cal D}N]$ has ghost-number -8.
Therefore the integration
of $\int [{\cal D}N]\,d^{16}d (\l)^{10}(d)^{14}\d^{10}(N)\d(J)\d'(N)$ in \fourzero\ with
ten powers of $\lambda$ has the net effect
of replacing $d_\a d_\b N^{mn}$ from the external vertices by a $\lambda$ bilinear. The
tensor structure is uniquely determined by group theory since the decomposition
of $d_\a \otimes d_\b \otimes N^{mn}$ contains only one component in the $SO(10)$
representation $(00002)$ of a chiral pure spinor bilinear:
\eqn\empir{
d_\a d_\b N^{mn} \longrightarrow (\l\g^{[m})_\a (\l\g^{n]})_\b 
}
Consequently, \contex\ leads
to the following kinematic factor for the four-point one-loop amplitude
\eqn\onefourkin{
K_4 = {1 \over 2} \langle V_1 (\l\g_m W_2)(\l\g_n W_3) \cF_4^{mn}\rangle + {\rm cyclic}(234)
}
whose BRST invariance one can easily check using the pure spinor constraint $(\lambda \gamma^m \lambda) = 0$
and elementary corollaries $(\lambda \gamma_m)_{\alpha} (\lambda \gamma^m)_{\beta}=0$ and
$(\lambda \gamma^m \gamma_{pq} \lambda) = 0$.

According to the arguments in appendix A, the replacement rule \empir\ still applies to
one-loop amplitudes with $n \geq 5$ legs. It passes the superspace kinematic factor built
from one unintegrated and $n-1$ integrated vertex operators to the tree-level zero mode
prescription $\langle \lambda^3 \theta^5 \rangle = 1$:
$$
K_n  \equiv \langle V^1(z_1) U^2(z_2) U^3(z_3) \ldots U^n(z_n) \rangle_{d_\a d_\b N^{mn} \longrightarrow (\l\g^{[m})_\a (\l\g^{n]})_\b}  
$$
Studying the interplay of \empir\ with the non-zero modes of the conformal fields in $U^j$
is the subject of the next section. Integrating out all but three weight one fields $d_\alpha
d_\beta N^{mn}$ obviously requires $n-4$ OPEs, and we will see that they give rise to new families
of BRST building blocks.

\newsec{BRST building blocks for loop amplitudes}
\seclab\BRSTbb

\noindent
As reviewed in section \revtree, tree-level BRST building blocks $T_{12\ldots k}$ are defined by
a two step procedure. Its starting point have been the residues of the single poles in iterated
OPEs of integrated vertex operators $U(z_j)$ with the unintegrated one $V(z_1)$. As a second step,
the BRST trivial components of these residues
had to be subtracted to obtain symmetry properties suitable for a diagrammatic interpretation. On
the genus zero worldsheet governing tree-level amplitudes, conformal fields of weight $+1$ have
no zero modes, so all of $d_\a$ and $N^{mn}$ are completely
integrated out in generating the residues entering BRST building blocks. However, this is no
longer the case at one-loop.

As seen in the previous section, the kinematic factor at one-loop comes from the
terms in the external vertices which contain two zero modes of $d_\a$ and one of $N^{mn}$.
Hence, we have to integrate out weight one fields from the $n-1$ integrated
vertex operators until we are left with the combination $(d)^2N$ which requires a total of $n-4$ OPE contractions. In doing so, one is naturally
led to define the composite superfields ${\tilde J}^{mn}_{12}, {\tilde K}_{12}^\a$ and higher rank generalizations ${\tilde J}^{mn}_{12 \ldots k}, {\tilde K}_{12 \ldots k}^\a$
as the remaining single-pole terms $\sim d_{\alpha}$ or $\sim N_{mn}$ in nested OPEs of multiple integrated vertex operators:
\eqnn\bbdefs
$$\eqalignno{
U_1(z_1) U_2(z_2)  & \longrightarrow {{\tilde J}^{mn}_{12} N_{mn}\over z_{21}} +
{d_\a {\tilde K}^\a_{12} \over z_{21}} + \cdots
 \cr
U_1(z_1) U_2(z_2) \ldots U_k(z_k)  &\longrightarrow {{\tilde J}^{mn}_{12 \ldots k} N_{mn}\over z_{k,k-1} \ldots z_{32}z_{21}} +
{d_\a {\tilde K}^\a_{12 \ldots k} \over z_{k,k-1} \ldots z_{32}z_{21}} + \cdots &\bbdefs
}$$
The ellipsis $\cdots$ indicates terms with $\Pi^m$ and $\partial \theta^\a$ as well as double poles in 
individual $z_{ij}$, they do not contribute to the end result for one-loop amplitudes. Given
the prescription $d_\a d_\b N^{mn} \mapsto (\l\g^{[m})_\a (\l\g^{n]})_\b$, the quantity of interest built from the ${\tilde K}^\a$ superfield is
\eqn\Ktdef{
{\tilde K}^m_{12 \ldots k} \equiv (\l\g^m)_\a {\tilde K}^\a_{12 \ldots k}.
} 
As a rank $k=2$ example, let us consider the OPE of two integrated vertices. It contains single and double poles
\eqnn \UUope
$$\eqalignno{
U_1(z_1) U_2(z_2)  \longrightarrow {1  \over z_{21}} \Big[ &(k_2\cdot A_1) U^2 + {1\over2 } (W_1 \g^m W_2) \Pi^m + (k_1 \cdot \Pi) (A_1 W_2) + \partial \theta^\alpha D_\alpha A_\beta^1 W_2^\beta \cr
+ &{1 \over 4} (d \gamma^{mn} W_2) {\cal F}_{mn}^1 +k_m^1 (W_1 \gamma_n W_2) N^{mn} -{1 \over 2} {\cal F}^1_{mp} {\cal F}^{2p}_{n} N^{mn} - (1 \leftrightarrow 2) \Big] \cr
+ &{ 1 +( k_1\cdot k_2) \over z^2_{21}}\Big[  (A_1 W_2) + (A_2 W_1) -(A_1 \cdot A_2) \Big] &\UUope
} $$
with $U^2 = \partial \theta^\alpha A^2_\alpha+ \Pi^m A_m^2 + d_\alpha W_2^\alpha + \half N_{mn} {\cal F}_2^{mn}$, and one can read off
\eqnn \Ktwo
\eqnn \Jtwo
$$\eqalignno{
\tilde K_{12}^m &= {1 \over 4} (\l \g^m \g^{pq} W_2)  {\cal F}^1_{pq}
+  (k_2 \cdot A_1)  (\l \g^m W_2) \ - \ (1 \leftrightarrow 2) 
  &\Ktwo \cr
\tilde J_{12}^{mn} &= {1 \over 2} \big[ (k_2 \cdot A_1) \, {\cal F}_2^{mn}
+ {\cal F}_{2 \, p}^{[m} \, {\cal F}_1^{n] p} + k_{12}^{[m} \, (W_1 \g^{n]} W_2)\big]
 \ - \ (1\leftrightarrow 2) & \Jtwo
} $$
from the superfields contracted with $d_\alpha$ and $N_{mn}$, respectively.

The definitions in \bbdefs\ lead to the following rank $\leq 3$ expressions
\eqnn\Kone\
\eqnn\Ktwo\
\eqnn\Ktildethree\
\eqnn\Jone\
\eqnn\Jtwo\
\eqnn\Jtildethree\
$$\eqalignno{
\tilde K_1^m & = (\l \g^m W_1)  &\Kone \cr
\tilde K^{m}_{12} &= {1 \over 4} (\l \g^m \g^{pq} W_2)  {\cal F}^1_{pq}
+  (k_2 \cdot A_1)  (\l \g^m W_2) \ - \ (1 \leftrightarrow 2) &\Ktwo\cr
\tilde K_{123}^m & =
- \half (k_{12}\cdot A_3) \,\tilde K^m_{12}
- (\l \g^m W_1) \, k_p^1 \, (W_2 \g^p W_3)  + \half (\l \g^m W_3) \, k_p^3 \, (W_1 \g^p W_2)  \cr
 &\quad + {1 \over 4} (k_1 \cdot A_2) \, \Big[ (\l \g^m \g^{pq} W_1) \, {\cal F}^3_{pq}
 -  (\l \g^m \g^{pq} W_3) \, {\cal F}^1_{pq} - 4 (k_3 \cdot A_1) \, (\l \g^m W_3) \Big]  \cr
&\quad +  {1 \over 2} (\l \g^m \g^{pq} W_2) \, k_p^1 \, (W_1 \g_q W_3)
  - {1\over 4} (\l \g^m \g^{pq} W_3) \, {\cal F}^{1 \, r}_p \, {\cal F}^2_{qr}\cr
&\quad +  {1 \over 16} \; (\l \g^m \g^{pq} \g^{rs} W_1) \, {\cal F}^{3}_{pq} \, {\cal F}^2_{rs}
 \ - \ (1 \leftrightarrow 2)  &\Ktildethree
\cr
\tilde J_1^{mn} &= {1 \over 2} \, \cF_1^{mn}  &\Jone \cr
 \tilde J^{mn}_{12} &= {1 \over 2} \big[\ (k_2 \cdot A_1) \, {\cal F}_2^{mn}
 +  {\cal F}_{2 \, p}^{[m} \, {\cal F}_1^{n] p}  +  k_{12}^{[m} \, (W_1 \g^{n]} W_2)\big]
 \ - \ (1\leftrightarrow 2) &\Jtwo\cr
 \tilde J_{123}^{mn} &=
 - \half \big[(k_{12}\cdot A_3)\,\tilde J_{12}^{mn}
 - (k_1 \cdot A_2) (k_3 \cdot A_1) \, {\cal F}^{mn}_3\big]
 + {\cal F}_{1\, \ p}^{[m} \, {\cal F}_{3 \,q}{}^{n]} \, {\cal F}_2^{pq}
 + (k_1 \cdot A_2) \, {\cal F}_{1\,r}^{[m} \, {\cal F}_3^{n]r}\cr
& + k^1_p \, {\cal F}_2^{p[m} \, (W_1 \g^{n]} W_3)
  - k_1^{[m} \, {\cal F}_2^{n]p} \, (W_1 \g_p W_3)
  + (k_2 \cdot A_1) \, k_{23}^{[m}\, (W_2 \g^{n]} W_3) \cr
& +  \half\big[
   (W_1 \g^{[m} W_2) \, {\cal F}_3^{n]p} \, k^{12}_p
 - k_{12}^{[m} {\cal F}_3^{n]p} \, (W_1 \g_p W_2)
 + k_2^p \, (W_1 \g_p W_3) \, {\cal F}_2^{mn}
\big] &\Jtildethree\cr
 &  + \ {1 \over 4} \big[ (k_{12}^{[m} (W_1 \g^{n]} \g^{pq}W_3) \, {\cal F}_{pq}^2
 + (W_2 \g^{pq} \g^{[m} W_3) \, {\cal F}_{pq}^1 + k_3^p \, (W_1 \g_p W_2) \, {\cal F}_3^{mn} \big]
\ - \ (1\leftrightarrow 2)\cr
 }$$
 where $k_{ij}^m:= k_i^m + k_j^m$. Expressions for the rank four building blocks $\tilde K^m_{1234}$ and $\tilde J^{mn}_{1234}$ are available from the authors upon request.

Similar to their tree-level counterparts $T_{12 \ldots k}$ \MSSI, the new composite superfields have two 
essential virtues: On the one hand, they have symmetry properties which reduce the independent rank $k$ 
components to $(k-1)!$ and thereby suggest an interpretation in terms of tree-level subdiagrams with 
one off-shell leg. On the other hand, they possess {\it covariant} BRST variations,
\eqnn\Qtildesnew
$$\eqalignno{
Q\tilde K^m_1 &= 0, &\Qtildesnew\cr
Q\tilde K^m_{12} &= s_{12}\(T_1 \tilde K_2^m - T_2 \tilde K_1^m \),\cr
Q\tilde K_{123}^m &= s_{13} \, L_{21} \, \tilde K^m_3 \ - \ s_{23} \, L_{12} \, \tilde K^m_3 \ + \ s_{12} \, \big[ \, L_{31} \, \tilde K^m_2 \ - \ L_{32} \, \tilde K^m_1 \, \big] \cr
&\qquad - \ (s_{13}+s_{23}) \, T_3 \, \tilde K_{12}^m \ + \ s_{12} \, \big[ \, T_1 \, \tilde K_{23}^m \ - \ T_2 \, \tilde K_{13}^m \, \big].
}$$
However, the appearance of the OPE residue $L_{21}$ in the right-hand side of $Q\tilde K_{123}^m$ instead of the BRST building block $T_{12}$
signals the need for a redefinition of $\tilde K^m_{123}$ analogous to the redefinitions
of $L_{2131 \ldots}$ to $\tilde T_{123 \ldots}$ at tree-level, see subsection \opetobb. In order to justify
this, let us recall the following general lesson from the tree-level analysis: Quantities whose $Q$ variation
contains BRST exact constituents such as $L_{(21)}=-{1\over2} Q(A_1\cdot A_2)$ combine to BRST trivial parts
of the amplitude. It is economic to remove these terms in an early step of the computation, i.e.\ to study the
BRST building block
\eqn\Kdef{
 K_{123}^{m} \equiv \tilde K_{123}^{m} \ + \ {1\over 2} \, \big[ \, (s_{13} - s_{23}) \, (A_{1}\cdot A_{2}) \, K_3^m \ + \ s_{12} \, ((A_1\cdot A_3) \, K_2^m \ - \ (A_2\cdot A_3) \, K_1^m) \, \big]
}
from now on whose BRST transformation gives rise to $T_{12}$ rather than $L_{21}$:
\eqn\QKthree{
Q K_{123}^m = (s_{13}+s_{23} ) (T_{12} \, K^m_3 - V_3 \, K^m_{12}) + s_{12} \big[ T_{13} \, K_2^m  -  V_2 \, K_{13}^m -  T_{23} \, K_1^m +  V_1 \, K_{23}^m  \big]
}
Also the higher rank cases $K_{12\ldots k}^m = \tilde K_{12\ldots k}^m + \ldots$ and $J_{12\ldots k}^{mn} = \tilde J_{12\ldots k}^{mn} + \ldots$ at $k \geq 4$ require modification to ensure BRST building blocks
$T_{12\ldots k}$ rather than the OPE residues $L_{21\ldots k1}$ (with BRST exact components) in
their $Q$ transformation. However, in contrast to the tree-level redefinitions $T_{12\ldots k} =
L_{21\ldots k1} + \ldots\,$, the symmetry properties of loop-specific building blocks are already present
in OPE residues $\tilde K^m$ and $\tilde J^{mn}$. For instance, we already have an antisymmetric residue $\tilde K_{12}^m = \tilde
K_{[12]}^m$ at rank two whereas the OPE residue $L_{21}$ has to be projected on its antisymmetric
part $T_{12} = L_{21} - L_{(21)}$.

Rank three is the first instance where modifications $Q \tilde K_{123}^m= s_{13} L_{21} \tilde K_3^m+ \ldots$ are
necessary to avoid BRST trivial admixtures $L_{(21)} = -{1 \over 2} Q(A_1 \cdot A_2)$ in the $Q$ variation and
to instead arrive at $Q K_{123}^m= s_{13} T_{12} \tilde K_3^m+ \ldots$ with $L_{21}\mapsto T_{12}$. Hence, the
loop-specific OPE residues $\tilde K_{12\ldots k}^m$ are more closely related to their BRST building blocks
$K_{12\ldots k}^m$ than the tree-level cousins $L_{21\ldots k1 }\leftrightarrow T_{12\ldots k}$.

The BRST variations of OPE residues $\tilde J^{mn}_{12 \ldots k}$ associated with $N^{mn}$ lead to similar
conclusions. Redefinitions $\tilde J^{mn}_{12 \ldots k} \rightarrow J^{mn}_{12 \ldots k}$ are needed in
order to trade $L_{ji\ldots}$ and $\tilde J^{mn}_{ij\ldots}$ present in $Q \tilde J^{mn}_{12\ldots k}$
for $T_{ij\ldots}$ and $J^{mn}_{ij\ldots}$ in $Q J^{mn}_{12\ldots k}$. However, when computing
their BRST variations one must take into account that the building blocks $J^{mn}_{12 \ldots k}$
(or $\tilde J^{mn}_{12 \ldots k}$) always
appear contracted with $(\l\g^m)_\a(\l\g^n)_\b$ because of the rule \empir. So even though one might
naively conclude $Q\tilde J^{mn}_1 = k_1^{[m} (\l\g^{n]}W_1) \neq 0$,
the effective contribution of its BRST variation to an amplitude is $(\l\g_m)_\a(\l\g_n)_\b Q\tilde J^{mn}_1 = 0$.
For any $Q \tilde J^{mn}_{\ldots}$ or $QJ^{mn}_{\ldots}$ displayed in the following, terms that vanish under
contraction with $K^m_{\ldots} K^n_{\ldots} \sim (\lambda \gamma^m)_{\alpha} (\lambda \gamma^n)_{\beta}$ are omitted.

In summary, the $Q$ variations of the BRST building blocks which will appear in loop amplitudes are given
by \Told\ and
\eqnn\QKos
\eqnn\QKts
\eqnn\QKtts
\eqnn\QJos
\eqnn\QJts
\eqnn\QJtts
$$\eqalignno{
%
Q  K^{m}_{12} &= s_{12}  \big(  T_1  K^{m}_2  - T_2 K^{m}_1   \big) &\QKos \cr
Q  K^{m}_{123} &= (s_{123}-s_{12})  \big(  T_{12}  K^{m}_3  - T_3  K^{m}_{12}  \big)  \cr
& \quad +  s_{12}  \big(  T_1  K^{m}_{23}  +    T_{13}   K^{m}_2 - T_{23}  K^{m}_1  - T_2  K^{m}_{13}  \big) & \QKts\cr
Q  K^{m}_{1234} &= (s_{1234} - s_{123})  \big(  T_{123}  K^{m}_4  - T_4  K^{m}_{123}  \big) \cr
& \quad +  (s_{123}-s_{12})  \big(  T_{12}  K^{m}_{34}  + T_{124}  K^{m}_3  - T_{34} K^{m}_{12} - T_3  K^{m}_{124} \big) \cr
& \quad +  s_{12} \big(  T_{134}  K^{m}_2  +  T_{13}  K^{m}_{24}  +  T_{14}  K^{m}_{23}  + T_1  K^{m}_{234}
- T_2 K^{m}_{134} \cr
& \qquad - T_{24}  K^{m}_{13}  - T_{23} K^{m}_{14}  - T_{234} K^{m}_1  \big) &\QKtts\cr
Q  J^{mn}_{12} &= s_{12}  \big(  T_1  J^{mn}_2  -  J^{mn}_1  T_2  \big) &\QJos\cr
Q  J^{mn}_{123} &=  (s_{123}-s_{12})  \big(  T_{12}  J^{mn}_3  -  J^{mn}_{12}  T_3  \big) \cr
&\qquad +  s_{12}  \big(  T_1  J^{mn}_{23} +   T_{13}   J^{mn}_2 - J^{mn}_1  T_{23} - J^{mn}_{13}  T_2  \big)&\QJts \cr
Q  J^{mn}_{1234} & =  (s_{1234} - s_{123})  \big(  T_{123}  J^{mn}_4 - J^{mn}_{123}  T_4  \big) \cr
&\quad  +  (s_{123}-s_{12})  \big(  T_{12}  J^{mn}_{34} + T_{124}  J^{mn}_3 - J^{mn}_{12}  T_{34} - J^{mn}_{124}  T_3  \big) \cr
&\quad  +  s_{12}  \big(  T_{134}  J^{mn}_2  +  T_{13}  J^{mn}_{24}  +  T_{14}  J^{mn}_{23}  + T_1  J^{mn}_{234}\cr
&\qquad -  J^{mn}_{134}  T_2  -  J^{mn}_{13}  T_{24}  -  J^{mn}_{14}  T_{23}  -  J^{mn}_1  T_{234} \big). &\QJtts\cr
}$$
The BRST variations $Q K^{m}_{12\ldots k}$ and $Q J^{mn}_{12\ldots k}$ of the new families can be obtained from $Q T_{12\ldots k}$ by
replacing either the first or the second $T_{ij\ldots }$ on the right hand side by the corresponding $K^{m}_{ij\ldots}$
or $J^{mn}_{ij \ldots }$. This doubles the number of terms in $QK^{m}_{12\ldots k}$ and $QJ^{mn}_{12\ldots k}$ compared
to $QT_{12\ldots k}$, and the two ways of replacing a $T_{\ldots}$ in the BRST variation by $K^m_{\ldots}$ or
$J^{mn}_{\ldots}$ enter with a relative minus sign (where the tree-level
building block $T_{\ldots}$ is always understood to be placed on the left of $K^m_{\ldots}$ and $J^{mn}_{\ldots}$).

The above variations generalizes as follows to rank $k$:
\eqnn\QKJsgen
$$\eqalignno{
Q T_{12{\dots} k} &= \sum_{j=2}^{k} \sum_{\a \in P(\beta_j)}  (s_{12 \ldots j} - s_{12 \ldots j-1})
T_{12{\ldots} j-1,\{\a\}} T_{j, \{\beta_j \backslash \a \} } &\QKJsgen \cr
Q K^m_{12{\dots} k} &= \sum_{j=2}^{k} \sum_{\a \in P(\beta_j)}  (s_{12 \ldots j} - s_{12 \ldots j-1})  \big(  T_{12{\ldots} j-1,\{\a\}} K^m_{j, \{\beta_j \backslash \a \} }  
-  T_{j, \{\beta_j \backslash \a \} } K^m_{12{\ldots} j-1,\{\a\}}\big)\cr
Q J^{mn}_{12{\dots} k} &= \sum_{j=2}^{k} \sum_{\a \in P(\beta_j)}  (s_{12 \ldots j} - s_{12 \ldots j-1})  \big(  T_{12{\ldots} j-1,\{\a\}} J^{mn}_{j, \{\beta_j \backslash \a \} } 
- T_{j, \{\beta_j \backslash \a \} }  J^{mn}_{12{\ldots} j-1,\{\a\}}\big)
}$$
where $V_i \equiv T_i$. The set $\b_j = \{j+1, j+2,{\ldots},k\}$ encompasses the $k-j$ labels
to the right of $j$, and $P(\beta_j)$ denotes its power set.

\subsec{Unified notation for one-loop BRST building blocks}
\subseclab\compactnot

\noindent
For each contraction pattern among integrated vertex OPEs, there are three kinematic factors associated with 
the same $z_i \rightarrow z_j$ singularity structure. This corresponds to the three ways of extracting the 
worldsheet fields $d_\alpha d_\beta N^{mn}$ from three nested $U^j$ OPEs a la \bbdefs. In other words, we have to sum
three different possibilities $(d,d,N)$, $(d,N,d)$ and $(N,d,d)$ to convert
the $U^j$ vertices after $n-4$ OPE fusions into building blocks $K^m K^n J_{mn}$
via $d_\alpha d_\beta N^{mn} \mapsto (\lambda \gamma^m)_\alpha (\lambda \gamma^n)_\beta$
\eqnn\Tijk
$$\eqalignno{
T^i_{a_1\ldots a_p} \, T^j_{b_1\ldots b_q} \, T^k_{c_1\ldots c_r} &\equiv  {2 \over 3}(K^m_{a_1\ldots a_p} \, K^n_{b_1\ldots b_q} \, J^{mn}_{c_1\ldots c_r}
+  K^m_{a_1\ldots a_p} \, J^{mn}_{b_1\ldots b_q} \, K^{n}_{c_1\ldots c_r} \cr
& \ \ \ \ \ \ \ \ \ \ \ \ \ \ +  J^{mn}_{a_1\ldots a_p} \, K^m_{b_1\ldots b_q} \, K^n_{c_1\ldots c_r})\,.&\Tijk
}$$
Note that \Tijk\ is completely symmetric in $i, j,k$ and under moving the $T^i, T^j$ and $T^k$ (which represent either $K^m_{\ldots }$ or $J^{mn}_{\ldots }$) 
across each other, i.e.\ $T^i_{a_1\ldots a_p} T^j_{b_1\ldots b_q} = T^j_{b_1\ldots b_q} T^i_{a_1\ldots a_p}$.
As can be seen from the $K^m_{\ldots} K^n_{\ldots} \sim (\lambda \gamma^m)_\alpha (\lambda \gamma^n)_\beta$ in the definition \Tijk, the combination $T^iT^jT^k$ has ghost-number two. In combination with the unintegrated vertex $V^1$ (or OPE contractions thereof with $U^j$), we arrive at the total ghost number three, as required by the $\langle \lambda^3\theta^5 \rangle =1$ prescription.

In the notation \Tijk, the BRST variations $Q K^m_{12{\dots} k}$ and $Q J^{mn}_{12{\dots} k}$ can be written in a unified way as
$$
Q\, T^{i}_{12{\dots} k} = \sum_{j=2}^{k}\, \sum_{\a \in P(\beta_j)} \, (s_{12 \ldots j} - s_{12 \ldots j-1}) \, 
\big( \, T_{12{\ldots} j-1,\{\a\}}\, T^{i}_{j, \{\beta_j \backslash \a \} } \ 
- \  T_{j, \{\beta_j \backslash \a \} }\, T^i_{12{\ldots} j-1,\{\a\}} \, \big) \ .
$$
Of course, it has to be kept in mind that only expressions containing a full triplet $T^i_{\ldots} T^j_{\ldots} T^k_{\ldots}$ of loop building blocks are well defined. Recall that the set
$\beta_j =  {j + 1,j + 2,...,n}$ encompasses $n-j$ labels to the right of $j$, and $P(\beta_j)$ denotes its power set.

\subsec{Diagrammatic interpretation of the loop building blocks}
\subseclab\diabbs

\noindent
According to our discussion above, the $T^i_{\ldots}$ share the symmetry properties and the structure of their $Q$
variation (in particular the Mandelstam variables therein) with the tree-level building blocks $T_{\ldots}$. So we
also think of $T^i_{12\ldots k}$ together with the $s_{12}^{-1},s_{123}^{-1},\ldots, s^{-1}_{12\ldots k}$ propagators
as representing a cubic tree subdiagram.

Since the conformal weight-one fields from $U_i$ can also be contracted with the $V_1$ vertex, the correlator of \onepresc\
additionally involves tree-level building blocks $T_{d_1 \ldots d_s}$.
Hence, every superspace constituent for
the open string loop amplitude encompasses four tree-level subdiagrams $T_{\ldots} T^i_{\ldots} T^j_{\ldots} T^k_{\ldots}$,
attached to a central vertex with four legs. As a reminder that this is the kinematic factor of a
stringy one-loop diagram, we represent this quartic vertex as a box, see \Tbox.

\ifig\Tbox{Interpretation of $\langle T_{d_1 \ldots d_s} \, T^i_{a_1\ldots a_p} \, T^j_{b_1\ldots b_q} \, T^k_{c_1\ldots c_r} \rangle$
 as the kinematic factor of a box diagram. The four tree subdiagrams at the corners are identified with building blocks $T$ and $T^i$.
}
{\epsfxsize=0.85\hsize\epsfbox{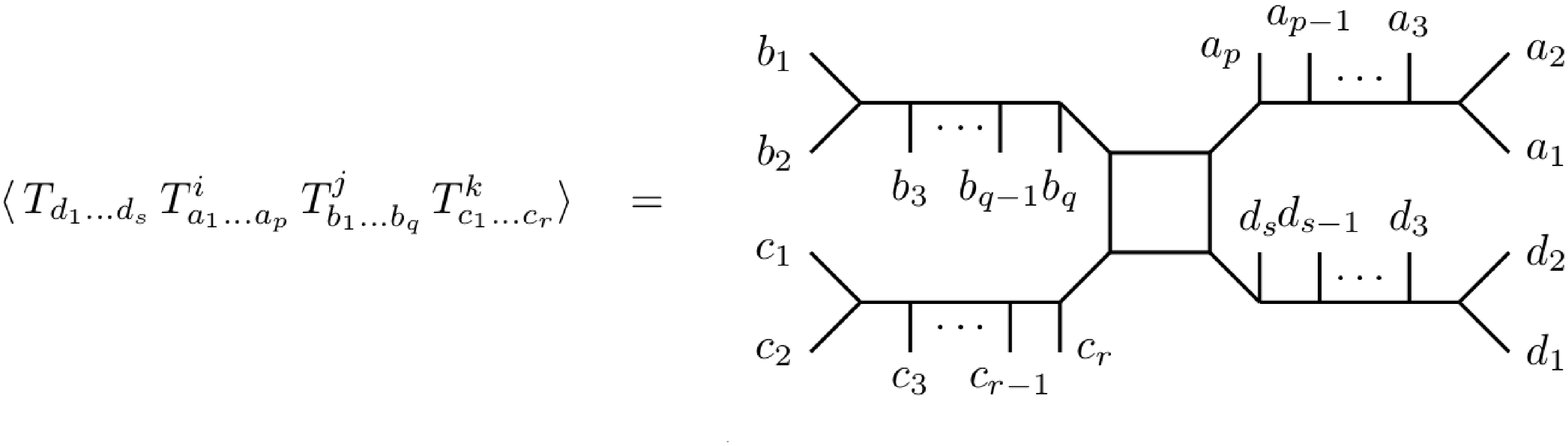}}

We should comment on the shortcoming of the diagrammatic representation \Tbox\ of
$\langle T_{d_1\ldots d_s} \, T^i_{a_1\ldots a_p} \, T^j_{b_1\ldots b_q} \, T^k_{c_1\ldots c_r} \rangle$
that it does not take the asymmetric role of the tree-level BRST building block $T_{d_1 \ldots d_s}$
into account, i.e.\ the lack of $(a_1\ldots a_p) \leftrightarrow (d_1 \ldots d_s)$ symmetry. Moving the one-loop building blocks (i.e.\ the $i,j,k$ superscripts) to different positions
amounts to reshuffling contact terms due to the quartic gluon vertex in the SYM action between cubic graphs.
For instance, the difference $\langle (T_{12} T_3^i - T_3 T_{12}^i) T_4^j T_5^k \rangle$ is proportional
to $s_{12}$ when evaluated in components and therefore cancels the propagator present in the common
diagram\foot{In order to see this, consider the two terms on the right hand side of
$$
0 = \langle Q M_{123}^i T_4^j T_5^k \rangle = {1 \over s_{12}} \langle (T_{12} T_3^i - T_3 T_{12}^i) T_4^j T_5^k \rangle + {1 \over s_{23}} \langle (T_{1} T_{23}^i - T_{23} T_{1}^i) T_4^j T_5^k \rangle  \ ,
$$
see section 4.3 for the definition of $M_{123}^i$. Cancellations between the first term $\sim s_{12}^{-1}$ and
the second one $\sim s_{23}^{-1}$ require that the numerators vanish on the residues of the poles,
i.e.\ $\langle (T_{12} T_3^i - T_3
T_{12}^i) T_4^j T_5^k \rangle \sim s_{12}$. This argument can be easily extended to higher
multiplicity by virtue of $0 = \langle Q M_{1234}^i T_5^j T_6^k \rangle$ and related
expressions.}.

A particular motivation for the suggestive box notation comes from the low energy limit of superstring
amplitudes. After dimensional reduction to four dimensions, they are supposed to reproduce amplitudes of $N=4$ SYM -- see e.g. \BGS\ for a derivation of the four-point box integral in field theory from a $D$-dimensional superstring computation in the $\ap \rightarrow 0$ limit. The fact that only quadruple $T_{\ldots}$ and no triple $T_{\ldots}$
enter the superspace kinematics in the string computation reminds of the ``no triangle'' property of the underlying
field theory \notriangle. In view of these matching structures in loop diagrams of SYM and kinematic constituents of string amplitudes, we found it natural to represent the central
tetravalent vertex gluing together the $T_{\ldots} T^i_{\ldots} T^j_{\ldots} T^k_{\ldots}$ as a box. However,
this does not claim a one-to-one correspondence between a particular superspace kinematic factor and a box
coefficient in field theory. The systematic reproduction of $N=4$ SYM amplitudes via $\ap \rightarrow 0$ limits
of the present results is not addressed in this paper and left for future work instead.

\subsec{Berends--Giele currents for loop amplitudes}

\noindent
As the next hierarchy level of building blocks we define loop-level Berends--Giele currents, $M^i_{12\ldots p}$
encompassing several tree subdiagrams described by $T^i_{a_1\ldots a_p}$.
They are closely related with the field-theory Berends--Giele currents of \BG, as
thoroughly explained (with examples) in \MSSI.
The collections of subdiagrams $M_{12\ldots p} = \sum T_{a_1\ldots a_p} / s^{p-1}$ which were
present in the superspace representations of tree-level amplitudes can be literally carried over
to the CFT ingredients of loop amplitudes. In other words, the tree-level formulae \treeelevena\
and \treetwelve\
directly translate into loop-level analogues
\eqnn\Mitwo
\eqnn\Mithree
\eqnn\Mifour
$$\eqalignno{
M_{12}^{i} &= \frac{ T^{i}_{12}}{s_{12}} &\Mitwo \cr
M_{123}^{i} &= \frac{ T^{i}_{123}}{s_{12} \, s_{123} } \ + \ \frac{ T^{i}_{321}}{s_{23} \, s_{123} } &\Mithree\cr
M_{1234}^{i} &= \frac{1 }{ s_{1234}} \; \bigg(\, \frac{T^{i}_{1234}}{ s_{12}s_{123} } 
\, + \, \frac{T^{i}_{3214}}{ s_{23}s_{123} }\, + \, \frac{T^{i}_{3421}}{ s_{34}s_{234} } \, + \, \frac{T^{i}_{3241} }{s_{23}s_{234} } 
\, + \, \frac{2T^{i}_{12[34]}}{ s_{12}s_{34} } \, \bigg)\, . &\Mifour
}$$
It is a necessary condition for BRST invariance that the kinematic factor in loop amplitudes combines the $T^i_{\ldots}$ to full-fledged 
Berends--Giele currents $M^i_{\ldots}$. This can be seen from their covariance under $Q$ with no additional Mandelstam factors
\eqn\QMijk{
\eqalign{
Q \, M^i_{a_1\ldots a_p} \, M^j_{b_1\ldots b_q} \, M^k_{c_1\ldots c_r}  &=
\sum_{\ell=1}^{p-1} \, \big( \, M_{a_1\ldots a_\ell} \, M_{a_{\ell+1} \ldots a_p}^i
-   M_{a_{\ell+1} \ldots a_p} \, M_{a_1\ldots a_\ell}^i \, \big) \, M^j_{b_1\ldots b_q} \, M^k_{c_1\ldots c_r}  \cr
&+ \ \sum_{\ell=1}^{q-1} \,\big( \, M_{b_1\ldots b_\ell} \, M_{b_{\ell+1} \ldots b_q}^j
-   M_{b_{\ell+1} \ldots b_q} \, M_{b_1\ldots b_\ell}^j \, \big) \, M^i_{a_1\ldots a_p} \, M^k_{c_1\ldots c_r}  \cr
&+ \ \sum_{\ell=1}^{r-1} \, \big( \, M_{c_1\ldots c_\ell} \, M_{c_{\ell+1} \ldots c_r}^k
-   M_{c_{\ell+1} \ldots c_r} \, M_{c_1\ldots b_\ell}^k \, \big) \, M^i_{a_1\ldots a_p}
\,M^j_{b_1\ldots b_q}
}}
in close analogy to \treefourteen\ at tree-level. Apart from their simple $Q$ variations \QMijk\ and their matching with the cubic diagram content of SYM trees, the definitions \Mitwo\ to \Mifour\ can be motivated by worldsheet integral manipulations \tradetree\ and \tradeloop. As detailed in section 5.2, the $M^i_{12\ldots p}$ naturally build up once the Green functions are arranged into a form which facilitates integration by parts.

One could also have
defined Berends--Giele currents $M^m_{12 \ldots k}$ and $M^{mn}_{12 \ldots k}$ for the individual building blocks $K^m_{12 \ldots k}$ and
$J^{mn}_{12 \ldots k}$ to later define $M^i_{12 \ldots k}$ by combining them following the
pattern seen in \Tijk:
\eqnn\Mijk
$$\eqalignno{
M^i_{a_1\ldots a_p} \, M^j_{b_1\ldots b_q} \, M^k_{c_1\ldots c_r} &\equiv{}  {2 \over 3} (M^m_{a_1\ldots a_p} \, M^n_{b_1\ldots b_q} \, M^{mn}_{c_1\ldots c_r}
 +  M^m_{a_1\ldots a_p} \, M^{mn}_{b_1\ldots b_q} \, M^{n}_{c_1\ldots c_r} \cr
 &\quad{}\quad{} \quad{}\quad{}+  M^{mn}_{a_1\ldots a_p} \, M^m_{b_1\ldots b_q} \, M^{n}_{c_1\ldots c_r}). &\Mijk
}$$
The combinatorics of zero mode saturation implies that the end result for amplitudes always involves a sum of all the three terms on the right hand side. That is why we will always use the notation on the left hand side of \Mijk\ in the rest of this work.

\subsec{BRST-invariant kinematics for loop amplitudes}
\subseclab\Csubsec

\noindent
Amplitudes computed with the pure spinor formalism give rise to superspace kinematic factors in the cohomology of the BRST operator. We have motivated $K$ and $J$ building blocks from their appearance in the iterated OPEs of integrated
vertex operators (along with the $d_\a$ and $N_{mn}$ worldsheet fields) and argued that their combinations $M^i_{a_1 \ldots a_p} M^j_{b_1 \ldots b_q}M^k_{c_1 \ldots c_r}$ have covariant BRST variations \QMijk\ connecting different pole channels. Given the strong constraints which BRST invariance
imposes on tree-level SYM amplitudes -- see subsection \cohoProbSec\ -- it is natural to explore the $Q$ cohomology
using the one-loop building blocks. In this subsection we will write down BRST invariants constructed from the above elements dictated by the minimal formalism. This amounts to anticipating the admissible kinematic structure in the result of the CFT computation of one-loop scattering amplitudes.

As mentioned in subsection \diabbs, the one-loop prescription \onepresc\ containing
one unintegrated vertex operator $V_1$ implies that one tree-level building block $T_{1 \ldots }$ (combined to a Berends--Giele current $M_{\ldots 1\ldots}$) has
to appear in these BRST invariants, in addition to three one-loop constituents $M^i_{\ldots}M^j_{\ldots }M^k_{ \ldots }$.
Hence, $Q$ invariant loop kinematics must be built from $M_{d_1\ldots d_s} M^i_{a_1 \ldots a_p} M^j_{b_1 \ldots b_q } M^k_{c_1 \ldots c_r}$
with $1 \in \{ d_1,\ldots, d_s\}$. The diagrammatic interpretation of such a term follows from the fact that Berends--Giele
currents represent color-ordered tree amplitudes with one off-shell leg, see \Mbox.

As explained before at the level of $T_{\ldots}T^i_{\ldots}T^j_{\ldots}T^k_{\ldots}$, this diagram does not take the asymmetry in $M_{d_1\ldots d_s} M^i_{a_1 \ldots a_p} \leftrightarrow M_{a_1 \ldots a_p}  M^i_{d_1\ldots d_s} $ into account. The difference between the two $(M_{\ldots},M^i_{\ldots})$ assignments corresponds to a reshuffling of contact terms in the cubic subdiagrams at the corners of the box.

\ifig\Mbox{Interpretation of $\langle \, M_{d_1 \ldots d_s} \, M^i_{a_1\ldots a_p} \, M^j_{b_1\ldots b_q} \, M^k_{c_1\ldots c_r} \, \rangle$
as four Berends--Giele currents (i.e. collections of tree subdiagrams guided by color-ordered tree-level amplitudes), glued together by a central quartic ``box''-vertex.
}
{\epsfxsize=0.80\hsize\epsfbox{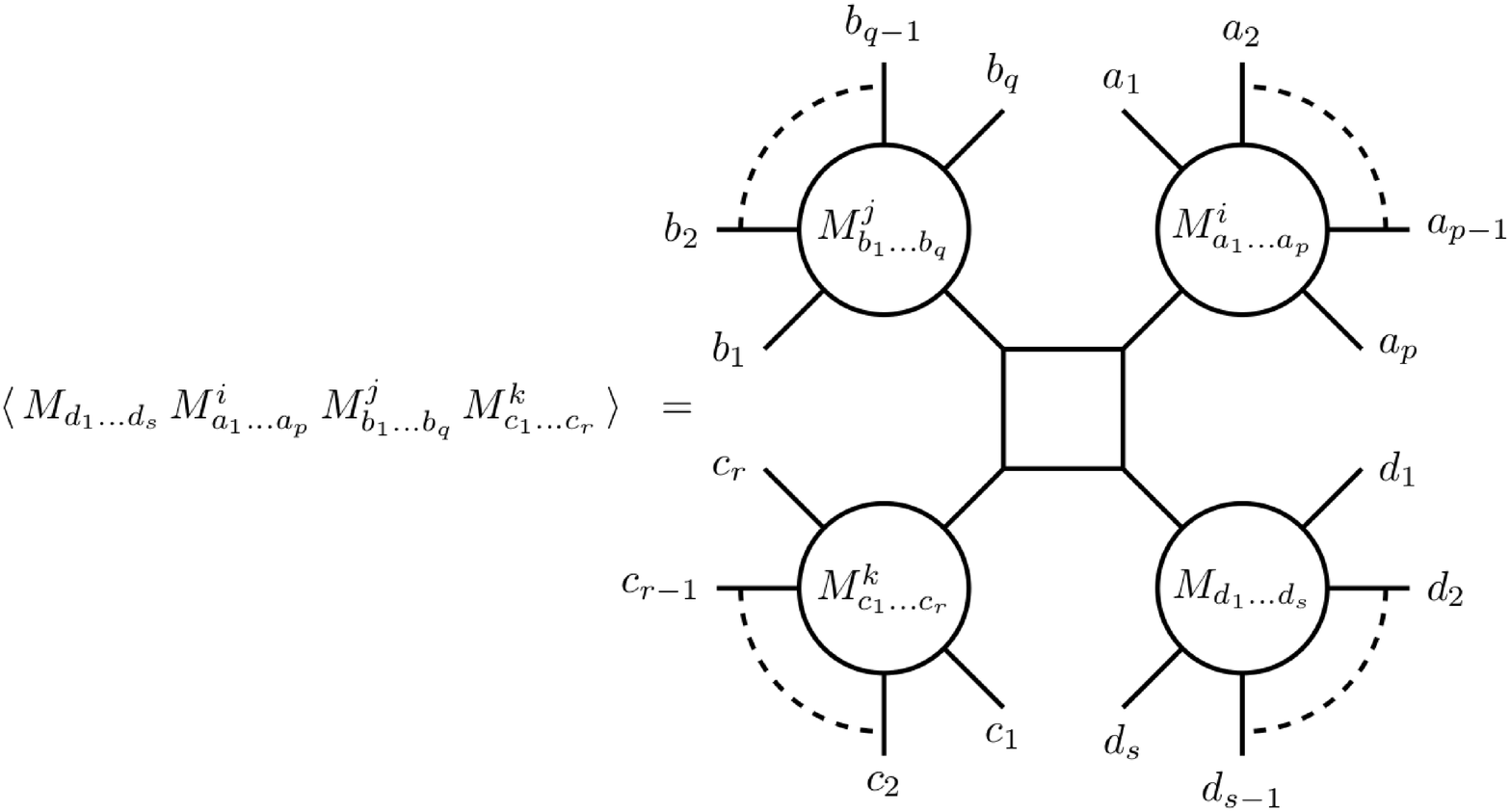}}

\noindent
In the following, we shall give a list of BRST invariants built from $M_{\ldots} M^i_{ \ldots }
M^j_{ \ldots } M^k_{ \ldots }$ up to seven-points. They are denoted by $C_{1,a_1 \ldots a_p,b_1
\ldots b_q,c_1 \ldots c_r}$ according to their first term $V^1 M^i_{a_1 \ldots a_p} M^j_{b_1
\ldots b_q } M^k_{c_1 \ldots c_r}$ where the unintegrated vertex is unaffected by OPEs:

\eqnn\Cooo
\eqnn\Cdoo
\eqnn\Ctoo
\eqnn\Cddo
\eqnn\Cfoo
\eqnn\Ctdo
\eqnn\Cddd
$$\eqalignno{
C_{1,2,3,4} &=   M_1 \, M_2^i \, M_3^j \, M_4^k  &\Cooo\cr
C_{1,23,4,5} &=   M_{1} \, M_{23}^i \, M_{4}^j \, M_{5}^k \ + \ M_{12} \, M_{3}^i \, M_{4}^j \, M_{5}^k \ + \ M_{31} \, M_{2}^i \, M_{4}^j \, M_{5}^k 
&\Cdoo \cr
C_{1,234,5,6} &=  M_{1} \, M_{234}^i \, M_{5}^j \, M_{6}^k \ + \ M_{123} \, M_{4}^i \, M_{5}^j \, M_{6}^k \ + \ M_{412} \, M_{3}^i \, M_{5}^j \, M_{6}^k  \cr
& \  + \ M_{341} \, M_{2}^i \, M_{5}^j \, M_{6}^k \ + \ M_{12} \, M_{34}^i \, M_{5}^j \, M_{6}^k \ + \ M_{41} \, M_{23}^i \, M_{5}^j \, M_{6}^k 
& \Ctoo \cr
C_{1,23,45,6} &=  M_{1} \, M_{23}^i \, M_{45}^j \, M_6^k \ + \ M_{12} \, M_{3}^i \, M_{45}^j \, M_{6}^k \ - \ M_{13} \, M_{2}^i \, M_{45}^j \, M_{6}^k  \cr
& \ + \ M_{14} \, M_{23}^i \, M_{5}^j \, M_{6}^k \ - \ M_{15} \, M_{23}^i \, M_{4}^j \, M_{6}^k \ + \ M_{413} \, M_{2}^i \, M_{5}^j \, M_{6}^k  \cr
& \ + \ M_{512} \, M_{3}^i \, M_{4}^j \, M_{6}^k \ - \ M_{412} \, M_{3}^i \, M_{5}^j \, M_{6}^k \ - \ M_{513} \, M_{2}^i \, M_{4}^j \, M_{6}^k 
&\Cddo \cr
%
%
%
C_{1,2345,6,7} &= M_1 \, M_{2345}^i \, M_6^j \, M_7^k \ + \ M_{512} \, M_{34}^i \, M_6^j \, M_7^k \ + \ \big[ \, M_{12} \, M_{345}^i  \ + \ M_{123} \, M_{45}^i  \cr
& \ + \  M_{1234} \, M_5^i  \ + \ M_{5123} \, M_4^i  \ - \ (2,3 \leftrightarrow 5,4) \, \big] \, M_6^j \, M_7^k& \Cfoo \cr
C_{1,234,56,7} &=  M_1 \, M_{234}^i \, M_{56}^j \, M_7^k \ + \ M_{214} \, M_3^i \, M_{56}^j \, M_7^k \ + \ (M_{15} \,  M_6^i \, - \, M_{16} \,  M_5^i) \, M_{234}^j \, M_7^k \cr
& \  + \ \big[ \, M_{12} \, M_{34}^i \ + \ M_{123} \, M_4^i \ + \ (2\leftrightarrow4) \, \big] \, M_{56}^j \, M_7^k \ + \ \big\{ \, \big[ \, M_{612} \, M_{34}^i \, M_5^j   \cr
& \  + \ M_{6123} \, M_4^i \, M_5^j   \ + \ M_{5124}  \, M_3^i \, M_6^j \ +  \ (2\leftrightarrow4) \, \big] \ - \ (5\leftrightarrow6) \, \big\} \, M_7^k  &\Ctdo \cr
%
C_{1,23,45,67} &=  M_1 \, M_{23}^i \, M_{45}^j \, M_{67}^k \ + \ \big[ \, ( M_{12} \, M_3^i \ - \ M_{13} \, M_2^i ) \, M_{45}^{j} \, M_{67}^k  \ + \ {\rm cyc}(23,45,67) \, \big]  \cr
& \  + \ \big\{ \, \big[ \, M_{217} \, M_3^i  \, M_{6}^j \ - \ (2\leftrightarrow3) \ - \ (6\leftrightarrow7) \, \big] \, M_{45}^k \ + \ {\rm cyc}(23,45,67) \, \big\}
 \cr
& \  + \ \big[ \, (M_{7135} + M_{7153} ) \, M_2^i \, M_4^j \, M_6^k \ - \ (2\leftrightarrow3) \ - \ (4\leftrightarrow5) \ - \ (6\leftrightarrow7) \, \big] \, . &\Cddd
}$$
Eight-point amplitudes contain four topologies $C_{1,23456,7,8}, \,C_{1,2345,67,8}, \, C_{1,234,56,78}$ and $C_{1,234,567,8}$ of BRST invariants. They are expanded in appendix \Chigher.

\subsec{Symmetry properties of the BRST invariants}
\subseclab\CsymSec

\noindent
In this subsection, we examine the symmetry properties of the BRST invariants of the previous subsection and determine
the number of independent permutations (at least under linear relations with constant coefficients). In
particular, we will argue that the $C_{1,\ldots}$ with label ``1'' in the first entry form a suitable basis. This ties
in with the one-loop prescription \onepresc\ for string amplitudes: The special role of the unintegrated vertex $V_1$
implies that only $C_{1,\ldots}$ can appear in the CFT computation, and these ingredients must be able to capture any
permutation $C_{i \neq1,\ldots}$ via linear combination.

In order to see that the reduction to $C_{1,\ldots}$ is possible, first note that the invariants
$C_{1,a_1\ldots a_p,b_1\ldots,b_q,c_1\ldots c_r}$ inherit the symmetry properties
of the Berends--Giele currents for each of individual three sets of labels $(a_1, \ldots,a_p)$, $(b_1, \ldots,b_q)$ and $(c_1, \ldots,c_r)$, i.e.
\eqn\BGsymnoC{
M_{12\ldots p}^i = (-1)^{p-1} M_{p\ldots 21} ,\qquad M_{\{ \beta\}1\{\alpha \}} =  (-1)^{n_{\beta}}\hskip-0.7cm 
\sum_{\sigma\, \in\, {\rm OP} ( \{\alpha \}, \{\beta^T \})} \hskip-0.4cm M^i_{1 \{\sigma \}}
}
directly carry over to
\eqnn\BGsymC
$$\eqalignno{
C_{1,a_1 a_2\ldots a_p,b_1\ldots b_q,c_1 \ldots c_r} &= (-1)^{p-1}C_{1,a_p \ldots a_2 a_1,b_1\ldots b_q,c_1 \ldots c_r} &\BGsymC  \cr
C_{1,\{ \beta\}i\{\alpha \},b_1\ldots b_q,c_1 \ldots c_r} &=\, (-1)^{n_{\beta}}\hskip-0.7cm 
\sum_{\sigma\, \in\, {\rm OP} ( \{\alpha \}, \{\beta^T \})} \hskip-0.4cm C_{1,i \{\sigma \},b_1\ldots b_q,c_1 \ldots c_r}\, . \cr
}$$
The notation for the sets $\a,\b,\s$ is the usual one appearing in the Kleiss--Kuijf relation \KK. The latter implies the subcyclic property (or photon decoupling identity)
\eqn\nonumbb{
\sum_{\sigma \in {\rm cyclic}} \!\!\!\! C_{1,\s(a_1a_2\ldots a_p),b_1\ldots b_q,c_1 \ldots c_r} =0 .
}
However, the above symmetries do not relate $C_{i,\ldots}$ to $C_{j\neq i,\ldots}$ (with different labels $i,j$ in the first slot). Equations of that type follow from the BRST cohomology of pure spinor superspace, i.e. from the vanishing of BRST exact terms at ghost number three,
\eqn\Bexacv{
\langle Q \, M^i_{a_1\ldots a_p} \,M^j_{b_1\ldots b_q}  \, M^k_{c_1\ldots c_r}\rangle = 0.
}
The left hand side is always organized into
linear combinations of $C$'s, let us illustrate this by examples: The four-point BRST invariant turns out to be totally symmetric,
\eqn\fptCsym{
0 = \langle Q \, M_{12}^i \, M_3^j \, M_4^k \rangle  \;\Rightarrow\;  C_{1,2,3,4} = C_{2,1,3,4} ,
}
and five-point invariants can be reduced to $C_{1,ij,k,l} = C_{[1,ij],k,l}$ by means of
\eqnn\Cfiveone
$$\eqalignno{
0 &= \langle Q \, M_{123}^i \, M_4^j \, M_5^k \rangle  \;\Rightarrow\;
C_{1,23,4,5} = -\,  C_{3,21,4,5} &\Cfiveone\cr
0 &= \langle Q \, M_{12}^i \, M_{34}^j \, M_5^k \rangle  \;\Rightarrow\;
C_{2,34,1,5}  -  C_{1,34,2,5}  +  C_{4,12,3,5}  -  C_{3,12,4,5} = 0 .&\cr
}$$
At six-points, there are three different topologies of BRST exact quantities
\eqnn\Csixone
$$\eqalignno{
0 &= \langle Q \, M_{1234}^i \, M_5^j \, M_6^k \rangle  \;\Rightarrow\;  C_{1,234,5,6} = C_{4,321,5,6} &\Csixone\cr
0 &= \langle Q \, M_{123}^i \, M_{45}^j \, M_6^k \rangle  \;\Rightarrow\;  C_{1,23,45,6}  +  C_{3,21,45,6}  +  C_{4,123,5,6} - C_{5,123,4,6} = 0 \cr
0 &= \langle Q \, M_{12}^i \, M_{34}^j \, M_{56}^k \rangle  \;\Rightarrow\;  C_{1,34,56,2}  -  C_{2,34,56,1}  +  C_{3,12,56,4} - C_{4,12,56,3} \cr
&\hskip3.8cm +  C_{5,12,34,6} - C_{6,12,34,5} = 0,
}$$
and the resulting equations are sufficient to decompose any given $C_{i,jk,lm,n}$ or $C_{i,jkl,m,n}$ into a basis of $C_{1, \ldots}$. A similar recursive argument applies at seven-points due to four types of equations:
\eqnn\Csevenone
$$\eqalignno{
0 &= \langle Q \, M_{12345}^i \, M_6^j \, M_7^k \rangle  \;\Rightarrow\;  C_{1,2345,6,7} = - \, C_{5,4321,6,7} &\Csevenone\cr
0 &= \langle Q \, M_{1234}^i \, M_{56}^j \, M_7^k \rangle  \;\Rightarrow\;  C_{5,1234,6,7} - C_{6,1234,5,7} + C_{1,234,56,7} - C_{4,321,56,7} = 0  \cr
0 &= \langle Q \, M_{123}^i \, M_{456}^j \, M_{7}^k \rangle  \;\Rightarrow\;  C_{1,456,23,7} + C_{3,456,21,7} + C_{4,123,56,7} + C_{6,123,54,7} = 0 \cr
0 &= \langle Q \, M_{123}^i \, M_{45}^j \, M_{67}^k \rangle  \;\Rightarrow\;  C_{1,23,45,67} + C_{3,21,45,67} + C_{4,123,67,5} - C_{5,123,67,4} \cr
&\hskip4cm + C_{6,123,45,7} - C_{7,123,45,6} = 0
}$$
In order to count the number of independent $C_{1,a_1\ldots a_p,b_1\ldots,b_q,c_1\ldots c_r}$, one should keep in mind that there are $(p-1)!$ independent components in $(a_1 \ldots a_p)$ due to Berends--Giele symmetries, the same number of cyclically
inequivalent configurations.
Hence, the number of independent $C_{1,a_1\ldots a_p,b_1\ldots,b_q,c_1\ldots c_r}$ in $n$-point amplitudes (where $p+q+r=n-1$) is
given by the number of ways to distribute $n-1$ elements to three cycles $(a_1\ldots a_p), (b_1\ldots,b_q)$ and $(c_1\ldots c_r)$. 
This is the defining property of the unsigned Stirling numbers of first kind $S^{n-1}_3$,
\eqnn\numCs
$$\eqalignno{
\# &(C_{1,a_1\ldots a_p,b_1\ldots,b_q,c_1\ldots c_r}) \, \Big|_{p+q+r=n-1} =  S^{n-1}_3 &\numCs \cr
 & = 1, 6, 35, 225, 1624, 13132, 118124, 1172700, 12753576, 150917976, \ldots \qquad n\ge 4,
 }$$
and the following table gathers examples of how individual topologies (i.e. different triplets of $p,q,r$ with constant sum) contribute to the Stirling numbers:

\bigskip
\moveright 2 cm
\vbox{\offinterlineskip
\halign{%
\strut \hfil\quad # \quad\hfil &\vrule \hfil \quad # \quad \hfil & \vrule \hfil  \quad \phantom{\Big|} # \quad \hfil & \quad # \hfil \cr
$n$ & $C$-topology & \# components \span\cr
\noalign{\hrule}
4 & $C_{1,2,3,4}$ & 1 & $= \ 1$ \cr
\noalign{\hrule}
5 & $C_{1,23,4,5}$ &  $4\choose2$ & $= \ 6$ \cr
\noalign{\hrule}
6 & $C_{1,234,5,6}$ & ${5\choose2} \cdot 2$ & $=\ 20$ \cr
6 & $C_{1,23,45,6}$ &  $5 \cdot 3$         & $=\ 15$ \cr
\noalign{\hrule}
7 & $C_{1,2345,6,7}$ & ${6\choose2} \cdot 6$ & $=\ 90$ \cr
7 & $C_{1,234,56,7}$ & ${6\choose3} \cdot 3 \cdot 2$ & $=\ 120$ \cr
7 & $C_{1,23,45,67}$ & $5!!$ & $=\ 15$ \cr
\noalign{\hrule}
8 & $C_{1,23456,7,8}$ & ${7\choose2} \cdot 24$ & $=\ 504$ \cr
8 & $C_{1,2345,67,8}$ & ${7\choose3} \cdot 3 \cdot 6$ & $=\ 630$ \cr
8 & $C_{1,234,56,78}$ & ${7\choose3} \cdot 3 \cdot 2$ & $=\ 210$ \cr
8 & $C_{1,234,567,8}$ & ${1\over 2} \cdot 7 \cdot {6\choose3} \cdot 2 \cdot 2$ & $=\ 280$\cr
}}
\medskip
{\leftskip=20pt\rightskip=20pt\noindent\ninepoint
{\bf Table 1.} The number of independent components of $C$ topologies up to $n=8$
\par }
\bigskip

\newsec{One-loop amplitudes in pure spinor superspace}
\seclab\OneLoopAmpSec

\noindent
The pure spinor BRST cohomology of building blocks will now be used
to deduce the form of the $n$-point one-loop open superstring amplitudes. Apart from the four- and five-point
amplitudes which were previously computed without explicit use of
building blocks \refs{\MPS,\oneloopM,\fiveptone}, the results for higher-points are strongly
guided by their cohomology properties.

From the discussion of section \secOnePS, the $n$-point kinematic factor for one-loop amplitudes
is given, up to OPE terms with the b-ghost, by the following correlator
\eqn\correl{
K_n = \langle V_1 U_2(z_2)U_3(z_3) \ldots U_n(z_n)\rangle_{ddN}
}
where the subscript $ddN$ is a reminder that the substitution rule \empir\ must be
applied.
It is easy to see that $n-4$ OPE contractions
among the vertex operators will have to be performed before the zero-mode combination $d_\a d_\b N^{mn}$
can be extracted. Throughout this section, we will immediately trade all the OPE residues $L_{2131\ldots \ell 1}$ and
$\tilde K^m_{\ell+1\ldots p} ,\tilde J^{mn}_{\ell+1\ldots p}$ for the corresponding BRST building
blocks $T_{12\ldots \ell}$ and $K^m_{\ell+1\ldots p} , J^{mn}_{\ell+1\ldots p}$. Experience with
the tree-level computation \refs{\MSSI,\towardsFT,\MSSTsix} shows that their difference can only contribute to BRST
trivial kinematics and drops out through total worldsheet derivatives.

The calculation of the kinematic factor will be
divided into three steps:
\medskip
\parindent=32pt
\item{1.} Express the correlator \correl\ in terms of BRST building blocks
\item{2.} Group these building blocks into Berends--Giele currents
\item{3.} Use integration by parts to combine different currents to BRST invariants $C_{1,\ldots}$
\medskip
\parindent=20pt
\noindent
Starting from six-points, we will use BRST invariance as an extra input in steps 1 and 2 to fix
certain parts of the correlator: This concerns the failure of $\eta_{ij} \eta_{jk}$ products to
obey the partial fraction identity $(z_{ij}z_{jk})^{-1} + {\rm cyc}(ijk) = 0$ from tree-level. This relation
plays an important role for the basis reduction of worldsheet integrals at tree-level, see \MSSII. After these
steps are performed the correlator \correl\ becomes a linear combination of the BRST
invariants $C_{1,\ldots}$ constructed in subsection \Csubsec, which we can regard as the one-loop
analogue of the tree-level subamplitudes ${\cal A}^{\rm YM}$. Hence, up to the aforementioned
partial fraction subtlety, the one-loop strategy follows the same logical step as the calculation
of the $n$-point tree amplitude in \MSSI.

Imposing BRST invariance from the beginning makes us blind to the hexagon anomaly in $D=10$
dimensions arising from the boundary of the $t$ integration \GSanom, so in our method we are not
able to reproduce the superspace anomaly computed in \BerkovitsBK. In other words, we compute
the non-anomalous or BRST-invariant part of the amplitude.

Although our final result for $K_n$ won't include leg one on the same footing as all the others, we will prove its
hidden total symmetry in subsection \auchnochmal. The basis choice $C_{1,\ldots}$ for the kinematic constituents
reflects the special role played by leg one entering the computation through the unintegrated $V^1$ vertex. New
cross-connections to color structures at tree-level will be pointed out in section \harmcki\ which trivialize the
outstanding symmetry proof.

\subsec{Step 1: CFT correlator in terms of building blocks}

\noindent
Using the definitions of the building blocks, the CFT correlator \correl\ will
encompass all possible combinations of building blocks
allowed by its total permutation symmetry in $(234 \ldots n)$.
As mentioned before, $n-4$ OPE contractions must be performed before the $ddN$ zero-modes
can be extracted and leave a triplet of building blocks $T^i_{\ldots} T^j_{\ldots} T_{\ldots}^k$ behind. 

As a trivial starting example, the four-point kinematic factor does not
require any OPE and can be written down immediately using the definitions \Kone, \Jone\ and \Tijk
\eqn\fourCFT{
K_4 = \langle V_1 U_2 U_3 U_4 \rangle_{ddN} =
\langle V_1 T_2^i T_3^j T_4^k \rangle\,.
}
The ten possible OPEs in the five-point kinematic factor give rise to two classes of
terms, depending on whether the contraction involves the unintegrated vertex or not:
\eqnn\fiveCFT
$$\eqalignno{
K_5 &= \langle V_1 U_2 U_3 U_4 U_5 \rangle_{ddN}\cr
& = \langle \, V_1 \underbrace{ U_2\,  U_3} \,  U_4 \, U_5  \, \rangle \ + \ 5 \, \rm{permutations}\, (23 \leftrightarrow 24,25,34,35,45) \cr
& \ \ \ \ \ + 
 \langle \underbrace{V_1 \, U_2} \, U_3 \, U_4 \, U_5 \rangle \ + \ 3 \, \rm{permutations} \, (2\leftrightarrow 3,4,5)
 &\fiveCFT
}$$
The resulting BRST building blocks are
\eqnn\fivo
\eqnn\fivt
$$\eqalignno{
\langle \underbrace{V_1 \, U_2} \, U_3 \, U_4 \, U_5 \rangle &=
\eta_{12}\, \langle T_{12}\, T^i_3 T^j_4 T^k_5\rangle  &\fivo\cr
\langle \, V_1 \underbrace{ U_2\,  U_3} \,  U_4 \, U_5  \, \rangle &=
 \eta_{23}\,\langle  V_{1}\,   T^i_{23}  T^j_4  T_5^k \rangle ,   &\fivt\cr
}$$
and the validity of the replacement $L_{21} \mapsto T_{12}$ follows from BRST-closedness of $T^i_3 T^j_4 T^k_5$.

Applying this kind of analysis to the six-point correlator leads to an ambiguity:
\eqnn\sixCFT
$$\eqalignno{
K_6 &= \langle V_1 U_2 U_3 U_4 U_5 U_6\rangle_{ddN}\cr
&=\langle \, \underbrace{V_1 \, U_2 \, U_3} \, U_4 \, U_5 \, U_6 \, \rangle +
\langle \, \underbrace{V_1 \, U_2} \, \underbrace{U_3 \, U_4} \, U_5 \, U_6 \, \rangle\cr
& \quad + \langle \, V_1 \, \underbrace{U_2 \, U_3 \, U_4} \, U_5 \, U_6 \, \rangle +
\langle \, V_1 \, \underbrace{U_2 \, U_3} \, \underbrace{U_4 \, U_5} \, U_6 \, \rangle
\cr
& \quad + \, {\rm permutations} \ + \ \eta_{ijk} (\ldots).
&\sixCFT
}$$
We firstly find those contractions which closely resemble the tree-level procedure (up to $z_{ij}^{-1} \mapsto \eta_{ij}$ and the new building blocks $T^i_{\ldots} T^j_{\ldots} T^k_{\ldots}$)
\eqna\topu
$$\eqalignno{
\langle \, \underbrace{V_1 \, U_2 \, U_3} \, U_4 \, U_5 \, U_6 \, \rangle &=
\eta_{12}\eta_{23}\, \langle T_{123}\, T^i_4 T^j_5 T^k_6\rangle &\topu a\cr
\langle \, \underbrace{V_1 \, U_2} \, \underbrace{U_3 \, U_4} \, U_5 \, U_6 \, \rangle 
&= \eta_{12}\eta_{34}\,\langle  T_{12}\,   T^i_{34}  T^j_5  T_6^k \rangle   &\topu b\cr
\langle \, V_1 \, \underbrace{U_2 \, U_3 \, U_4} \, U_5 \, U_6 \, \rangle 
&= \eta_{23}\eta_{34}\, \langle V_1 \,  T_{234}^i  T_5^j  T_6^k\rangle    &\topu c\cr
\langle \, V_1 \, \underbrace{U_2 \, U_3} \, \underbrace{U_4 \, U_5} \, U_6 \, \rangle 
&=  \eta_{23}\eta_{45}\, \langle V_1 \, T_{23}^i T_{45}^j T_6^k\rangle. &\topu d\cr
}$$
But in addition to that, the correlator could contain terms with worldsheet functions
\eqn\etaijkdef{
\eta_{ijk} = \eta_{ij} \, \eta_{ik} +  \eta_{ji} \, \eta_{jk} +  \eta_{ki} \, \eta_{kj},
}
which are invisible in the $z_i \rightarrow z_j$ limit since $(z_{ij} z_{ik})^{-1} + {\rm
cyc}(ijk) =0$. These parts of the CFT correlator cannot be fixed on the basis of the
leading OPE singularity and symmetry arguments in $(23\ldots n)$. Instead, we will keep
them undetermined for the moment and use BRST invariance in the following subsections to argue
their absence in the end result.
The precise way to combine permutations will be
discussed in the next subsection.

Similarly, the seven-point kinematic factor receives contributions from
\eqnn\sevCFT
$$\eqalignno{
K_7 &= \langle V_1 U_2 U_3 U_4 U_5 U_6 U_7\rangle_{ddN}\cr
&=\langle \, \underbrace{V_1 \, U_2 \, U_3  \, U_4} \, U_5 \, U_6  U_7\, \rangle +
\langle \, \underbrace{V_1 \, U_2\,  U_3} \, \underbrace{U_4 \, U_5} \, U_6 \, U_7 \, \rangle  + \langle \, \underbrace{V_1 U_2}\, \underbrace{U_3 \, U_4 \, U_5} \, U_6 \, U_7 \, \rangle \cr
& \quad +
\langle \, \underbrace{ V_1 U_2}\, \underbrace{U_3 \, U_4} \, \underbrace{U_5 \, U_6} \, U_7 \, \rangle + \langle \, V_1 \underbrace{ U_2 U_3 \, U_4 \, U_5} \, U_6 \, U_7 \, \rangle
+ \langle \, V_1 \underbrace{ U_2 U_3 \, U_4} \underbrace{U_5 \, U_6} \, U_7 \, \rangle\cr
&\quad + \langle \, V_1 \underbrace{ U_2 U_3} \underbrace{U_4 \, U_5} \, \underbrace{U_6 \, U_7} \, \rangle
+ \, {\rm permutations} \ + \ \eta_{ijk} (\ldots), &\sevCFT\cr
}$$
where the seven different types of OPEs yield
\eqna\sevtops
$$\eqalignno{
\langle \, \underbrace{V_1 \, U_2 \, U_3  \, U_4} \, U_5 \, U_6  U_7\, \rangle &= \eta_{12}
\eta_{23} \eta_{34} \langle T_{1234} T_5^i T_6^j T_7^k \rangle &\sevtops a\cr
\langle \, \underbrace{V_1 \, U_2\,  U_3} \, \underbrace{U_4 \, U_5} \, U_6 \, U_7 \, \rangle &=
\eta_{12} \eta_{23} \eta_{45} \langle T_{123} T_{45}^i T_6^j T_7^k \rangle &\sevtops b\cr
\langle \, \underbrace{V_1 U_2}\, \underbrace{U_3 \, U_4 \, U_5} \, U_6 \, U_7 \, \rangle &=
\eta_{12} \eta_{34} \eta_{45} \langle T_{12} T_{345}^i T_6^jT_7^k \rangle &\sevtops c\cr
\langle \, \underbrace{ V_1 U_2}\, \underbrace{U_3 \, U_4} \, \underbrace{U_5 \, U_6} \, U_7 \, \rangle&=
\eta_{12} \eta_{34} \eta_{56} \langle T_{12} T_{34}^i T_{56}^j T_7^k \rangle &\sevtops d\cr
\langle \, V_1 \underbrace{ U_2 U_3 \, U_4 \, U_5} \, U_6 \, U_7 \, \rangle & =
\eta_{23} \eta_{34} \eta_{45} \langle V_1 T_{2345}^i T_6^j T_7^k \rangle &\sevtops e\cr
\langle \, V_1 \underbrace{ U_2 U_3 \, U_4} \underbrace{U_5 \, U_6} \, U_7 \, \rangle &=
\eta_{23} \eta_{34} \eta_{56} \langle V_1 T_{234}^i T_{56}^j T_7^k \rangle &\sevtops f\cr
\langle \, V_1 \underbrace{ U_2 U_3} \underbrace{U_4 \, U_5} \, \underbrace{U_6 \, U_7} \, \rangle &=
\eta_{23} \eta_{45} \eta_{67} \langle V_1 T_{23}^i T_{45}^j T_{67}^k \rangle . &\sevtops g\cr
}$$
These six- and seven-point cases give an idea of the general pattern for the $n$-point correlator: The kinematic factor $K_n$ encompasses all tree-level
building blocks involving the unintegrated vertex $\eta_{12} \eta_{23} \ldots \eta_{\ell-1,\ell} T_{12\ldots \ell}$,
multiplied with all the possible topologies of $(\eta^{p-\ell-1}
T_{\ell+1\ldots p}^i)(\eta^{q-p-1} T^j_{p+1\ldots q })( \eta^{n-q-1} T^k_{q+1 \ldots n})$ of the remaining $n-\ell$ legs where zero modes of $d_\a d_\b N^{mn}$ are extracted:
\eqnn\generalCFT
$$\eqalignno{
\langle V_1(z_1) &U_2(z_2) U_3(z_3) \ldots U_n(z_n)\rangle_{ddN}\cr
&= \langle \, \sum_{\ell=1}^{n-3} (\eta_{12} \, \ldots \, \eta_{\ell-1,\ell} \, T_{12\ldots \ell} ) \! \! \! \sum_{\ell+1\leq p<q<n}
\! \! \! (\eta_{\ell+1,\ell+2} \, \ldots \, \eta_{p-1,p} \, T_{\ell+1 \ldots p}^i) \cr
 & \ \ \ \ \ \ \ \ \times \ (\eta_{p+1,p+2} \ldots  \eta_{q-1,q} \, T^j _{p+1 \ldots q}) \  (\eta_{q+1,q+2}  \ldots \eta_{n-1,n}\, T^k_{q+1 \ldots n}) \cr
 & \ \ \ \ + \  {\rm permutations} \ + \ \eta_{ijk} (\ldots) \, \rangle. &\generalCFT\cr
}$$
The next tasks to be addressed in the following subsections are to trade the BRST building blocks for Berends--Giele currents and to resolve the
ambiguity about the $\eta_{ijk}$ terms.

\subsec{Step 2: Berends--Giele currents}

\noindent
In the $n$-point tree amplitude computations of \MSSI\ the worldsheet integrands combine the
BRST building blocks with $z_{ij}$ poles via $T_{12\ldots p} \leftrightarrow (z_{12} z_{23}\ldots z_{p-1,p})^{-1}$. The essential step for further simplification
lies in trading $T_{\ldots}$ for Berends--Giele currents $M_{\ldots}$ using the identity
\eqn\tradetree{
{T_{12\ldots p} \over z_{12} \, z_{23} \ldots  z_{p-1,p}} + {\rm sym}(23\ldots p) = 
M_{12\ldots p} \; \prod_{k=2}^p \sum_{m=1}^{k-1} \frac{ s_{mk}}{z_{mk}} + {\rm sym}(23\ldots p)
}
It has already been proven at tree-level \MSST\ that the Berends--Giele currents are the natural objects
to describe the SYM amplitudes. The identity \tradetree\ was the key step in identifying
the $n$-point superstring amplitude as sum of $(n-3)!$ SYM amplitudes \MSSI\ dressed by
hypergeometric worldsheet integrals \MSSII. To what extent can the tree-level identity \tradetree\ and its
corollaries be generalized to one-loop?

In order to answer this question note that
the tree-level proof of \tradetree\ required two assumptions: the symmetries of the building blocks
and the partial fraction identities $(z_{ij}z_{jk})^{-1} + {\rm cyc}(ijk) = 0$. As the loop building blocks $T^{i,j,k}_{\ldots}$
obey the same symmetry identities as their tree-level counterparts,
the only obstacle against a direct one-loop generalization of \tradetree\ comes from the fact
that the functions $\eta_{ij}\eta_{ik}$ do not obey a similar partial fraction identity in general. That is why we have defined
the totally symmetric function $\eta_{ijk}=\eta_{ij}  \eta_{ik} +  \eta_{ji}  \eta_{jk} +  \eta_{ki} \eta_{kj}$ measuring
the failure of the tree-level partial fraction identity to hold at higher genus. With this definition at hand, the one-loop generalization of \tradetree\ is
\eqnn\tradeloop
$$\eqalignno{
\eta_{12} \, \eta_{23} \ldots \eta_{p-1,p} \, T_{12\ldots p} \ + \ {\rm sym}(23\ldots p)& \ \ = \ \
M_{12\ldots p} \; \prod_{k=2}^p \sum_{m=1}^{k-1}  s_{mk}\, \eta_{mk} \cr
 + \ {\rm sym}&(23\ldots p) \ + \
\eta_{ijk}(\ldots)\ . &\tradeloop\cr
}$$
Of course, the same identity holds for the loop cousins $(T_{\ldots},M_{\ldots}) \mapsto
(T^i_{\ldots},M^i_{\ldots})$ since the $T_{\ldots}^i$ enjoy the same symmetry properties as the
tree-level building blocks $T_{\ldots}$ and the definition of $M_{\ldots}^i$ in terms of
$T^i_{\ldots}$ incorporates the same functional dependence as $M_{\ldots}$ expressed in terms of
$T_{\ldots}$.

We will show in the following subsection that discarding $\eta_{ijk}$ corrections in both
\tradeloop\ and \generalCFT\ yields BRST-invariant kinematic factors describing
non-anomalous terms in the amplitude\foot{
In fact, $\eta_{ijk}$ terms are related to the gauge anomaly discussion of \GSanom\ in a subtle way.
As pointed out by Michael Green \ref\privDisc{M.B. Green, private communication} and elaborated in
future work, the associated BRST non-invariant kinematic factors responsible for the hexagon anomaly
might leave gauge-invariant fingerprints in certain subsectors of the amplitude.}.
Let us see some examples. For the four-point correlator \fourCFT\ the trading identity is
trivial in view of $V_1 = T_1 = M_1$ and $T_2^i = M_2^i$,
\eqn\fourtrade{
K_4 = \langle V_1 T_2^i T_3^j T_4^k \rangle = \langle M_1 M_2^i M_3^j M_4^k \rangle.
}
In order to prevent overcrowding in the formul{\ae} below the following shorthand notation will
be used
\eqn\Xdef{
X_{ij} \ \equiv \ s_{ij}\eta_{ij}.
}
The five-point correlator \fiveCFT\ is also rather trivially converted to Berends--Giele currents
$M_{12} = T_{12} / s_{12}$ and $M_{23}^i = T_{23}^i / s_{23}$. The permutations generated by
\fivo\ and \fivt\ combine to ten terms
\eqnn\fiveex
$$\eqalignno{
K_5 &= \langle X_{12} \, M_{12} \, M_3^i \, M_4^j \, M_5^k
+  X_{13} \, M_{13} \, M_2^i \, M_4^j \, M_5^k
+  X_{14} \, M_{14} \, M_2^i \, M_3^j \, M_5^k \cr
&\quad + X_{15} \, M_{15} \, M_2^i \, M_3^j \, M_4^k 
+ X_{23} \, M_{1} \, M_{23}^i \, M_4^j \, M_5^k
+ X_{24} \, M_{1} \, M_{24}^i \, M_3^j \, M_5^k\cr
&\quad + X_{25} \, M_{1} \, M_{25}^i \, M_3^j \, M_4^k
+ X_{34} \, M_{1} \, M_{34}^i \, M_2^j \, M_5^k 
+ X_{35} \, M_{1} \, M_{35}^i \, M_2^j \, M_4^k\cr
&\quad + X_{45} \, M_{1} \, M_{45}^i \, M_2^j \, M_3^k\, \rangle. &\fiveex
}$$
The six-point amplitude is the first instance where the identity \tradeloop\ finds non-trivial
application. Dropping the terms proportional to $\eta_{ijk}$ in lines with the BRST reasoning, the
six-point topologies \topu a --- \topu d\ give rise to
\eqnn\sixex
$$\eqalignno{
K_6 = &\langle  \, 10 \ {\rm terms}\ \big[ \, M_{123} \, X_{12} (X_{13} + X_{23}) +  (2 \leftrightarrow 3) \,\big] \, M_4^i \, M_5^j \, M_6^k \cr
& + \ 30 \ {\rm terms}\ X_{12} \, M_{12} \, X_{34} \, M_{34}^i \, M_5^j \, M_6^k \cr
& + \ 15 \ {\rm terms}\ M_1 \, X_{23} \, M_{23}^i \, X_{45} \, M_{45}^j \, M_6^k  \cr
& + \ 10 \ {\rm terms}\ M_1 \, \big[ \, M_{234}^i \, X_{23} (X_{24} + X_{34}) 
+  (3 \leftrightarrow 4) \, \big] \, M_5^j \, M_6^k \, \rangle . &\sixex
}$$
At this point, we shall be more explicit about the permutations within the correlator. As
mentioned before, the correlator must be symmetric in all the legs $(23\ldots n)$ of integrated
vertices, but the last term in $K_6$ only contains $2\times 10$ out of the 60 possible terms
$M_{pqr}^i X_{pq} (X_{pr}+X_{qr})$ with $p,q,r \in \{ 2,3,4,5,6\}$. It turns out that by the
symmetry properties of Berends--Giele currents (e.g. $M_{234}^i = M_{432}^i$ and $M_{234}^i + {\rm
cyc}(234) = 0$ in the rank three case at hand), the expression
\eqn\secret{
M_{23 \ldots p}^i \, \left( \, \prod_{k=3}^p \sum_{m=2}^{k-1} X_{mk} \, \right) \ + \ {\rm sym}(34\ldots p)
}
is secretly totally symmetric\foot{The same hidden symmetry occurs in the representation \dixonduca
$$
{\cal M}^{\rm YM}_n  \ \ \sim \ \ \sum_{\sigma \in S_{n-2}} f^{1\sigma(2)a} \,f^{a \sigma(3)b } \,
f^{b\sigma(4)c} \, \ldots \, f^{y\sigma(n-2) z } \, f^{z\sigma(n-1)n} \,
{\cal A}^{\rm YM}\big( 1,\sigma(2,3,\ldots,n-1),n \big)
$$
of the color-dressed SYM amplitude: The structure constant contractions $(f^{bcd})^{n-2}$ share the
symmetry properties of the integrand $\prod_{k=3}^p \sum_{m=2}^{k-1} X_{mk}$ and the
rank $p$ Berends--Giele current taking the role of a $(p+1)$-point SYM amplitude with one off-shell leg guarantees that the
color-ordered ${\cal A}^{\rm YM}_n$ have the same symmetry properties as the $M^i_{\ldots}$.
Hence, the total symmetry of ${\cal M}^{\rm YM}_n$ implies that of \secret\ by virtue of the
dictionary explained above.} in $(23\ldots p)$ even though only the smaller symmetry in $(34\ldots
p)$ is manifest. That is why each of the ten choices to single out three legs from $\{
2,3,4,5,6\}$ realizes two out of six possible terms only, without spoiling the overall $(23456)$
symmetry.

It is crucial to note the symmetry properties of the two sides of the $T_{\ldots} \leftrightarrow
M_{\ldots}$ trading identity \tradetree. The left-hand side
is totally symmetric at tree-level, even in trading leg one for one of the others. But this makes
use of partial fraction relations that cause extra terms $\sim \eta_{ijk}$ at loops. The $z_i$
dependence on the right hand side, however, is built from combinations $s_{ij}/z_{ij}$ where it is
obvious from the Mandelstam factors that there are no partial fractions at work to see the
symmetries. Only the right hand side of \tradetree\ stays totally symmetric in $(12\ldots p)$ under the loop-conversion $s_{ij}/z_{ij} \rightarrow s_{ij}\eta_{ij} = X_{ij}$ of worldsheet functions.

For these reasons, the following expression for the seven-point kinematic factor,
\eqnn\sevex
$$\eqalignno{
K_7 = & \langle \, 15 \ {\rm terms} \ M_1 \, \big[ \,  M_{2345}^i \, X_{23} \,
(X_{24} + X_{34}) (X_{25} + X_{35}  + X_{45})  +  {\rm sym}(345) \, \big] \, M_6^j \, M_7^k \cr
& + \ 60 \ {\rm terms} \ M_1 \, \big[ \, M_{234}^i \, X_{23} \, (X_{24} + X_{34})  +  (3\leftrightarrow 4) \, \big] \, X_{56} \, M_{56}^j \,M_7^k \cr
& + \ 15 \ {\rm terms} \ M_1 \, X_{23} \, M_{23}^i \, X_{45} \, M_{45}^j \, X_{67} \, M_{67}^k \cr
& + \ 20 \ {\rm terms} \ \big[ \, M_{1234} \, X_{12} \, (X_{13} + X_{23}) (X_{14} + X_{24} + X_{34})  +  {\rm sym}(234) \, \big] \, M_5^i \, M_6^j \, M_7^k \cr
& + \ 90 \ {\rm terms} \ \big[ \, M_{123} \, X_{12} \, (X_{13} + X_{23})  +  (2\leftrightarrow 3) \, \big]  \, X_{45} \, M_{45}^i \, M_6^j \, M_7^k \cr
&  + \ 60 \ {\rm terms} \ X_{12} \, M_{12} \, \big[ \, M_{345}^i \, X_{34} \, (X_{35} + X_{45})  +  (4\leftrightarrow 5) \, \big] \, M_6^j \, M_7^k \cr
& + \ 90 \ {\rm terms} \ X_{12} \, M_{12} \, X_{34} \,  M_{34}^i \, X_{56} \, M_{56}^j \, M_7^k \rangle &\sevex
}$$
is totally symmetric even though only those six $M_{\sigma(2345)}^i$ permutations $\sigma \in S_4$ with fixed point $\sigma(2) = 2$ occur. 

The $n$-point generalization of the above patterns is given by
\eqnn\BGgeneral
$$\eqalignno{
K_n  & = \langle \, \sum_{\ell=1}^{n-3} M_{12\ldots \ell} \, \left( \, \prod_{k=2}^\ell \sum_{m=1}^{k-1} X_{mk} \, \right) \! \sum_{\ell+1\leq p<q<n}
\! \! \!  M_{\ell+1 \ldots p}^i \, \left( \, \prod_{k=\ell+2}^p \sum_{m=\ell+1}^{k-1} X_{mk} \, \right) \cr
 & \ \ \ \ \ \ \ \ \times \ M^j _{p+1 \ldots q} \left( \, \prod_{k=p+2}^q \sum_{m=p+1}^{k-1}
 X_{mk} \right) \, M^k_{q+1 \ldots n} \left( \, \prod_{k=q+2}^n \sum_{m=q+1}^{k-1} X_{mk} \right) \cr
 & \ \ \ \ + \ {\rm permutations}\, \rangle\,. &\BGgeneral
}$$

\subsec{Step 3: Integration by parts}
\subseclab\IBP

\noindent
In this step the number of one-loop worldsheet integrals will be reduced using partial integration identities. These manipulations have been crucial in the computation of the $n$-point disk amplitude \refs{\MSSI,\MSSII} and had already found appearance in the string inspired rules towards field theory amplitudes \refs{\BernUX, \BernAQ}. As emphasized in the references, integration by parts allows to eliminate double derivatives of the bosonic Green function.

After this reduction is performed the
kinematic factor for the one-loop amplitude becomes a sum over manifestly BRST invariant
objects multiplied by $n-4$ powers of $X_{ij}$; schematically, this means $K_n = \sum X^{n-4}\,\langle C_{1,\ldots}\rangle$.

In order to see how these partial integrations can be performed note that
the worldsheet integrands at any loop order contain a universal factor
proportional to the correlation function of the plane wave exponential factors, the so-called Koba--Nielsen factor
\eqn\KN{
{\rm KN} \ \equiv \ \Big\langle \, \prod_{i=1}^n \, e^{i k_i \cdot x(z_i,\bar z_i) } \, \Big\rangle \ \propto \
\exp \Big( \, \sum_{i<j}^n \, s_{ij} \,\langle  x(z_i,\bar z_i) \, x(z_j,\bar z_j)  \rangle
\, \Big)\, .
}
The precise form of the bosonic Green's function $\langle x(z_i,\bar z_i) x(z_j,\bar
z_j) \rangle$ in terms of Jacobi theta functions is irrelevant for the analysis
in the following. What matters is its appearance in the Koba--Nielsen factor and the antisymmetry
of its derivative $\eta_{ij}= {\p\over \p z_i} \langle x(z_i,\bar z_i) x(z_j,\bar z_j) \rangle = -
\eta_{ji}$ which can be viewed as the one-loop generalization of the $1/z_{ij}$ single pole at tree-level. The form \KN\ of the Koba--Nielsen
factor implies that the combinations $X_{ij} = s_{ij} \eta_{ij}$ can be integrated by parts
\eqn\inpt{
0 = \int {\p\over\p z_i} \; {\rm KN} = \int \sum_{j \neq i} s_{ij} \, \eta_{ij} \; {\rm KN}.
}
Boundary terms at $z_i = z_{i+1}$ do not contribute to \inpt\ because the Koba--Nielsen factor vanishes for real and positive $s_{i,i+1}$ as $|z_i - z_{i+1}|^{s_{i,i+1}}$. Analytic continuation in the complex $s_{i,i+1}$ plane allows to extend this argument to generic kinematic regimes.

This identity still holds in presence of further $\eta_{pq}$ factors in the integrand as long as none of the $p,q$ labels coincides with the differentiation leg $i$, for instance
\eqnn\ibp
$$\eqalignno{
\int {\rm KN} \ X_{12} \, (X_{13} + X_{23}) = &\int {\rm KN} \ (X_{34} + X_{35} + \cdots + X_{3n}) \, (X_{23} + X_{24} + \cdots + X_{2n})\cr
\int {\rm KN} \ \prod_{k=2}^p \sum_{m=1}^{k-1}  X_{mk} = &\int {\rm KN} \  \prod_{k=2}^p \sum_{m=k+1}^{n} X_{km} .&\ibp\cr
}$$
The ubiquitous $\prod_{k=2}^p \sum_{m=1}^{k-1}  X_{mk}$ products in equation \BGgeneral\ for $K_n$ turn out to be maximally partial-integration-friendly. This has already been exploited in tree-level computations \MSSI.

Once we have removed any appearance of $z_{1}$ from $X_{ij}$ via integration by parts \ibp, the remaining terms in the correlator
will build up various BRST invariants $C_{1,\ldots}$. This is a trivial statement in the four-point correlator \Cooo,
\eqn\fourC{
K_4 = \langle M_1 M_2^i M_3^j M_4^k \rangle = \langle C_{1,2,3,4} \rangle
}
whereas the five-point kinematic factor requires $X_{12}=X_{23}+ X_{24} + X_{25}$ and $(2345)$ permutations thereof (which is valid under integration against KN only). After eliminating the $X_{1j}$ at $j=2,3,4,5$ in \fiveex, we find the manifestly BRST-invariant expression
\eqnn\fivefinal
$$\eqalignno{
K_5 &= X_{23} \langle M_1 M_{23}^i M_4^jM_5^k + M_{12} M_3^i M_4^j M_5^k + M_{31} M_2^i M_4^j M_5^k \rangle \cr
&\quad + X_{24} \langle M_1 M_{24}^i M_3^jM_5^k + M_{12} M_4^i M_3^j M_5^k + M_{41} M_2^i M_3^j M_5^k \rangle \cr
&\quad + X_{25} \langle M_1 M_{25}^i M_3^jM_4^k + M_{12} M_5^i M_3^j M_4^k + M_{51} M_2^i M_3^j M_4^k \rangle \cr
&\quad + X_{34} \langle M_1 M_{34}^i M_2^jM_5^k + M_{13} M_4^i M_2^j M_5^k + M_{41} M_2^i M_3^j M_5^k \rangle \cr
&\quad + X_{35} \langle M_1 M_{35}^i M_2^jM_4^k + M_{13} M_5^i M_2^j M_4^k + M_{51} M_2^i M_3^j M_4^k \rangle \cr
&\quad + X_{45} \langle M_1 M_{45}^i M_2^jM_3^k + M_{14} M_5^i M_2^j M_3^k + M_{51} M_2^i M_3^j M_4^k \rangle \cr
&= X_{23} \langle C_{1,23,4,5}\rangle
+ X_{24} \langle C_{1,24,3,5}\rangle
+ X_{25} \langle C_{1,25,3,4}\rangle\cr
&\quad + X_{34} \langle C_{1,34,2,5}\rangle
+ X_{35} \langle C_{1,35,2,4}\rangle
+ X_{45} \langle C_{1,45,2,3}\rangle  &\fivefinal
}$$
which agrees with the expression from \fiveptone\
when its component expansion is evaluated \PSS. The total worldsheet derivatives are suppressed in
\fivefinal\ and in all subsequent kinematic factors.

The general lesson to learn from the five-point computation concerns the choice of integral basis and the role of the $M_{12\ldots p}$ terms in
\BGgeneral\ with leg one attached and $p \geq 2$. Once we eliminate $z_1$ from every
$X_{rs}$ in the integrand, the remaining $X^{n-4}$ polynomials are guaranteed to be minimal under \inpt\ and
the superfield prefactors must be BRST closed $C_{1,\ldots}$. The superfields associated with the integrands $X_{1j}$
outside the desired basis have in common that leg one is attached to a rank $p \geq 2$ Berends--Giele current $M_{12\ldots p}$.
After integration by parts, the worldsheet dependence will be transformed into
$z_1$ independent $X_{rs}$ combinations ($r,s\neq 1$), so the associated $V_1 M_{23\ldots p}^i M_{p+1\ldots q}^j
M_{q+1\ldots n}^k$ permutations will receive corrections containing $M_{12\ldots p}$ at $p\geq 2$. Hence, the
job of all the $M_{12\ldots p}$ is to provide the BRST invariant completion of $V_1 M_{23\ldots
p}^i M_{p+1\ldots q}^j M_{q+1\ldots n}^k$ to form $C_{1,23\ldots p, p+1\ldots q,q+1\ldots n}$.

Let us consider the six-point amplitude to see these mechanisms in action. The first two lines in \sixex\ require integration
by parts in the form $X_{12} X_{34} = X_{34}(X_{23}+X_{24}+ X_{25}+X_{26})$ and $X_{12}(X_{13}+X_{23}) = (X_{23}+X_{24}+ X_{25}+X_{26})(X_{34}+X_{35}+X_{36})$
in order to eliminate all the $X_{1j}$. The remaining two lines already involve integrands in
the $z_1$ independent basis, and the associated kinematics receive corrections
\eqnn\corrections{
$$\eqalignno{
X_{23} X_{45} &M_1 M_{23}^i M_{45}^j M_6^k \ \ \mapsto \ \ X_{23} X_{45} \,
\Big( \, M_{1} \, M_{23}^i \, M_{45}^j \, M_6^k \ + \ M_{12} \, M_{3}^i \, M_{45}^j \, M_{6}^k \cr
 & - \ M_{13} \, M_{2}^i \, M_{45}^j \, M_{6}^k  \ + \ M_{14} \, M_{23}^i \, M_{5}^j \, M_{6}^k \ - \ M_{15} \, M_{23}^i \, M_{4}^j \, M_{6}^k \cr
 &  + \ M_{413} \, M_{2}^i \, M_{5}^j \, M_{6}^k \ + \ M_{512} \, M_{3}^i \, M_{4}^j \, M_{6}^k \cr
 & - \ M_{412} \, M_{3}^i \, M_{5}^j \, M_{6}^k \ - \ M_{513} \, M_{2}^i \, M_{4}^j \, M_{6}^k \, \Big)\cr
= & \ \ X_{23} X_{45} C_{1,23,45,6} &\corrections
}$$
due to the $X_{23} X_{45}$ on the right hand side of integration by parts formulae. By carefully gathering
all $X_{23} X_{45}$ corrections, the superfield expressions can be seen to build up the full-fledged $C_{1,23,45,6}$.
So the net effect of integrating $z_1$-dependent $X_{pq}$ by parts is the replacement
$M_1 M_{23}^i M_{45}^j M_6^k \mapsto C_{1,23,45,6}$ and $M_1 M_{234}^i M_5^j M_6^k \mapsto C_{1,234,5,6}$:
\eqnn\sixfinal
$$\eqalignno{
K_6 &= X_{23} (X_{24}+X_{34}) \langle C_{1,234,5,6}\rangle \ + \ X_{24} (X_{23}+X_{43}) \langle C_{1,243,5,6}\rangle \cr
&\quad + \ X_{23} (X_{25}+X_{35}) \langle C_{1,235,4,6}\rangle \ + \ X_{25} (X_{23}+X_{53}) \langle C_{1,253,4,6}\rangle \cr
&\quad + \ X_{23} (X_{26}+X_{36}) \langle C_{1,236,4,5}\rangle \ + \ X_{26} (X_{23}+X_{63}) \langle C_{1,263,4,6}\rangle \cr
&\quad + \ X_{24} (X_{25}+X_{45}) \langle C_{1,245,3,6}\rangle \ + \ X_{25} (X_{24}+X_{54}) \langle C_{1,254,3,6}\rangle \cr
&\quad + \ X_{24} (X_{26}+X_{46}) \langle C_{1,246,3,5}\rangle \ + \ X_{26} (X_{24}+X_{64}) \langle C_{1,264,3,5}\rangle \cr
&\quad + \ X_{25} (X_{26}+X_{56}) \langle C_{1,256,3,4}\rangle \ + \ X_{26} (X_{25}+X_{65}) \langle C_{1,265,3,4}\rangle \cr
&\quad + \ X_{34} (X_{35}+X_{45}) \langle C_{1,345,2,6}\rangle \ + \ X_{35} (X_{34}+X_{54}) \langle C_{1,354,2,6}\rangle \cr
&\quad + \ X_{34} (X_{36}+X_{46}) \langle C_{1,346,2,5}\rangle \ + \ X_{36} (X_{34}+X_{64}) \langle C_{1,364,2,5}\rangle \cr
&\quad + \ X_{35} (X_{36}+X_{56}) \langle C_{1,356,2,4}\rangle \ + \ X_{36} (X_{35}+X_{65}) \langle C_{1,365,2,4}\rangle \cr
&\quad + \ X_{45} (X_{46}+X_{56}) \langle C_{1,456,2,3}\rangle \ + \ X_{46} (X_{45}+X_{65}) \langle C_{1,465,2,3}\rangle \cr
&\quad + \ X_{23} \, X_{45} \, \langle C_{1,23,45,6}\rangle  \ + \ X_{23} \, X_{46} \, \langle C_{1,23,46,5}\rangle \ + \ X_{23} \, X_{56} \, \langle C_{1,23,56,4}\rangle  \cr
&\quad + \ X_{24} \, X_{35} \, \langle C_{1,24,35,6}\rangle  \ + \ X_{24} \, X_{36} \, \langle C_{1,24,36,5}\rangle \ + \ X_{24} \, X_{56} \, \langle C_{1,24,56,3}\rangle  \cr
&\quad + \ X_{25} \, X_{34} \, \langle C_{1,25,34,6}\rangle  \ + \ X_{25} \, X_{36} \, \langle C_{1,25,36,4}\rangle \ + \ X_{25} \, X_{46} \, \langle C_{1,25,46,3}\rangle  \cr
&\quad + \ X_{26} \, X_{34} \, \langle C_{1,26,34,5}\rangle  \ + \ X_{26} \, X_{35} \, \langle C_{1,26,35,4}\rangle \ + \ X_{26} \, X_{45} \, \langle C_{1,26,45,3}\rangle  \cr
&\quad + \ X_{34} \, X_{56} \, \langle C_{1,34,56,2}\rangle  \ + \ X_{35} \, X_{46} \, \langle C_{1,35,46,2}\rangle \ + \ X_{36} \, X_{45} \, \langle C_{1,36,45,2}\rangle   &\sixfinal
}$$
The above patterns lead to a seven-point kinematic factor given by
\eqnn\sevenfinal
$$\eqalignno{
K_7 &=  \hbox{15 terms} \  \big[ \, X_{23} \, (X_{24} + X_{34}) \, (X_{25} + X_{35}  + X_{45}) \,
\langle C_{1,2345,6,7}\rangle \ + \ {\rm sym}(345) \, \big]  \cr
& \quad + \hbox{60 terms} \  \big[ \,  X_{23} \, (X_{24} + X_{34}) \, \langle C_{1,234,56,7}\rangle \ + \
(3\leftrightarrow 4) \, \big] \, X_{56} \cr
& \quad + \hbox{15 terms} \ \big[ X_{23} \,   X_{45} \,   X_{67} \, \langle C_{1,23,45,67}\rangle \, \big]. &\sevenfinal
}$$
In order to make the permutations in \sevenfinal\ more precise and to compactly write down its $n$-point generalization,
we shall introduce some notation that facilitates the bookkeeping of the $S_3^{n-1}$ terms in $K_n$.

\subsec{The closed-form $n$-point kinematic factor}
\subseclab\Stirnot

\noindent
We have argued in subsection \CsymSec\ that the symmetries \BGsymC\ of the BRST invariants yield a basis with $S_{3}^{n-1}$
elements under relations with constant coefficients. It became evident from the examples \fivefinal, \sixfinal\ and \sevenfinal\ that
each independent $C_{1,\ldots}$ occurs in $K_{n}$ where \secret\ determines the associated worldsheet function to be
$$
\langle C_{1,23 \ldots p, p+1 \ldots q,q+1 \ldots n}\rangle \ \ \leftrightarrow \ \
\left( \prod_{k=3}^p \sum_{m=2}^{k-1} X_{mk} \right)
\left( \prod_{k=p+2}^q \sum_{m=p+1}^{k-1} X_{mk} \right) \left(\prod_{k=q+2}^n \sum_{m=q+1}^{k-1} X_{mk}  \right) .
$$
Writing down the kinematic factor $K_n$ in a closed form for general multiplicity $n$ is a matter
of notation. That is why we shall now introduce a set ${\cal S}_3^{k}$ with $S^k_3$ elements which takes
care of the $C_{1,\ldots}$ bookkeeping. It compasses all the partitions of $k$ elements $12\ldots k$ into
three indistinguishable cycles, say $(a_1\ldots a_p), \, (b_1\ldots b_q),\,(c_1\ldots c_r)$, where $p+q+r = k$
and none of the cycles remains empty, i.e.\ $p,q,r \neq 0$. For given sets $\{a_1\ldots a_p\}, \, \{b_1\ldots b_q\}$
and $\{c_1\ldots c_r\}$, only cyclically inequivalent configurations are considered as distinct ${\cal S}_3^{k}$
elements. Fixing the first entry $a_1,b_1,c_1$ of each cycle is one convenient way to implement this, we are
then left with permutations
$$
(a_1 \sigma(a_2\ldots a_p) ), \, (b_1 \pi (b_2\ldots b_q)),\,(c_1 \rho(c_2\ldots c_r)), \quad \sigma \in S_{p-1},\quad \pi \in S_{q-1},\quad \rho \in S_{r-1}
$$
for a partition characterized by $p,q,r$. Of course, we have to avoid overcounting due to the
indistinguishable cycles, i.e.\ $(2,3),\, (4,5), \,(\ldots)$ is identified with $(4,5),\, (2,3),\, (\ldots)$ in ${\cal S}_3^{k}$. A formal way to summarize these properties of ${\cal S}_3^{k}$ is
\eqnn\defstirling
$$\eqalignno{
{\cal S}_3^{k} &=  \bigcup_{ { p \geq q \geq r \geq 1 \atop p+q+r=k}} {\Xi_{p,q,r}( S_k)  \over Z_p \times Z_q \times Z_r \times S_{\nu(p,q,r)} }   &\defstirling \cr
\Xi_{p,q,r}(12\ldots k)& = (12\ldots p)\times (p+1\ldots p+q) \times (p+q+1 \ldots k) \cr
\nu(p,q,r)&= 1+\delta_{p,q} + \delta_{q,r} \ .
}$$
The map $\Xi_{p,q,r}$ cuts a given $S_k$ permutation of $(12\ldots k)$ into three tuples $(12\ldots p), \, (p+1\ldots p+q)$ and $(p+q+1 \ldots k)$ of cardinality, $p$, $q$ and $r$, respectively. Each of them is modded out by the corresponding cyclic group $Z_p$, $Z_q$, $Z_r$, and in case of coinciding cardinalities ($p=q$ or $q=r$ or both), we divide by permutations $S_{\nu(p,q,r)}$ of these tuples of equal size. Indeed, we can check that the number of elements in the individual $(p,q,r)$ contributions to \defstirling,
$$
\left| {  S_k \over Z_p \times Z_q \times Z_r \times S_{\nu(p,q,r)} } \right|  = { k! \over pqr \cdot  \nu(p,q,r)!}
$$
reproduce the entries of table 1.

The structure of the $n$-point kinematic factor is described by
\eqnn\nnfinal
$$\eqalignno{
K_n &= \sum_{p,q} \biggl\{ \langle C_{1,23 \ldots p, p+1 \ldots q,q+1 \ldots n}\rangle
\left( \prod_{k=3}^p \sum_{m=2}^{k-1} X_{mk} \right)
\left( \prod_{k=p+2}^q \sum_{m=p+1}^{k-1} X_{mk} \right) \left(\prod_{k=q+2}^n \sum_{m=q+1}^{k-1} X_{mk}  \right)
\cr
&\quad  \ + \ {\rm sym}(34\ldots p)
\ + \ {\rm sym}(p+2,\ldots q) \  + \ {\rm sym}(q+2,\ldots n) \ + \ {\rm permutations} \biggr\} \,.&\nnfinal
}$$
The definition \defstirling\ of ${\cal S}^k_3$ allows to make the permutations involved very precise:
\eqnn\nfinal
$$\eqalignno{
K_n &= \sum_{\sigma \times \pi \times \rho \in {\cal S}_3^{n-1}}  \langle C_{1,\sigma(23 \ldots p), \pi(p+1 \ldots q), \rho(q+1 \ldots n)}\rangle \cr
& \sigma
\left( \prod_{k=3}^p \sum_{m=2}^{k-1} X_{mk} \right) \ \pi 
\left( \prod_{k=p+2}^q \sum_{m=p+1}^{k-1} X_{mk} \right) \ \rho \left(\prod_{k=q+2}^n \sum_{m=q+1}^{k-1} X_{mk}  \right) .
&\nfinal
}$$
The variables $p,q$ are related to the cardinality of the permutations $\sigma,\pi,\rho$ via
$p=|\sigma |+1$ and $q-p = |\pi|$ and should not be confused with the summation variables in
\defstirling.

We shall conclude this section with a comment on the rigid $s_{ij} \eta_{ij} = X_{ij}$
combinations in the worldsheet integrand \nfinal. The $z_i \rightarrow z_j$ singularities from
$\eta_{ij} = z_{ij}^{-1} + {\cal O}(z_{ij})$ in connection with the Koba Nielsen factor \KN\ give
rise to kinematic poles in the corresponding Mandelstam variable, at least for some choices of the
integration region. The connection between worldsheet poles and massless propagators was
thoroughly explored at tree-level \MSSII, and since the $z_i \rightarrow z_j$ singularities are
local effects on the worldsheet regardless of its global properties, we expect the pole analysis
to carry over to higher genus.

The fact that short distance singularities on the worldsheet always occur in the combination $X_{ij}=s_{ij}
\eta_{ij}$, any potential kinematic pole is immediately smoothed out by the Mandelstam numerator
$s_{ij}$. That is why the $z_i$ integrals do not introduce any poles in kinematic invariants\foot{This does not exclude massless poles from the modular $t$ integration due to closed string exchange in non-planar cylinder diagrams \GSanom.},
i.e.\ that all massless open string propagators enter through the BRST invariants $C_{1,\ldots}$. However,
this does not rule out branch cut singularities in $s_{ij}$ as they are expected from the
polylogarithms in field theory loop amplitudes. Systematic study of the non-analytic momentum
dependence is a rewarding challenge which we leave for future work.

\newsec{One-loop kinematic factors built from tree-level data}

\noindent
In this section, we will show that the BRST invariant constituents $C_{1,\ldots}$ of the one-loop
kinematic factor $K_n$ can be expanded in terms of SYM tree amplitudes. More
precisely, these kinematic building blocks for one-loop amplitudes are local linear combinations
of the $\ap^2$ correction ${\cal A}^{F^4}$ to color-ordered superstring tree amplitudes, defined by
\eqn\treeAF{
{\cal A}^{\rm tree}(1,2, \ldots,n;\ap) = {\cal A}^{\rm YM}(1,2, \ldots,n) + \zeta(2) \ap^2 {\cal A}^{F^4}(1,2, \ldots,n) + {\cal O}(\ap^3) \ .
}
The notation ${\cal A}^{F^4}$ is motivated by the fact that the first string correction to \treeAF\ at order\foot{Higher dimensional operators such as $D^{2n}F^4$ and $F^{4+n}$ with $n \geq 1$ contribute to \treeAF\ at orders $\alpha'^{2+n}$ and are not reflected in $\langle C_{1,\ldots} \rangle$ which carries the same mass dimension as ${\cal A}^{F^4}$.} $\alpha'^2$ can be attributed to a supersymmetrized $F^4$ operator in the low energy effective action \medinaD, see later remarks. Comparing with the central result of \refs{\MSSI,\MSSII}
$$
{\cal A}^{\rm tree}(1,2, \ldots,n;\ap) = \sum_{\sigma \in S_{n-3}}{\cal A}^{\rm YM}(1,\sigma(2, \ldots,n-2),n-1,n)  F^\sigma(\ap)
$$
for the disk amplitude, one can identify the ${\cal O}(\ap^2)$ power of the functions
$F^\sigma$ as the expansion coefficients of ${\cal A}^{F^4}$ in terms of $(n-3)!$ field theory
subamplitudes:
\eqn\treeAFdim{
\zeta(2) \ap^2 {\cal A}^{F^4}(1,2,\ldots,n) = \sum_{\sigma \in S_{n-3}} F^\sigma(\ap) \big|_{\ap^2}
{\cal A}^{\rm YM}(1,\sigma(2, \ldots,n-2),n-1,n)  
}
The first examples up to multiplicity $n=6$ read
\eqnn \exples
$$\eqalignno{
{\cal A}^{F^4}(1,2,3,4)&= s_{12} s_{23} {\cal A}^{\rm YM}(1,2,3,4)   
\cr
{\cal A}^{F^4}(1,2,3,4,5) &= (s_{12}s_{34}-s_{34}s_{45}-s_{12}s_{51})   {\cal A}^{\rm YM}(1,2,3,4,5) +s_{13}  s_{24}  {\cal A}^{\rm YM}(1,3,2,4,5) \cr
{\cal A}^{F^4}(1,2,3,4,5,6) &= -(s_{45}s_{56}+s_{12}s_{61}-s_{45}s_{123}-s_{12}s_{345}+s_{123}s_{345})  {\cal A}^{\rm YM}(1,2,3,4,5,6) \cr
& \ \  - s_{13} (s_{23}-s_{61}+s_{345})  {\cal A}^{\rm YM}(1,3,2,4,5,6) - s_{14}s_{25} {\cal A}^{\rm YM}(1,4,3,2,5,6) \cr
& \ \ +  s_{14} s_{35}  {\cal A}^{\rm YM}(1,4,2,3,5,6) - s_{35} (s_{34}-s_{56}+s_{123})  {\cal A}^{\rm YM}(1,2,4,3,5,6)  \cr
& \ \ + s_{13} s_{25}  {\cal A}^{\rm YM}(1,3,4,2,5,6), & \exples
} $$
and a ${\cal O}(\ap^3)$ momentum expansion for the $n=7$ functions $F^\si$ -- i.e. the defining data for ${\cal A}^{F^4}(1,2,3,4,5,6,7)$ -- can be found in the appendix of \MSSII.

As we will show, our BRST invariants governing the one-loop kinematics
\eqn\promise{
\langle C_{1,\ldots} \rangle = \sum_{\rho} {\cal A}^{F^4}(1,\rho(2,3,\ldots,n)) = \sum_{\pi} p_\pi(s_{ij})  {\cal A}^{\rm YM}(1,\pi(2,3,\ldots,n))
}
are linear combinations of SYM trees, accompanied by fine-tuned quadratic polynomials
$p_\pi(s_{ij})$ in Mandelstam variables. The summation ranges for the $S_{n-1}$ permutations
$\rho,\pi$ will be made precise soon.

Since the three-point tree does not receive any $\ap$ corrections, higher-point disk amplitudes do
not factorize on exclusively cubic vertices. Hence, the role of the Mandelstam bilinears
$p_\pi(s_{ij})$ lies in avoiding $n-3$ simultaneous poles in any ${\cal A}^{F^4}$. One can
attribute these $\ap^2$ corrections to a quartic contact interaction $\sim {\rm Tr} \{ F^4\}$
(formed by the non-linearized gluon field strength $F$) in the low energy effective action \medinaD\
(which explains the terminology ${\cal A}^{F^4}$). The diagrams of ${\cal A}^{F^4}$ having one
quartic vertex and cubic SYM vertices otherwise require $n-4$ propagators (instead of the $n-3$
propagators in cubic ${\cal A}^{\rm YM}$ diagrams).

In fact, the appearance of the tree-level kinematics ${\cal A}^{F^4}$ due to the (supersymmetric
completion of the) operator $\sim {\rm Tr} \{ F^4 \}$ in one-loop amplitudes can be
explained by supersymmetry: Naive power counting shows that BRST invariants $C_{1,\ldots}$ are generated by
a term of mass dimension eight in the effective action. The vertex $\sim {\rm Tr} \{ F^4 \}$ is
the unique mass dimension eight operator compatible with 16 supercharges, i.e. $N=1$ supersymmetry in
ten spacetime dimensions or $N=4$ supersymmetry in four dimensions. Cubic operators of type $\sim {\rm
Tr} \{ D^{2k} F^3 \}$ can be ruled out since none of them is supersymmetrizable. That is why
one-loop kinematics in maximally supersymmetric theories have no other choice than reproducing the
${\cal A}^{F^4}$ which have firstly been observed at tree-level.

The organization of this section proceeds as follows: We will first develop a pure spinor
superspace representation for ${\cal A}^{F^4}$ in terms of quadruple Berends--Giele currents
$M_{d_1 \ldots d_s} M^i_{a_1 \ldots a_p} M^j_{b_1 \ldots b_q} M^k_{c_1 \ldots c_r}$ using their diagrammatic
interpretation from figure \Mbox. The central box in these diagrams is then identified with the aforementioned
quartic contact interaction vertex $\sim {\rm Tr} \{ F^4 \}$. We exploit the Berends--Giele representation
to identify the ${\cal A}^{F^4}$ as linear combinations of the one-loop BRST invariants
$C_{1,\ldots}$. Finally, the $S^{n-1}_3$ basis for $C_{1,\ldots}$ can be used to explain amplitude
relations between ${\cal A}^{F^4}$ permutations and (closely related) finite one-loop amplitudes
in pure (non-supersymmetric) Yang--Mills theory.

\subsec{Diagrammatic expansion of tree-level $\ap^2$ corrections}

\noindent
Following the ideas of \BCJ, a method which associates pure spinor building
blocks to cubic tree diagrams of SYM amplitudes in $D=10$ was reviewed in
section \revtree\ on the basis of \refs{\towardsFT,\MSSTsix,\MSST}.
The pure spinor superfield method of \MSST\ rests on two basic assumptions:
\medskip
\item{1.} the kinematic numerator of a cubic graph can only contain BRST building blocks whose $Q$ variation cancels one of the kinematic poles 
\item{2.} the sum of the expressions associated to all cubic graphs must be in the
pure spinor BRST cohomology.
\medskip\noindent
Now we are interested in an analogous diagrammatic
method for constructing the tree-level $\ap^2$ corrections and relating them to one-loop kinematic
structures.

\ifig\afdiags{
The building block prescription for the four- and five-point ${\cal A}^{F^4}$ diagrams.
The rule is that the Berends--Giele current with leg one is always to the left, carries no $i,j,k$ labels
and the combination of superfields must contain the same kinematic poles of the graph.
}
{\epsfxsize=0.85\hsize\epsfbox{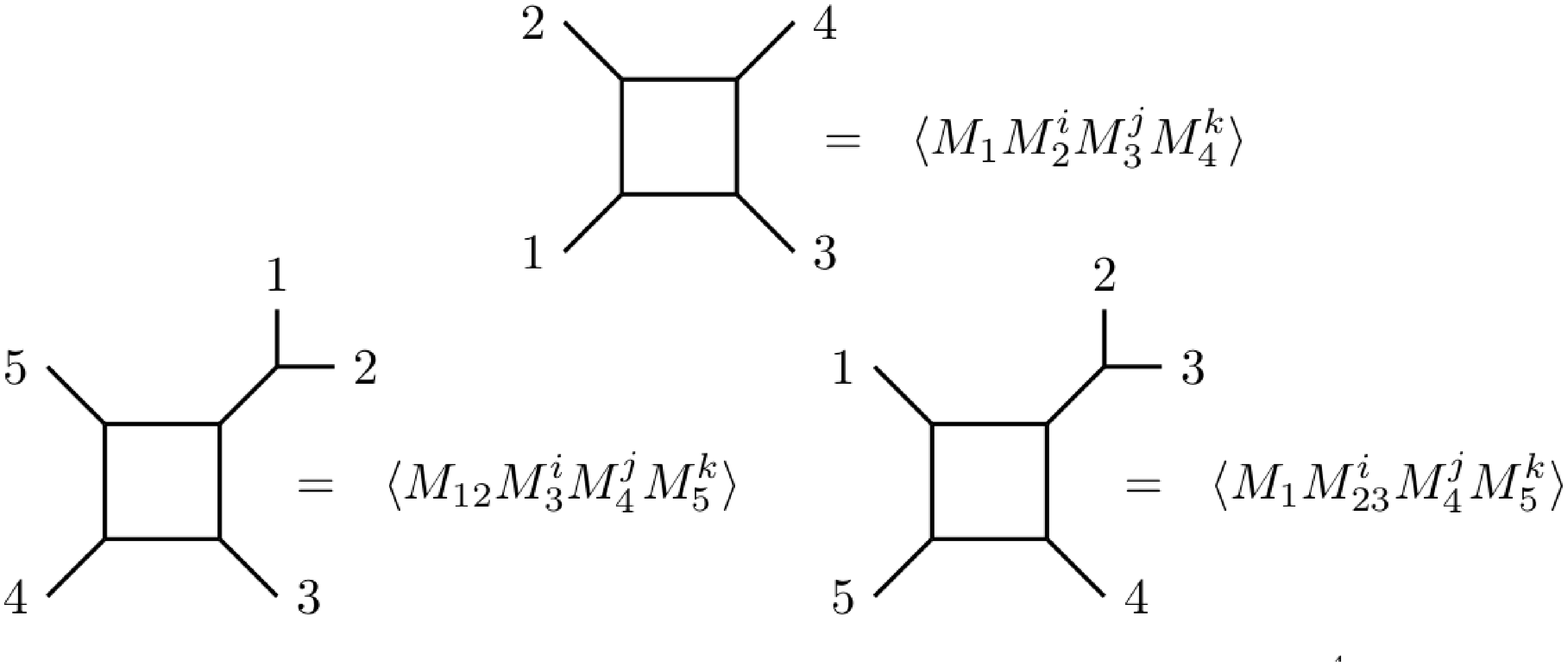}}

At $n$-points, ${\cal A}^{F^4}(1,2,\ldots,n)$ has $n-4$ simultaneous poles corresponding to
diagrams with $n-4$ cubic vertices and one quartic vertex. Since we are using the same superspace
ingredients $M_{d_1 \ldots d_s} M^i_{a_1 \ldots a_p} M^j_{b_1 \ldots b_q} M^k_{c_1 \ldots c_r}$
present in one-loop BRST invariants, the box notation introduced in subsections \diabbs\ and
\Csubsec\ will be kept and can be identified with the tree-level quartic vertex\foot{Even though the diagrammatic rules in this subsection might suggest an association of the kinematic factors ${\cal A}^{F^4}$ with box integrals in the field theory limit, they will also find appearance along with pentagons and higher $n$-gons. The $\alpha' \rightarrow 0$ limit of the worldsheet integrals (to be analyzed in later work) will determine the kinematic coefficient of higher $n$-gons in terms of ${\cal A}^{F^4}$ -- at least up to anomalous terms.} $\sim {\rm Tr} \{
F^4 \}$. The unified diagrammatic language for both $\ap^2$ corrected trees and loop-level
kinematic factors emphasizes that they can be represented by the same class of subamplitudes
${\cal A}^{F^4}$. As mentioned before, this can be traced back to the uniqueness of $N=1$
supersymmetric dimension eight operators in $D=10$.

The four- and five-point diagrams associated with the tree-level $\ap^2$ correction are depicted in \afdiags,
together with their pure spinor superfield mapping. The expression 
\eqn\aFfour{
{\cal A}^{F^4}(1,2,3,4) = \langle M_1 M_2^i M_3^j M_4^k\rangle
}
correctly reflects the absence of poles in ${\cal A}^{F^4}(1,2,3,4)$ and is BRST closed.

The five-point ${\cal A}^{F^4}(1,2,3,4,5)$ has two kinds of Berends--Giele-constituents. They are characterized by
the position of the leg with label one -- it can either enter through the cubic vertex $(\rightarrow M_{1j})$ or
as a standalone corner $(\rightarrow M_{1})$ of the box. The superfield mapping
is slightly different for each possibility, and the rule is that leg number one is never associated
with loop-specific Berends--Giele currents $M^{i,j,k}_{ \ldots}$. The dictionary of \afdiags\ leads
to the following $Q$ closed expression
\eqnn\aFfive
$$\eqalignno{
{\cal A}^{F^4}(1,2,3,4,5) & =
\langle M_{12} M_3^i M_4^j M_5^k\rangle
+ \langle M_{1} M_{23}^i M_4^j M_5^k\rangle &\aFfive\cr
& \quad + \langle M_{1} M_{2}^i M_{34}^j M_5^k\rangle
+ \langle M_{1} M_{2}^i M_3^j M_{45}^k\rangle
+ \langle M_{51} M_{2}^i M_3^j M_4^k\rangle\,.
}$$
In the four-point case, it was shown in \MafraAR\ on superfield level that $\langle M_1 M_2^i M_3^j M_4^k\rangle$ agrees
with the SYM tree representation ${\cal A}^{F^4}(1,2,3,4) = s_{12} s_{23} {\cal A}^{\rm YM}(1,2,3,4)$. This requires
the pure spinor superspace expression \treeseventeen\ for the latter,
\eqn\RandomNameTwo{
 \langle M_1 M_2^i M_3^j M_4^k\rangle = s_{23} \langle T_{12} V_3 V_4 \rangle + s_{12} \langle V_1  T_{23} V_4 \rangle = s_{12} s_{23} {\cal A}^{\rm YM}(1,2,3,4).
}
However, we could not find a superspace proof for \aFfive\ to agree with the ${\cal A}^{\rm YM}$ combination
$(s_{12}s_{34}-s_{34}s_{45}-s_{12}s_{51}) {\cal A}^{\rm YM}(1,2,3,4,5) +s_{13} s_{24} {\cal A}^{\rm YM}(1,3,2,4,5)$
required by \exples. Instead, we have checked that this combination of five gluon trees matches the bosonic terms of
the component expansion of \aFfive. The agreement of the gluonic components extends to the full supermultiplet because
the $\langle \lambda^3 \theta^5 \rangle =1$ prescription respects supersymmetry.

For six-points the story is the same, and the mappings between diagrams and superfields depend on the position of leg
number one.

Following the mapping rules
depicted in \AFdiags\ the 21 graphs which compose the six-point ${\cal A}^{F^4}$ are represented by
the following 15 terms in pure spinor superspace,
\eqnn\AFsix
$$\eqalignno{
{\cal A}^{F^4}(1,2,\ldots,6) &= \langle M_{123}  M_4^i  M_5^j  M_6^k\rangle 
+ \langle M_{612}  M_3^i  M_4^j  M_5^k \rangle + \langle M_{561}  M_{2}^i  M_3^j  M_4^k \rangle \cr
& \quad + \langle M_{1}  M_{234}^i  M_{5}^j  M_6^k \rangle + \langle M_{1}  M_{2}^i 
M_{345}^j  M_6^k \rangle + \langle M_{1}  M_2^i  M_3^j  M_{456}^k\rangle \cr
& \quad + \langle M_{12}  M_{34}^i  M_5^j  M_6^k \rangle + \langle M_{61}  M_{23}^i 
M_4^j  M_5^k \rangle + \langle M_{12}  M_3^i  M_4^j  M_{56}^k\rangle \cr
& \quad + \langle M_{61}  M_2^i  M_3^j  M_{45}^k \rangle + \langle M_1  M_2^i  M_{34}^j
 M_{56}^k \rangle + \langle M_1  M_{23}^i  M_{45}^j  M_6^k\rangle \cr
& \quad + \langle M_{12}  M_3^i  M_{45}^j  M_6^k \rangle + \langle M_{61}  M_2^i 
M_{34}^j  M_5^k \rangle + \langle M_1  M_{23}^i  M_4^j  M_{56}^k\rangle, &\AFsix \cr
}$$
which one can check to be BRST-closed using the formul{\ae} given in the
previous section. The first six terms in \AFsix\ altogether describe
twelve graphs whereas each of the last nine terms describe a single graph. 

\ifig\AFdiags{Pure spinor diagrammatic rules for the six-point $\ap^2$ correction ${\cal A}^{F^4}$. The leg number one
is never associated with a loop-specific Berends--Giele current $M^{i,j,k}$ and the labels in the superfields
are arranged according the order in which they appear in the diagrams.
}
{\epsfxsize=0.80\hsize\epsfbox{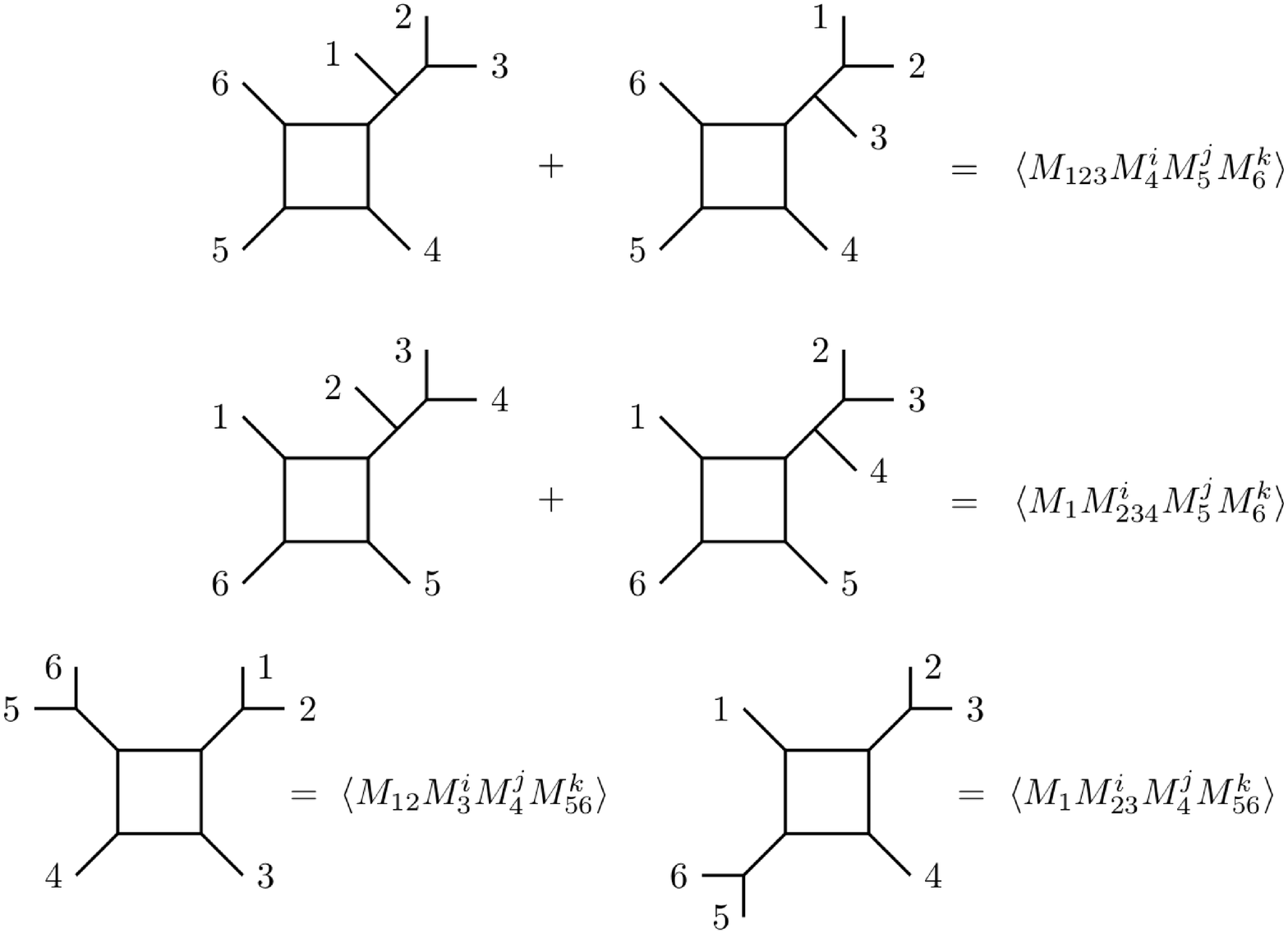}}

Let us state the general rule for proceeding beyond six-points: Our superspace proposal for ${\cal
A}^{F^4}(\ldots)$ encompasses all distinct planar diagrams (with unit relative weight) which are
\medskip
\item{1.} made of one totally
symmetric quartic vertex and otherwise antisymmetric cubic vertices from SYM
\item{2.} compatible with the cyclic ordering of the external legs in ${\cal
A}^{F^4}(\ldots)$.
\medskip\noindent
The sum of their
$M_{\ldots}M_{\ldots}^i M_{\ldots}^j M_{\ldots}^k$ superspace representatives can be checked to
enjoy BRST invariance. Up to multiplicity nine, we have
\eqnn\sevenafq
\eqnn\eightafq
\eqnn\nineafq
$$\eqalignno{
{\cal A}^{F^4}(1,2,\ldots,7) &= \langle M_{1234}  M_5^i  M_6^j  M_7^k \rangle
+ \langle M_{123}M_{45}^i  M_{6}^j  M_7^k \rangle
+ \langle M_{123}  M_{4}^i  M_{56}^j  M_7^k \rangle \cr
& \quad + \langle M_{123}  M_{4}^i  M_{5}^j  M_{67}^k \rangle
+ \langle M_{12}  M_{34}^i  M_{56}^j  M_7^k \rangle
+ {\rm cyc}(1234567) \cr
& \quad = \hbox{35 terms and 84 diagrams}&\sevenafq\cr
{\cal A}^{F^4}(1,2,\ldots,8) &= \langle M_{12345}  M_6^i  M_7^j  M_8^k \rangle
+ \langle M_{123}  M_{456}^i  M_7^j  M_8^k \rangle \cr
& \quad + \langle M_{1234}  \big(   M_{56}^i  M_7^j  M_8^k
+  M_5^i  M_{67}^j  M_8^k + M_5^i  M_6^j  M_{78}^k  \big)\rangle \cr
& \quad  + \langle M_{123}  \big( M_{45}^i  M_{67}^j  M_8^k 
+ M_{45}^i  M_{6}^j  M_{78}^k + M_4^i  M_{56}^j  M_{78}^k  \big) \rangle
+  {\rm cyc}(12345678) \cr
  & \quad + \langle M_{123}   M_4^i  M_{567}^j  M_8^k \rangle
  + {\rm cyc}(1234) \cr
  & \quad + \langle M_{12}  M_{34}^i  M_{56}^j  M_{78}^k \rangle
  + \langle M_{81}  M_{23}^i  M_{45}^j  M_{67}^k \rangle \cr
& \quad = \hbox{70 terms and 330 diagrams} &\eightafq\cr
{\cal A}^{F^4}(1,2,\ldots,9) & = \langle M_{123456}  M_7^i  M_8^j  M_9^k \rangle
+ \langle M_{123}  M_{45}^i  M_{67}^j  M_{89}^k \rangle \cr
& \quad + \langle M_{12345}  \big(   M_{67}^i  M_8^j  M_9^k
+ M_6^i  M_{78}^j  M_9^k + M_6^i  M_7^j  M_{89}^k  \big)\rangle \cr
& \quad  + \langle M_{1234}  \big(   M_{56}^i  M_{78}^j  M_9^k
+ M_{56}^i  M_{7}^j  M_{89}^k  + M_5^i  M_{67}^j  M_{89}^k  \big) \rangle \cr
& \quad  + \langle M_{1234}  \big(   M_{567}^i  M_8^j  M_9^k
+ M_5^i  M_{678}^j  M_9^k + M_5^i  M_6^j  M_{789}^k  \big) \rangle \cr
& \quad  + \langle M_{123}  \big(   M_{456}^i  M_{78}^j  M_9^k
+ M_{45}^i  M_{678}^j  M_9^k + M_{456}^i  M_7^j  M_{89}^k  \big)\rangle \cr
& \quad +  {\rm cyc}(123456789) \cr
& \quad = \hbox{126 terms and 1287 diagrams} &\nineafq
}$$
Summing over cyclic permutations in the specified labels
slightly abuses notation in view of the rule
that leg number one is always attached to the tree-level current $M_{\ldots}$ rather than to $M^{i,j,k}_{\ldots}$.
For example, the cyclic orbit of $M_{123}M_4^iM_5^jM_6^k$ shall be understood as
$$\eqalignno{
M_{123}M_4^iM_5^jM_6^k + {\rm cyc}(123456) &= M_{123}M_4^iM_5^jM_6^k + M_1 M_{234}^i M^j_5 M^k_6
+ M_1 M_{2}^i M^j_{345} M^k_6 \cr
&\quad  + M_1 M_{2}^i M^j_{3} M^k_{456}
+ M_{612} M_{3}^i M^j_{4} M^k_5
+ M_{561} M_{2}^i M^j_{3} M^k_4.
}$$
Using this refined cyclic summation, one can verify BRST closure of the above expressions as well as the correct
diagrammatic content to represent the $\ap^2$ corrections ${\cal A}^{F^4}$ to
tree amplitudes. Moreover, as a sufficient condition, we have explicitly checked their agreement
up to $n=6$ by computing the bosonic component expansions \PSS\ and comparing with \exples. It is
highly plausible that the (well-tested) experimental rule
\medskip
{\narrower\it BRST-closed objects with the same kinematic pole structure are proportional\par}
\medskip\noindent
persists for $n\geq 7$ legs.

The above expressions for ${\cal A}^{F^4}(1,2, \ldots,n)$ are not manifestly cyclic invariant in $(1,2, \ldots,n)$ because
the leg number one is treated differently. This is an artifact of the one-loop
prescription from section \secOnePS\ which associates only leg number one with the unintegrated vertex operator $V^1$.
But it can be shown that the difference to another choice of $V^{i \neq 1}$ is BRST-exact and therefore zero,
\eqn\diffexact{
{\cal A}^{F^4}(1,2,\ldots,n)  - {\cal A}^{F^4}(2,3,\ldots,n,1) = \langle Q\, {\cal X}_n\rangle = 0,
}
for example\foot{The general formula for ${\cal X}_n$ can be conveniently written using the definition
\eqn\Eijdef{
E^{ij}_{12 \ldots p} = \sum_{k=1}^{p-1} M^i_{12 \ldots k}M^j_{k+1 \ldots p}
}
as
\eqn\GenX{
{\cal X}_n = \sum_{p=2}^{n-2} M^i_{12 \ldots p} E^{jk}_{p+1 \ldots n} + {\rm tcyc}(12 \ldots n)
- \sum_{p=3}^{n-1} E^{ij}_{23 \ldots p}M^k_{p+1, \ldots n,1}
}
and ${\rm tcyc}(12 \ldots n)$ means the {\it truncated} cyclic permutations of the enclosed
indices. It is defined such that the permutations which lead to the leg number one
not being in the ``first'' $M^i$ are dropped. For example,
$M_{123}^i M_4^j M_5^k + {\rm tcyc}(12345)  = M_{123}^i M_4^j M_5^k 
+ M_{512}^i M_3^j M_4^k + M_{451}^i M_2^j M_3^k$.},
\eqnn\Xfour
\eqnn\Xfive
\eqnn\Xsix
$$\eqalignno{
{\cal X}_4 & = M_{12}^i M^j_3 M^k_4 \cr
{\cal X}_{5} & = M_{123}^i  M_4^j  M_5^k + M_{512}^i  M_3^j  M_4^k
+ M_{12}^i  (M_3^j  M_{45}^k + M_{34}^j  M_5^k) &\Xfive\cr
{\cal X}_{6} & = M_{1234}^i  M_5^j  M_6^k +  M_{6123}^i  M_4^j  M_5^k
+ M_{5612}^i  M_3^j  M_4^k
+ M_{123}^i  (M_4^j  M_{56}^k + M_{45}^j  M_6^k)\cr
& \quad + M_{612}^i  (M_3^j  M_{45}^k + M_{34}^j  M_5^k)
+ M_{12}^i  (M_{3}^j  M_{456}^k + M_{34}^j  M_{56}^k + M_{345}^j  M_6^k). &\Xsix
}$$

\subsec{Tree-level $\ap^2$ corrections versus one-loop kinematics}
\subseclab\AFvloopSec

\noindent
This subsection builds a bridge between tree-level $\ap^2$ corrections ${\cal A}^{F^4}$ and the
one-loop kinematics $C_{1,\ldots}$. Both of them have a superspace representation in terms of
$M_{d_1 \ldots d_s} M^i_{a_1 \ldots a_p} M^j_{b_1 \ldots b_q} M^k_{c_1 \ldots c_r}$ -- see the
previous subsection for ${\cal A}^{F^4}$ and \Cooo\ --- \Cddd\ for $C_{1,\ldots}$. Using the
symmetry properties \BGsymnoC\ of Berends--Giele currents, we find
\eqn\AFasC{
{\cal A}^{F^4}(1,2, \ldots,n) = \sum_{2 \leq p < q \leq n-1} \langle C_{1,23 \ldots p,p+1 \ldots q, q+1\ldots n} \rangle,
}
where legs $23\ldots n$ are distributed in all possible ways among the three slots of the BRST
invariants $C_{1, \ldots}$ which preserve their order. This leads to $(n-2)(n-3)/2$ terms in the
$C_{1,\ldots}$ expansion of the color-ordered ${\cal A}^{F^4}(1,2,\ldots,n)$, let us display examples up to
multiplicity $n=8$:
\eqnn\AFfourC
\eqnn\AFfiveC
\eqnn\AFsixC
\eqnn\AFsevenC
\eqnn\AFeightC
$$\eqalignno{
{\cal A}^{F^4}(1,2,\ldots,4) &= \langle C_{1,2,3,4}\rangle &\AFfourC\cr
{\cal A}^{F^4}(1,2,\ldots,5) & = \langle C_{1,23,4,5} \rangle
+ \langle C_{1,34,2,5} \rangle
+ \langle C_{1,45,2,3} \rangle &\AFfiveC\cr
{\cal A}^{F^4}(1,2,\ldots,6) & = \langle C_{1,234,5,6} \rangle
+ \langle C_{1,345,2,6} \rangle
+ \langle C_{1,456,2,3} \rangle
+ \langle C_{1,23,45,6} \rangle \cr
& \quad + \langle C_{1,23,56,4} \rangle
+ \langle C_{1,34,56,2}\rangle &\AFsixC\cr
{\cal A}^{F^4}(1,2,\ldots,7) & =
\langle C_{1,2345,6,7} \rangle
+ \langle C_{1,3456,2,7} \rangle
+ \langle C_{1,4567,2,3} \rangle
+ \langle C_{1,234,56,7} \rangle \cr
&\quad + \langle C_{1,234,67,5} \rangle
 + \langle C_{1,23,456,7} \rangle
+ \langle C_{1,23,567,4} \rangle
+ \langle C_{1,345,67,2} \rangle\cr
&\quad + \langle C_{1,34,567,2} \rangle
+ \langle C_{1,23,45,67}\rangle &\AFsevenC\cr
{\cal A}^{F^4}(1,2,\ldots,8) & = 
\langle C_{1,23456,7,8} \rangle 
+ \langle C_{1,34567,2,8} \rangle 
+ \langle C_{1,45678,2,3} \rangle 
+ \langle C_{1,2345,67 ,8} \rangle\cr
&\quad + \langle C_{1,2345,78,6} \rangle
+ \langle C_{1,234,567,8} \rangle 
+ \langle C_{1,234,678,5} \rangle 
+ \langle C_{1,23,4567,8} \rangle\cr 
&\quad + \langle C_{1,23,5678,4} \rangle
 + \langle C_{1,3456,78,2} \rangle
+ \langle C_{1,345,678,2} \rangle
+ \langle C_{1,34,5678,2} \rangle\cr
&\quad+ \langle C_{1,23,45,678} \rangle
+ \langle  C_{1,23,456,78} \rangle
+ \langle C_{1,234,56,78}\rangle &\AFeightC
}$$
We will argue in the next subsection that the representation \AFasC\
of ${\cal A}^{F^4}$ in terms of $C_{1,\ldots}$ is invertible, i.e. that one can express any individual
BRST invariant $C_{1,\ldots}$ in terms of ${\cal A}^{F^4}$ permutations with rational coefficients.
As promised in \promise, this implies that all kinematic ingredients $C_{1,\ldots}$ of one-loop amplitudes can
be written in terms of tree-level kinematics ${\cal A}^{F^4}$, and the latter can in principle be
expressed through ${\cal A}^{\rm YM}$ permutations (with bilinear Mandelstam coefficients). The
reduction of five-point $C_{1,ij,k,l}$ to SYM trees proceeds as follows,
\eqnn\CfiveAsAF
$$\eqalignno{
\langle C_{1,24,3,5} \rangle & =
{\cal A}^{F^4}(1,3,2,4,5)
- {\cal A}^{F^4}(1,4,3,2,5)
-  {\cal A}^{F^4}(1,3,4,5,2).&\CfiveAsAF\cr
&= {\cal A}^{\rm YM}(1,2,3,4,5)s_{34}\big[ s_{12} - s_{45} - {s_{12}s_{23}\over s_{14}}\big]\cr
&\quad + {\cal A}^{\rm YM}(1,3,2,4,5)\big[(s_{23}+s_{34})(s_{12}-s_{45})
+{s_{23}\over s_{14}}(s_{12}s_{24}- s_{51}(s_{34} + s_{45}))\big] \, ,
}$$
and we shall finally give a six-point example:
\eqnn\CtooAF
\eqnn\CtooAFd
$$\eqalignno{
3 \langle C_{1,345,2,6} \rangle &=
        {\cal A}^{F^4}(1,2,3,6,4,5)
       + {\cal A}^{F^4}(1,2,5,4,3,6)
       + {\cal A}^{F^4}(1,2,5,4,6,3)\cr
       & \quad + {\cal A}^{F^4}(1,2,6,3,4,5)
       + {\cal A}^{F^4}(1,3,2,4,5,6)
       + {\cal A}^{F^4}(1,3,2,4,6,5)\cr
       & \quad - {\cal A}^{F^4}(1,3,2,5,4,6)
       + {\cal A}^{F^4}(1,3,4,2,5,6)
       + {\cal A}^{F^4}(1,3,4,2,6,5)\cr
       & \quad + {\cal A}^{F^4}(1,3,4,6,2,5)
       + {\cal A}^{F^4}(1,6,2,3,4,5)
       + {\cal A}^{F^4}(1,6,4,3,2,5)&\CtooAF\cr
3 \langle C_{1,23,45,6} \rangle &=
       - {\cal A}^{F^4}(1,2,3,4,5,6)
       + {\cal A}^{F^4}(1,2,3,5,4,6)
       + {\cal A}^{F^4}(1,2,3,6,4,5)\cr
       & \quad - {\cal A}^{F^4}(1,2,3,6,5,4)
       + {\cal A}^{F^4}(1,2,5,3,4,6)
       - {\cal A}^{F^4}(1,2,5,6,4,3)\cr
       & \quad + {\cal A}^{F^4}(1,2,6,3,4,5)
       - {\cal A}^{F^4}(1,2,6,3,5,4)
       + {\cal A}^{F^4}(1,2,6,4,3,5)\cr
       & \quad + {\cal A}^{F^4}(1,2,6,4,5,3)
       - {\cal A}^{F^4}(1,2,6,5,3,4)
       - {\cal A}^{F^4}(1,3,2,5,4,6)\cr
       & \quad + {\cal A}^{F^4}(1,3,2,5,6,4)
       - 2{\cal A}^{F^4}(1,3,2,6,4,5)
       + 2{\cal A}^{F^4}(1,3,2,6,5,4)\cr
       & \quad + {\cal A}^{F^4}(1,3,4,5,2,6)
       + {\cal A}^{F^4}(1,6,5,2,3,4)\,. &\CtooAFd\cr
}$$
It is not difficult to verify the above relations using the explicit
expressions \AFfiveC\ and \AFsixC\ together with the symmetries obeyed
by the invariants $C_{1,\ldots}$.

\subsec{KK-like identities for ${\cal A}^{F^4}$ and finite QCD amplitudes}
\subseclab\secKKAF

\noindent
We have argued in subsection \CsymSec\ that the symmetries \BGsymC\ of the $C_{1,\ldots}$ align
them into a $S^{n-1}_3$-dimensional basis under relations with rational coefficients. This
subsection focuses on amplitude relations between ${\cal A}^{F^4}$ following from their expansion
\AFasC\ in terms of $C_{1,\ldots}$. Cyclic symmetry and $(-1)^n$ parity leave $(n-1)!/2$
potentially independent ${\cal A}^{F^4}(1,2,\ldots,n)$ permutations, but since they are all linear
combinations of $S^{n-1}_3$ independent $C_{1,\ldots}$, there must be lots of identities with
rational coefficients among them. Following the terminology of \copenhagen, we will refer to these relations
as ``Kleiss Kuijf-like'' (KK-like). The first example is ${\cal A}^{F^4}(1,2,3,4) = {\cal
A}^{F^4}(1,3,2,4)$ being totally symmetric. Examples at five-points are
\eqnn\afid
\eqnn\afit
$$\eqalignno{
0 &= {\cal A}^{F^4}(1,2,3,4,5) - {\cal A}^{F^4}(1,4,2,3,5) -  {\cal A}^{F^4}(1,2,4,5,3)\cr
& \quad + {\cal A}^{F^4}(1,2,4,3,5) - {\cal A}^{F^4}(1,3,2,4,5) -  {\cal A}^{F^4}(1,2,3,5,4),&\afid\cr
0 &= {\cal A}^{F^4}(1,2,3,4,5) + {\rm sym}(2,4,5) , &\afit
}$$
they can be easily checked using ${\cal A}^{F^4}(1,2,3,4,5) = \langle C_{1,23,4,5} + C_{1,2,34,5} + C_{1,2,3,45} \rangle$.

The basis dimension $S^{n-1}_3$ for $C_{1,\ldots}$ furnishes an upper bound on the number of independent
${\cal A}^{F^4}$ under KK-like relations (e.g.\ one has at most six independent ${\cal
A}^{F^4}(1,\si(2,3,4,5))$ under \afid\ and \afit). If this bound is saturated, then the equation
system
\eqn\AFperm{
{\cal A}^{F^4}(1,\si(2, \ldots,n)) = \sum_{2 \leq \si(p) <  \si(q) \leq n-1} \langle C_{1,\si(23 \ldots p),\si(p+1 \ldots q), \si(q+1\ldots n)} \rangle ,
}
is invertible and we can solve it for $C_{1,\ldots}$ in terms of ${\cal A}^{F^4}$ permutations. We
will now give an indirect argument that this is indeed the case.

Relations of type \afid\ and \afit\ have already been observed in \copenhagen\ for finite one-loop amplitudes
in four-dimensional pure Yang Mills theory involving gluons of positive helicity only. Using the all
multiplicity formula from \BernQK\foot{The expression \QCD\ for pure Yang Mills amplitudes $A_{n;1}^{(1)}$ was observed in \BernJA\ to agree with dimension-shifted one-loop amplitudes of $N=4$ SYM in $D \mapsto D+4$ dimensions.}
\eqn\QCD{
A_{n;1}^{(1)}(1^+,2^+,\ldots,n^+) = - {i \over 48\pi^2} { \sum_{i<j<k<l} \langle i j \rangle [jk] \langle kl \rangle [li] \over \langle 12 \rangle \langle 23 \rangle \ldots \langle n1 \rangle}
}
(with $\langle i j \rangle$ and $[i j]$ denoting products of the momentum spinors of gluons $i$ and $j$)
the authors of \copenhagen\ derive amplitude relations between different $A_{n;1}^{(1)}$
permutations and also find the basis dimension $S^{n-1}_3$ under KK-like relations. Moreover, they
develop a diagrammatic method to handle the symmetries using graphs with one quartic vertex and
otherwise cubic vertices. This strongly resembles our diagrammatic interpretation of one-loop
building blocks $\langle T_{d_1\ldots d_s} T^i_{a_1\ldots a_p} T^j_{b_1\ldots b_q} T^k_{c_1\ldots
c_r} \rangle$. Reference \BoelsMN\ puts the idea to derive relations between box coefficients from quartic expressions in Berends--Giele currents into a more general context.

The coincidence between $\ap^2$ corrections to superstring tree amplitudes and four-dimensional
pure Yang Mills amplitudes was firstly noticed in \refs{\StiebergerBH, \stienotation}. The authors
point out that the four-dimensional reduction of gluonic ${\cal A}^{F^4}$ amplitudes\foot{The
${\cal A}^{\rm YM}$ representation \treeAFdim\ of ${\cal A}^{F^4}$ is dimension-agnostic -- the
functional dependence of SYM trees on gluon polarization vectors is the same in any number of
dimensions, and one can use spinor helicity variables and the Parke Taylor formula \ParkeGB\ in
the four-dimensional MHV situation.} in MHV helicity configurations are proportional to \QCD,
\eqn\QCDF{
{\cal A}^{F^4}(1^-,2^-,3^+,4^+\ldots,n^+) \sim \langle 12 \rangle^4 \ A_{n;1}^{(1)}(1^+,2^+,\ldots,n^+)
}
up to the permutation-insensitive ``MHV--dressing'' $\langle 12 \rangle^4$. This explains why
four-dimensional MHV representatives of ${\cal A}^{F^4}$ fall into the same $S^{n-1}_3$-dimensional
basis found in \copenhagen\ under KK-like relations. In other words, the MHV components of the
${\cal A}^{F^4}$ saturate the upper bound of $S^{n-1}_3$ basis amplitudes found through our
reasoning above based on $C_{1,\ldots}$ expansions. Generalizations of four-dimensional MHV
${\cal A}^{F^4}$ to other helicities, to other supermultiplet members and to higher dimensions can
only require a larger basis than the MHV specialization, but exceeding the $S_3^{n-1}$ is incompatible with the upper
bound found for ten-dimensional superamplitudes\foot{This is not
a strict proof that the non-MHV ${\cal A}^{F^4}$ obey the {\it same} $(n-1)!/2 - S^{n-1}_3$
KK-like relations as their MHV cousins and the pure Yang Mills amplitudes \QCD, but we take strong
confidence from the fact that our ${\cal A}^{F^4}$ in their helicity-agnostic $C_{1,\ldots}$
representation obey all the $A_{n;1}^{(1)}$ amplitude relations explicitly written in
\copenhagen.}. This completes our argument why
\AFperm\ admits to express any BRST invariant $C_{1,\ldots}$ as a linear combination of
${\cal A}^{F^4}$.

To conclude this section, let us display higher order examples for KK-like identities between ${\cal A}^{F^4}$. At six-points, for instance, one can check
$$\eqalignno{
0 & ={\cal A}^{F^4}(1,5,4,3,6,2)  -  {\cal A}^{F^4}(1,5,4,2,6,3)  - {\cal A}^{F^4}(1,5,4,6,2,3)
+ {\cal A}^{F^4}(1,5,4,6,3,2) \cr
& + {\cal A}^{F^4}(1,5,6,4,2,3)  -  {\cal A}^{F^4}(1,5,6,4,3,2)  +  {\cal A}^{F^4}(1,6,2,3,4,5)
-  {\cal A}^{F^4}(1,6,3,2,4,5) \cr
& - {\cal A}^{F^4}(1,6,4,2,3,5)  -  {\cal A}^{F^4}(1,6,4,2,5,3)  +  {\cal A}^{F^4}(1,6,4,3,2,5)
+  {\cal A}^{F^4}(1,6,4,3,5,2) \cr
& - {\cal A}^{F^4}(1,6,4,5,2,3)  +  {\cal A}^{F^4}(1,6,4,5,3,2)  +  {\cal A}^{F^4}(1,6,5,4,2,3)
-  {\cal A}^{F^4}(1,6,5,4,3,2).
}$$
using \AFsixC, and a neat form for all-multiplicity relations is given in \copenhagen:
$$
0 = 2 {\cal A}^{F^4}(1,2,\ldots,n) - (-1)^n \sum_{\sigma \in {\rm OP}(\{ 4 \} \cup \{ \beta \} )} {\cal A}^{F^4}(3, \{ \sigma \},5) + {\rm sym}(45\ldots n).
$$
Similar to KK relations, the notation ${\rm OP}(\{ 4 \} \cup \{ \beta \} )$ includes all ways to shuffle
leg four into the set $\{ \beta \} = \{2,1,n,n-1, \ldots,6\}$ while preserving the order of the latter.

\subsec{BCJ-like identities for ${\cal A}^{F^4}$}

\noindent
We always pointed out that the basis dimensions $S^{n-1}_3$ for both $C_{1,\ldots}$ and ${\cal A}^{F^4}$ only
take relations with constant, rational coefficients into account which we call
KK-like. So far, we completely neglected the fact that ${\cal A}^{F^4}$ decompose into ${\cal A}^{\rm YM}$
permutations (weighted by bilinears in $s_{ij}$) which are well-known to have a
$(n-3)!$ basis under KK- and BCJ relations \BCJ. Starting from $(n=5)$-points, the ${\cal A}^{\rm
YM}$ basis contains strictly less elements than the ``KK-like'' basis of ${\cal A}^{F^4}$ since
$(n-3)! < S^{n-1}_3 $ for $n \geq 5$. Hence, there must be extra relations with Mandelstam coefficients between
${\cal A}^{F^4}$ that are independent under KK-like relations.

At five-points, extra identities with bilinear coefficients in Mandelstam variables reduce the
${\cal A}^{F^4}$ or pure Yang-Mills amplitudes $A_{n;1}^{(1)}$ to two independent ones (in
agreement with the $(n-3)!$ basis of ${\cal A}^{\rm YM}$). Examples on the $A_{n;1}^{(1)}$ side
are shown in equation (5.2) of \copenhagen, we have checked that they are also satisfied by ${\cal
A}^{F^4}$. However, the most compact relations we could find between five-point BRST invariants
involve the $C_{1,\ldots}$ rather than ${\cal A}^{F^4}$. Let $P_{i} = \sum_{j=1}^5 x_{ij} s_{j}$
denote linear polynomials in Mandelstam variables $s_i = s_{i,i+1}$ with constants $x_{ij}$, then
the ansatz
\eqnn\cansa\
$$
\eqalignno{
  s_{23} P_{1} C_{1, 23, 4, 5} 
 &+  s_{24} P_{2} C_{1, 24, 3, 5} 
 +  s_{25} P_{3} C_{1, 25, 3, 4} \cr
 +  s_{34} P_{4} C_{1, 34, 2, 5} 
 &+  s_{35} P_{5} C_{1, 35, 2, 4} 
 +  s_{45} P_{6} C_{1, 45, 2, 3} = 0 &\cansa
}$$
is sufficient to find a two-dimensional
basis of BRST invariants. The ansatz \cansa\ is motivated by the fact that the ${1 \over s_{23}}$ pole in
$C_{1,23,4,5}$ does not appear in any other $C_{1,\ldots}$, so it must be cancelled by a $s_{23}$ prefactor
for $C_{1,23,4,5}$. Plugging in the polynomials $P_i = \sum_{j=1}^5 x_{ij} s_{j}$ and solving the system of equations
which follow from a component evaluation of \cansa\ using \PSS\ lead to four independent
quadratic relations between $C_{1,\ldots}$. As a result, we can express $\{C_{1,24,3,5},C_{1,25,3,4},
C_{1,34,2,5},C_{1,35,2,4}\}$ in terms of a $\{C_{1,23,4,5},C_{1,45,2,3}\}$ basis
\eqn\solu{
C_{1,ij,k,l} = \pm {s_{1k}\over s_{ij}}\(s_{1i}s_{1l} - s_{jl}s_{ij}\) {C_{1,23,4,5}\over (s_{23}-s_{45})s_{45}}
 \pm  {s_{1l}\over s_{ij}}\(s_{1j}s_{1k} - s_{ik}s_{ij}\) {C_{1,45,2,3}\over (s_{45}-s_{23})s_{23}}
}
where the signs are given by $\{(+-),(++),(--),(-+)\}$, respectively.

\newsec{Harmony between color, kinematics and worldsheet integrands}
\seclab\harmcki

\noindent
In this section, we will explore the common combinatorial structures that govern on the one hand
the kinematic building blocks $C_{1,\ldots}$ of one-loop amplitudes and the corresponding
worldsheet integrands $X_{ij} = s_{ij} \eta_{ij}$, on the other hand also the color factors from
the $\ap^2$ corrections to tree amplitudes. In all the three cases, the basis dimensions are given
by the unsigned Stirling number $S_3^{n-1}$. It can be viewed as the one-loop analogue of the
magic number $(n-3)!$ omnipresent in tree-level bases of worldsheet integrals as well as color-ordered string- and
SYM amplitudes. This coincidence has led us to a harmonious duality between
color-dressed tree amplitudes at order $\ap^2$ and the integrand of one-loop amplitudes in open
superstring theory.

In the open string sector, the color-dressed tree amplitude is given by
\eqn\cdressed{
{\cal M}_n^{\rm tree}(\ap) = \!\!\!\!\sum_{\s \in S_{n-1}/Z_2}\!\!\!
\buildrel{\leftrightarrow}\over{{\rm Tr}}
\big[
\, T^{a_{\si(1)}} \, T^{a_{\si(2)}} \cdots T^{a_{\si(n-1)}} \, T^{a_n} \, \big]
\, {\cal A}^{\rm tree}(\si(1,2,\ldots,n-1),n;\ap)
}
where the summation includes all cyclically inequivalent permutations of the labels
modded out by the $(-1)^n$ parity of color-stripped $n$-point amplitudes. The $T^{a_i}$ denote the 
Chan--Paton factors\foot{Our normalization conventions are fixed by ${\rm Tr} [T^a T^b] = \delta^{ab}/2$
and $[T^a,T^b] = i f^{abc} T^c$.} in the fundamental representation of the gauge group, and parity weighting is represented as
\eqn\ptrace{
\buildrel\leftrightarrow\over{\rm Tr} \big[ \, T^{a_1} \, T^{a_2} \cdots  T^{a_n} \, \big] \
:= \ {\rm Tr} \big[ \, T^{a_1} \, T^{a_2} \, \ldots \,  T^{a_n}
\ + \ (-1)^n \,  T^{a_n} \, T^{a_{n-1}} \cdots  T^{a_1} \, \big]\,.
}
A convenient basis for these parity weighted traces involves structure constants $f^{abc}$ and symmetrized
traces $d^{a_1a_2a_3 \ldots a_{2n}}$ of even rank only, the latter being defined as \Vcolor,
\eqn\symtrace{
d^{a_1a_2\ldots a_{2n}} \ := \ \frac{1}{(2n-1)!} \, \sum_{\si \in S_{2n-1}} \, {\rm Tr}
\big[ \, T^{a_{\si(1)}}  \cdots T^{a_{\si(2n-1)}} \, T^{a_{2n}} \, \big] \,.
}
We will use shorthands $f^{123} \equiv f^{a_1 a_2 a_3}$ and $d^{12\ldots k} \equiv d^{a_1a_2\ldots a_{k}}$ for the (adjoint) color degrees of freedom.

As mentioned in \Vcolor, the explicit computation of symmetrized traces is tedious to perform by
hand but it is also well-suited for a computer implementation. The first non-trivial relations
are relatively compact \refs{\bilal,\Hunter}
\eqnn\ptracethree
\eqnn\ptracefour
\eqnn\ptracefive
$$\eqalignno{
\buildrel{\leftrightarrow}\over{\rm Tr}  (T^{1} \, T^{2} \, T^{3})
&= {i \over 2} f^{123} &\ptracethree \cr
\buildrel{\leftrightarrow}\over{\rm Tr}  (T^{1} \, T^{2} \, T^{3} \, T^{4})
&= 2d^{1234}
+  \frac{1}{6}\bigl(  f^{23n} \, f^{14n} - f^{12n} \, f^{34n} \bigr), &\ptracefour\cr
\buildrel{\leftrightarrow}\over{\rm Tr}(T^1 T^{2} T^{3} T^{4} T^{5})
&=  {i \over 12} \big[\, f^{12 a}f^{a 4 b}f^{b 3 5} -  3 f^{12 a}f^{a 3 b}f^{b 4 5}  +  f^{13 a}f^{a 2 b}f^{b 4 5} \cr
 &\qquad +  f^{13 a}f^{a 4 b}f^{b 2 5}  +  f^{14 a}f^{a 2 b}f^{b 3 5}  -  f^{14 a}f^{a 3 b}f^{b 2 5} \big] \cr
 &\quad + i\big[\, f^{23 a}d^{a 145}  +  f^{24 a}d^{a 135}  +  f^{25 a}d^{a 134}  \cr
 &\qquad +  f^{34 a}d^{a 125}  +  f^{35 a}d^{a 124}  +  f^{45 a}d^{a 123}  \big],&\ptracefive
}$$
but the lenghty relations for $n=6$ and $7$ were computed
using the {\tt color} package of {\tt FORM} \FORM\ and the $n=6$ case can be found in the Appendix~\ColorApp.
Note that \ptracefour\ and \ptracefive\ have been cast into a minimal form in the sense that all the generalized Jacobi relations
\eqnn\Jacobi
\eqnn\genJacobi
$$\eqalignno{
f^{a[ij}f^{k]ab} &= 0 &\Jacobi\cr
d^{a(ijk}f^{l)ab} & = 0 &\genJacobi
}$$
are taken into account. For the color structures involving a symmetrized trace, this amounts to placing leg number one to the $d^{1\ldots}$.

Once the color-dressed disk amplitude \cdressed\ is rewritten in this color basis, the subamplitude
relations at various orders in $\ap$ impose selection rules on what kinds of tensor contribute to ${\cal M}_n^{\rm tree}(\ap)$
at the order in question. Keeping the first two terms in \treeAF\ $\sim \ap^0,\ap^2$, the KK identities \KK\ between
${\cal A}^{\rm YM}$ select those color tensors with $n-2$ powers of structure constants and project out any
symmetrized trace \symtrace. The subamplitudes ${\cal A}^{F^4}$ associated with the first string corrections,
on the other hand, satisfy another set of relations which we called KK-like in the discussion of subsection \secKKAF.
They select those color tensors made of $n-4$ structure constants $f^{abc}$ and one symmetrized four-trace
$d^{1234} := {1\over 6}  \sum_{\si \in S_{3}}  {\rm Tr}
\big[ T^{\si(1)} T^{\si(2)} T^{\si(3)} T^{4}  \big]$. This ties in with the symbolic vertices $D^{abcd}$
and $F^{abc}$ used in \copenhagen\ to gain intuition for the amplitude relation between finite one-loop
pure Yang Mills amplitudes $A_{n;1}^{(1)}$. The color tensors $d^{a_1a_2a_3a_4} (f^{bcd})^{n-4}$ selected
by ${\cal A}^{F^4}$ are another manifestation of their intimate connection to the $A_{n;1}^{(1)}$.

As a first example, let us consider the four-point color-dressed amplitude up to ${\cal O}(\ap^2)$,
\eqnn\fourdr
$$\eqalignno{
{\cal M}_4^{\rm tree}(\ap) = - \, \frac{1}{2} \, &\big( \, f^{12a} \, f^{a34} \, {\cal A}^{\rm YM}(1,2,3,4)
\ + \ f^{13a} \, f^{b24} \, {\cal A}^{\rm YM}(1,3,2,4) \, \big) \cr
& \ + \ 6\zeta(2) \ap^2  \, d^{1234} \, {\cal A}^{F^4}(1,2,3,4) \ + \ {\cal O}(\ap^3) , &\fourdr
}$$
see \bilal\ for the color structure at higher order in $\ap$. The notation for higher multiplicity versions of \cdressed\ shall be lightened using
\eqn\cnot{
{\cal M}^{\rm tree}_n(\ap) \ \equiv \ {\cal M}_n^{\rm YM} \ + \ \zeta(2) \, \ap^2 {\cal M}_n^{F^4}
\ + \ {\cal O}(\ap^3) ,
}
and the $\ap^2$ correction ${\cal M}_n^{F^4}$ will be the object of main interest in this section where we show its tight connection to the one-loop integrand \nfinal.

Before looking at the color tensor structure at order $\ap^2$ and their interplay with ${\cal A}^{F^4}$
symmetries, let us review the color organization at the SYM level $\ap^0$. At five-points, the
KK relations for the field theory subamplitudes yield
$$\eqalignno{
 {\cal M}_5^{\rm YM} & =  -  {i \over 2} {\cal A}^{\rm YM}(1,2,3,4,5) f^{12 a}f^{a 3 b}f^{b 4 5} +
 {\rm sym}(234)\cr
}$$
in agreement with the color-decomposition proposed by \dixonduca. More generally, this reference suggests the following $(n-2)!$ element Kleiss--Kuijf bases 
\eqn\KKbases{
\big\{ f^{1\si(2)a} \,f^{a \si(3)b } \, \cdots \, f^{z\si(n-1)n}, \ \sigma \in S_{n-2} \big\} , \quad \big\{
{\cal A}^{\rm YM}\big( 1,\si(2,3,\ldots,n-1),n \big) , \ \sigma \in S_{n-2} \big\}
}
for the color factors $(f^{bcd})^{n-2}$ and the SYM subamplitudes (using Jacobi identities for the former
and KK relations for the latter). In this setting, one can reproduce the $(n-2)!$ color-decomposition proven in \dixonduca
\eqn\ftcolor{
{\cal M}^{\rm YM}_n = {i^{n-2}\over 2}  \sum_{\si \in S_{n-2}}
f^{1\si(2)a} \,f^{a \si(3)b } \, \cdots \, f^{z\si(n-1)n} \,
{\cal A}^{\rm YM}\big( 1,\si(2,3,\ldots,n-1),n \big) \,,
}
starting from \cdressed, and the cancellation of $d^{12\ldots 2k}$ contributions at order $\ap^0$ becomes
manifest due to KK relations. In the remainder of this section, we will find remnants of \ftcolor\ in
${\cal M}^{F^4}_n$, in particular the basis choice \KKbases\ for $(f^{bcd})^{n-2}$ color factors is
path-breaking for the organization of the color tensors $d^{a_1a_2a_3a_4} (f^{bcd})^{n-4}$ relevant at $\ap^2$ order.

\subsec{The color-dressed $(n \leq 7)$-point disk amplitude at order $\ap^2$}
\subseclab\sevenone

\noindent
Keeping the dual bases \KKbases\ for ${\cal M}^{\rm YM}_n$ in mind, we shall next give $n=5,6,7$-point
results for ${\cal M}^{F^4}_n$. According to \ptracefive, the five-point color tensors $d^{aijk} f^{lab}$
are brought into a (six element) basis of $d^{aij1} f^{lab}$ (with leg one attached to the $d$ tensor) via
generalized Jacobi identity $d^{a(ijk} f^{l)ab}=0$. This leads to a compact result for ${\cal M}^{F^4}_5$:
\eqnn\fivedress
$$\eqalignno{
{\cal M}_5^{F^4} & =   \!\!\sum_{\s \in S_{4}/Z_2}\!
\buildrel{\leftrightarrow}\over{{\rm Tr}}
\big[
\, T^{a_{\si(1)}} \, T^{a_{\si(2)}} T^{a_{\si(3)}} T^{a_{\si(4)}} \, T^{a_5} \, \big]
\, {\cal A}^{F^4}(\si(1,2,3,4),5) \cr
&=6 i \langle C_{1, 23, 4, 5}\, f^{23 a}d^{a 145}
 +   C_{1, 24, 3, 5}\, f^{24 a}d^{a 135}
 +   C_{1, 25, 3, 4}\, f^{25 a}d^{a 134} \cr
& \qquad +   C_{1, 34, 2, 5}\, f^{34 a}d^{a 125}
 +   C_{1, 35, 2, 4}\, f^{35 a}d^{a 124}
 +   C_{1, 45, 2, 3}\, f^{45 a}d^{a 123} \rangle &\fivedress
}$$
First of all, the symmetries of ${\cal A}^{F^4}$ imply that color factors of type $f^{1\si(2)a} f^{a\si(3)b} f^{b\si(4)5}$
drop out, see the first two lines of \ptracefive. Secondly, the expansion
${\cal A}^{F^4}(1,2,3,4,5) = \langle C_{1,23,4,5} + C_{1,2,34,5} + C_{1,2,3,45} \rangle$ makes all contributions
to a fixed (basis) color tensor $f^{ija}d^{a1kl}$ collapse to one single term $C_{1,ij,k,l}$\foot{
Going through the calculation reveals that the terms
proportional to $f^{23 a}d^{a 145}$ are
$$\eqalignno{
&       {\cal A}^{F^4}(1,2,3,4,5)
       + {\cal A}^{F^4}(1,2,3,5,4)
       + {\cal A}^{F^4}(1,2,4,3,5)
       + {\cal A}^{F^4}(1,2,4,5,3)\cr
       + &{\cal A}^{F^4}(1,2,5,3,4)
       + {\cal A}^{F^4}(1,2,5,4,3)
       - {\cal A}^{F^4}(1,3,2,4,5)
       - {\cal A}^{F^4}(1,3,2,5,4)\cr
       - &{\cal A}^{F^4}(1,3,4,2,5)
       - {\cal A}^{F^4}(1,3,5,2,4)
       + {\cal A}^{F^4}(1,4,2,3,5)
       - {\cal A}^{F^4}(1,4,3,2,5)\cr
}$$
and yet they collapse to a single term $6C_{1,23,4,5}$ once the relation \AFfiveC\ and the
symmetries of the one-loop BRST invariants are used. }. The precise correspondence $C_{1,ij,k,l} \leftrightarrow f^{ij a}d^{a1kl}$ between kinematic and
color basis elements is a non-trivial
reorganization when looked from the perspective of the composing ${\cal A}^{F^4}$ terms, especially
at higher orders.

Even more striking cancellations occur
when the symmetrized trace decompositions of the Appendix \ColorApp\ are used
to evaluate the six- and seven-point color-dressed amplitudes at order $\ap^2$: The $35=S^{5}_3$ term sum in the six-point case
\eqnn\sixdress
\eqnn\sevendress
$$\eqalignno{
- \, \frac{1}{6} \, {\cal M}_6^{F^4}
 & = \langle C_{1, 23, 45, 6} f^{23 a}d^{a 16 b}f^{b 45}  \ + \  C_{1, 23, 46, 5} f^{23 a}d^{a 15 b}f^{b 46}  
 \ + \  C_{1, 23, 56, 4} f^{23 a}d^{a 14 b}f^{b 56}  \cr & 
 + \ C_{1, 24, 35, 6} f^{24 a}d^{a 16 b}f^{b 35}  \ + \ C_{1, 24, 36, 5} f^{24 a}d^{a 15 b}f^{b 36}  \ + \ 
  C_{1, 24, 56, 3} f^{24 a}d^{a 13 b}f^{b 56}   \cr & 
 +  \ C_{1, 25, 34, 6} f^{25 a}d^{a 16 b}f^{b 34}  \ + \  C_{1, 25, 36, 4} f^{25 a}d^{a 14 b}f^{b 36}   \ + \ 
 C_{1, 25, 46, 3} f^{25 a}d^{a 13 b}f^{b 46}  \cr &
 + \ C_{1, 26, 34, 5} f^{26 a}d^{a 15 b}f^{b 34}  \ + \  C_{1, 26, 35, 4} f^{26 a}d^{a 14 b}f^{b 35}   \ + \
   C_{1, 26, 45, 3} f^{26 a}d^{a 13 b}f^{b 45}  \cr &
 + \  C_{1, 34, 56, 2} f^{34 a}d^{a 12 b}f^{b 56} \ + \ C_{1, 35, 46, 2} f^{35 a}d^{a 12 b}f^{b 46}  
 \ + \  C_{1, 36, 45, 2} f^{36 a}d^{a 12 b}f^{b 45} \cr
&+ \ \big[ \, C_{1, 234, 5, 6} f^{23 a}f^{a 4 b} \ + \ C_{1, 243, 5, 6} f^{24 a}f^{a 3 b} \, \big]  d^{b 156} \cr
 & + \ \big[ \, C_{1, 235, 4, 6} f^{23 a}f^{a 5 b} \
 + \  C_{1, 253, 4, 6} f^{25 a}f^{a 3 b} \, \big] d^{b 146}  
   \cr & 
 + \ \big[  \, C_{1, 245, 3, 6} f^{24 a}f^{a 5 b} \ + \ C_{1, 254, 3, 6} f^{25 a}f^{a 4 b} \, \big] d^{b 136}  \cr
 &+ \ \big[ \, C_{1, 236, 4, 5} f^{23 a}f^{a 6 b} \ + \ C_{1, 263, 4, 5} f^{26 a}f^{a 3 b} \, \big] d^{b 145} \cr
 &+ \ \big[ \, C_{1, 246, 3, 5} f^{24 a}f^{a 6 b} \ + \ C_{1, 264, 3, 5} f^{26 a}f^{a 4 b} \, \big]
 d^{b 135}  \cr 
 & + \ \big[ \, C_{1, 256, 3, 4} f^{25 a}f^{a 6 b} \ + \ C_{1, 265, 3, 4} f^{26 a}f^{a 5 b} \, \big] d^{b 134}  \cr
  & + \ \big[ \, C_{1, 345, 2, 6} f^{34 a}f^{a 5 b} \ + \ C_{1, 354, 2, 6} f^{35 a}f^{a 4 b} \, \big] d^{b 126}   \cr
& + \ \big[ \, C_{1, 346, 2, 5} f^{34 a}f^{a 6 b} \ + \ C_{1, 364, 2, 5} f^{36 a}f^{a 4 b} \, \big] d^{b 125}   \cr 
& +  \ \big[ \, C_{1, 356, 2, 4} f^{35 a}f^{a 6 b} \ + \ C_{1, 365, 2, 4} f^{36 a}f^{a 5 b} \, \big] d^{b 124}  \cr
& + \ \big[ \, C_{1, 456, 2, 3} f^{45 a}f^{a 6 b} \ + \ C_{1, 465, 2, 3} f^{46 a}f^{a 5 b} \, \big] d^{b 123} \rangle  
&\sixdress\cr
}$$
exhibits color-kinematic-correspondence
$$
C_{1, 23, 45, 6} \leftrightarrow f^{23 a}d^{a 16 b}f^{b 45},\quad C_{1, 234, 5, 6} \leftrightarrow f^{23 a}f^{a 4 b} d^{b156} .
$$
Likewise, the $225 = S^6_{3}$ terms in
$$\eqalignno{
\frac{i}{6} \, {\cal M}_7 ^{F^4}  &= \langle 15 \ {\rm terms} \ \big[ \, \langle C_{1,2345,6,7}f^{23a} f^{a4b}f^{b5c} + {\rm
 sym}(345) \, \big] d^{c167} \cr
  &+ \ 60 \ {\rm terms} \ \big[ \, C_{1,234,56,7} f^{23a} f^{a4b} \ + \ (3 \leftrightarrow 4) \, \big]  f^{56c} d^{bc17} \cr
& +\ 15 \ {\rm terms} \ 
C_{1, 23, 45, 67} f^{23 a}f^{45 b}f^{67 c}d^{1 abc}   \rangle
  &\sevendress
}$$
allow to read off the dictionary
\eqnn\sevendict
$$\eqalignno{
C_{1, 23, 45, 67}  &\leftrightarrow f^{23 a}f^{45 b}f^{67 c}d^{1 abc} \cr
C_{1,234,56,7} &\leftrightarrow f^{23a} f^{a4b} f^{56c} d^{bc17} &\sevendict\cr
C_{1,2345,6,7} &\leftrightarrow f^{23a} f^{a4b}f^{b5c}d^{c167} .
}$$
In the next subsection, we shall put these observations into a more general context. Note that $d^{123456}$ and
$d^{12345a} f^{a67}$ tensors (or more generally $d^{a_1 \ldots a_6} (f^{bcd})^{n-6}$ and $d^{a_1 \ldots a_{2k}} (f^{bcd})^{n-2k}$
at $k \geq 3$) from the rank $\geq 6$ traces do not contribute at ${\cal O}(\ap^2)$ because of the
KK-like amplitude relations between ${\cal A}^{F^4}$.

\subsec{Dual bases in color and kinematic space}
\subseclab\seventwo

\noindent
We conclude from the calculations above that the BRST invariants $C_{1,\ldots}$ are natural objects to
appear not only in the one-loop integrand but also in color-dressed tree-level amplitudes. According
to \AFasC, they are related to subamplitudes ${\cal A}^{F^4}$ at order $\ap^2$ by a change of ($S^{n-1}_3$ element)
basis with coefficients $\pm 1$ and automatically solve their KK-like relations. Moreover, the
$C_{1,a_1\ldots a_p,b_1\ldots,b_q,c_1\ldots c_r}$ inherit all symmetry properties of the Berends--Giele
current triplet $M^i_{a_1\ldots a_p} M^j_{b_1\ldots,b_q} M^k_{c_1\ldots c_r}$, see subsection \CsymSec.
This makes their $S_{3}^{n-1}$ basis under relations with constant coefficients manifest and leads to the
observed harmony with the symmetries of color tensors $d^{a_1a_2a_3a_4} (f^{bcd})^{n-4}$.

In fact, arriving at the simple results \fivedress, \sixdress\ and \sevendress\ for the $\ap^2$ correction to the color-dressed amplitude
crucially relies on the fact that the dimension of the basis for color factors and the kinematics
matches. This fact has been exploited to choose ``compatible'' bases
of color structures and corresponding kinematics, generalizing the tree-level correspondence \KKbases\ between
color factors $(f^{bcd})^{n-2}$ and ${\cal A}^{\rm YM}$ in their $(n-2)!$ KK bases. In the SYM case,
the reduction to $(n-2)!$ bases makes use of Jacobi identities on the color side and the KK relations for the subamplitudes.

We shall now explain why also the $d^{a_1a_2a_3a_4} (f^{bcd})^{n-4}$ color factors align into a basis
of $S^{n-1}_3$ elements. The reduction algorithm consists of two steps:
\medskip
\parindent=32pt
\item{1.} Move label number one to the symmetrized four-trace, i.e.\
$d^{ijk \ldots} (f^{bcd})^{n-4} \mapsto \sum d^{1pq \ldots} (f^{bcd})^{n-4}$, by repeated use of generalized
Jacobi identities \genJacobi. At five-points, one applications is enough,
\eqn\jacenough{
d^{234a}f^{1ab} = - d^{123a}f^{4ab} - d^{412a}f^{3ab} - d^{341a}f^{2ab}.
}
For six-points there are two possibilities for the color factors which do not
contain the label number one inside the symmetrized trace $d^{ijkl}$. The number of color space propagators $\delta^{ab}$ between leg number one and $d^{ijkl}$ is either one (as in
$f^{34a} f^{1ab} d^{b246}$) or two (as in
$f^{12a}f^{a3b}d^{b456}$). For the one-propagator-link one uses the generalized Jacobi identity
\jacenough, whereas in the two-propagator-case the identity
\eqn\firstorder{
f^{12a}f^{a3b}d^{b456} =
- f^{12a} d^{a45b} f^{b63}
- f^{12a} d^{a64b} f^{b53}
- f^{12a} d^{a56b} f^{b43}
}
reduces it to terms of one-propagator form where \jacenough\ can be applied. The
analysis for higher-points is analogous, with more successive applications of \genJacobi\ needed. The possibility
to reduce $d^{ijkl} (f^{bcd})^{n-4} \mapsto \sum d^{1pqr} (f^{bcd})^{n-4}$ is the color dual of the
finding in subsection \CsymSec\ that any $C_{i,\ldots}$ with $i\neq 1$ can be expressed as a
sum of $C_{1,\ldots}$ in the BRST cohomology.
\item{2.} Mod out the $d^{1pqr} (f^{bcd})^{n-4}$ by Jacobi identities \Jacobi\ among the $f^{n-4}$ factors:
Consider a generic color structure $d^{1\ldots}$ passing the first selection rule,
\eqnn\ctype
$$\eqalignno{
d^{1x_py_{q}z_{r}} \, f^{a_1 a_2 x_2} \, f^{x_2 a_3 x_3} \, \cdots \, f^{x_{p-2} a_{p-1} x_{p-1}}
\, &f^{x_{p-1} a_{p} x_{p}} \cr
\times \, f^{b_1b_2 y_2}  \, f^{y_2 b_3 y_3} \, \cdots \, f^{y_{q-2} b_{q-1} y_{q-1}} \, &f^{y_{q-1} b_q y_q} \cr
 \times \, f^{c_1 c_2 z_2} \, f^{z_2 c_3 z_3} \, \cdots \, f^{z_{r-2} c_{r-1} z_{r-1}} \,&f^{z_{r-1}c_r z_r}.
 &\ctype
}$$
Each of the remaining three slots of $d^{1x_py_qz_r}$ can adjoin a tree subdiagram with $p,q$ and
$r$ external legs, respectively, such that $p+q+r=n-1$. Within the color tensors of each subdiagram like
$f^{a_1 a_2 x_2} f^{x_2 a_3 x_3} \ldots f^{x_{p-2} a_{p-1} x_{p-1}}f^{x_{p-1} a_{p}
x_{p}}$, we can eliminate the Jacobi redundancy in analogy to the tree-level KK basis \KKbases. This amounts to
the convention that $a_1$ is kept fixed at the ``outmost'' structure constant $f^{a_1 \si(a_2) x_2}$ whereas
the remaining free indices $\{ a_2,a_3,\ldots , a_p\}$
can appear in any of the $(p-1)!$ possible permutations. Then, the collection of $f^{a_1 \si(a_2)
x_2} f^{x_2 \si(a_3) x_3} \ldots f^{x_{p-2} \si(a_{p-1}) x_{p-1}}f^{x_{p-1} \si(a_{p}) x_{p}}$
with $\si \in S_{p-1}$ exhaust all Jacobi-independent half-ladder diagrams with fixed endpoints $a_1$ and $x_p$.
The kinematical dual is the reduction of $C_{1,\s(a_1 \ldots a_p), \ldots}$ to the smaller set of
$C_{1,a_1\s(a_2 \ldots a_p), \ldots}$.
\medskip
\parindent=20pt

\noindent
After this two--step reduction, the basis dimension for the color factors $d^{ijkl} (f^{bcd})^{n-4}$ is manifestly
equal to the unsigned Stirling number $S^{n-1}_3$, the same number which governs the
number of independent BRST invariants $C_{1,\ldots}$.

A more intuitive understanding of the interplay between kinematic- and color basis can be found by inspecting the unique term $V_1 
M^i_{a_1\ldots a_p} M^j_{b_1\ldots b_q} M^k_{c_1\ldots c_r}$ in $C_{1, a_1\ldots a_p, b_1\ldots b_q, c_1\ldots c_r}$
with the standalone unintegrated vertex operator $V_1$, see the explicit expression in subsection \Csubsec. The
ellipsis in $C_{1, a_1\ldots a_p, b_1\ldots b_q, c_1\ldots c_r} = V_1 
M^i_{a_1\ldots a_p} M^j_{b_1\ldots b_q} M^k_{c_1\ldots c_r} + \ldots$ obeys the same
symmetry properties, so the first term is a valid representative for the symmetry analysis. Recall that the
Berends--Giele currents $M^i_{a_1\ldots a_p}$ are color-ordered $(p+1)$-point amplitudes with leg
number $p+1$ off-shell (corresponding to the color index $x_p$ which is contracted into the box
$d^{1x_py_qz_r}$), see \Cbox. Within each of these three off-shell subamplitudes $M^i_{a_1\ldots a_p} $, we pick a Kleiss--Kuijf
basis where, again, $a_1$ is kept fixed as the first subscript of $M^i_{a_1 \ldots}$ and $a_{\geq 2}$ can appear in any permutation.

\ifig\Cbox{The diagram associated with the leading term
of $$\abovedisplayskip=2pt C_{1, a_1\ldots a_p, b_1\ldots b_q, c_1\ldots c_r} =  V_1 \, M^i_{a_1\ldots a_p} \, M^j_{b_1\ldots b_q} \, M^k_{c_1\ldots c_r}
+ {\rm BRST \ invariant \ completion}.\belowdisplayskip=2pt$$
The Kleiss--Kuijf relations for the tree
subdiagrams represented by $M^i_{a_1\ldots a_p}, \ M^j_{b_1\ldots b_q}$ and $M^k_{c_1\ldots c_r}$
yield all identities between the permutations $\si,\rho,\pi$ in $C_{1, \si(a_1\ldots a_p), \rho(b_1\ldots b_q), \pi( c_1\ldots c_r)}$.
}
{\epsfxsize=0.50\hsize\epsfbox{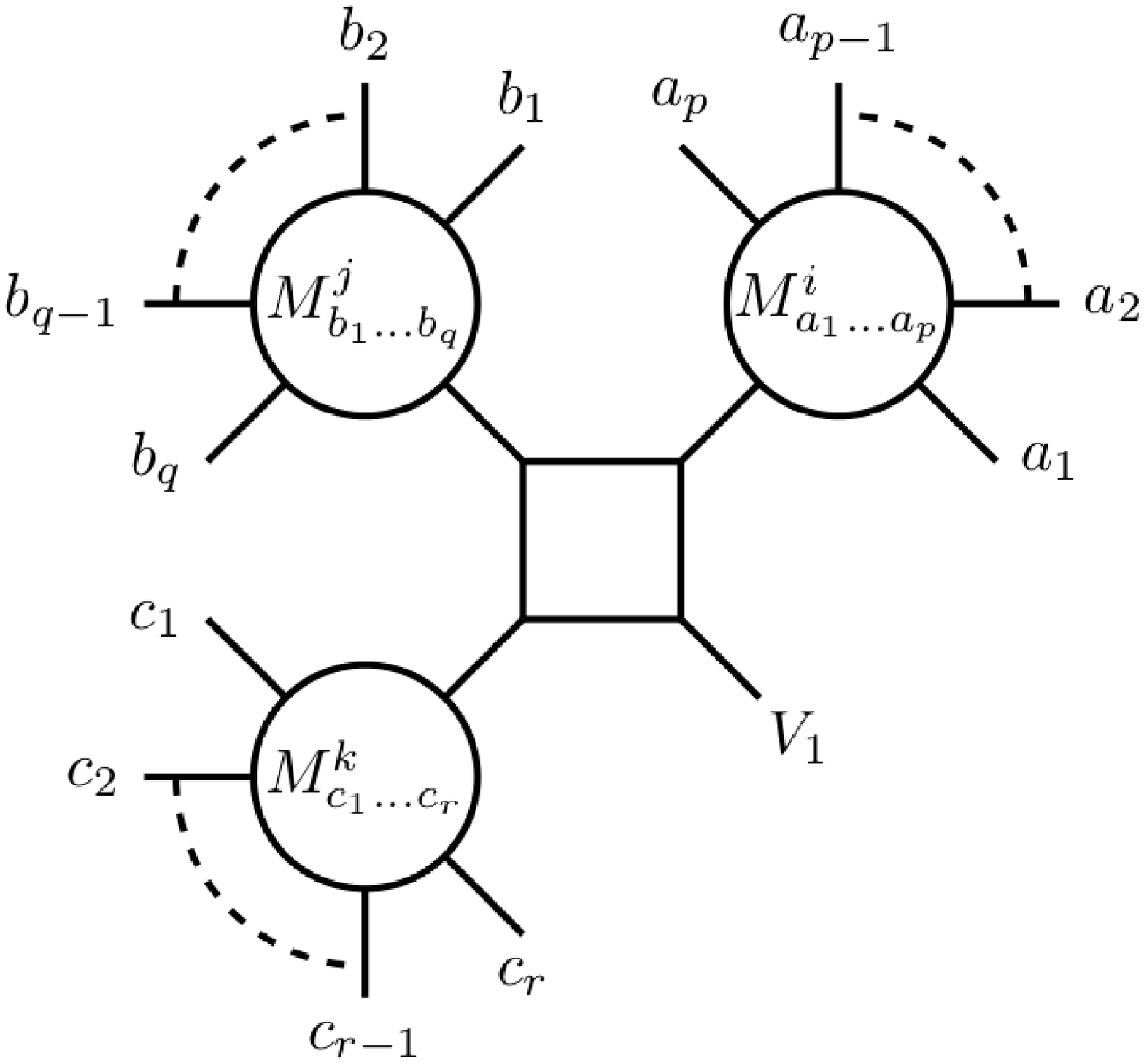}}

Each of the KK basis elements $M^i_{a_1 \si(a_2 \ldots a_p)}$ is accompanied by a $f^{p-1}$ color factor which is adapted to
the permutation $\si \in S_{p-1}$ according to the tree-level rule \ftcolor:
$$
M^i_{a_1 \si(a_2 \ldots a_p)} \ \ \leftrightarrow \ \ f^{a_1 \si(a_2) x_2} \, f^{x_2 \si(a_3) x_3}
\, \cdots \, f^{x_{p-1} \si(a_{p}) x_{p}}\,.
$$
The three chains of $f$ corresponding to the $M_{\ldots}^i, M_{\ldots}^j$ and $M_{\ldots}^k$ are then contracted with
the $x_p,y_{q},z_{r}$ indices of $d^{1x_py_{q}z_{r}}$, i.e. glued to the three corners of the box where leg one is not
attached to. This amounts to the following rule how the dual $S_3^{n-1}$ element bases for color- and kinematic factors
enter ${\cal M}_{n}^{F^4}$: Permutations of $C_{1, a_1\ldots a_p, b_1\ldots b_q, c_1\ldots c_r}$
for fixed sets $\{ a_1,a_2,\ldots,a_p\}, \ \{ b_1,b_2,\ldots,b_q\}$ and $\{ c_1,c_2,\ldots,c_r\}$
always appear in the combination
\eqnn\cpatt
$$\eqalignno{
 \sum_{\si \in S_{p-1}} &\sum_{\rho \in S_{q-1}}
 \sum_{\pi \in S_{r-1}} C_{1,a_1 \si(a_2\ldots a_p), b_1 \rho(b_2\ldots b_q), c_1 \pi (c_2\ldots c_r)}
  \times   f^{a_1 \si(a_2) x_2} \, f^{x_2 \si(a_3) x_3} \, \cdots \, f^{x_{p-1} \si(a_{p}) x_{p}} \cr
&\times  f^{b_1\rho(b_2) y_2}  \, f^{y_2 \rho(b_3) y_3} \, \cdots \,  f^{y_{q-1} \rho(b_q) y_q}
 \times  f^{c_1 \pi(c_2) z_2} \, f^{z_2 \pi(c_3) z_3} \, \cdots \,f^{z_{r-1}\pi(c_r) z_r}\, d^{1x_p y_q z_r},
}$$
in agreement with our results \fivedress, \sixdress\ and \sevendress\ for ${\cal M}_{n\leq7}^{F^4}$. This can be
recognized as sum over the $S_{3}^{n-1}$ partitions of legs $23\ldots n$ into three cycles, see subsection \Stirnot\
for the associated set ${\cal S}_3^{n-1}$. Using the latter notation defined in \defstirling, we can
compactly write the $n$-point color-dressed amplitude as
\eqn\nCf{
\eqalign{
 &\, {\cal M}_n^{F^4}   = 6i^n \! \! \! \! \! \sum_{\sigma \times \pi \times \rho \in {\cal S}_3^{n-1}} \! \! \! \! \!  \langle C_{1,\sigma(23 \ldots p), \pi(p+1 \ldots q), \rho(q+1 \ldots n)}\rangle \, d^{1x_py_q z_n} \, \sigma
\left(f^{23x_3} \,
f^{x_3 4 x_4} \, \cdots \, f^{x_{p-1} p x_p}  \right) \cr
&  \pi
\left( f^{p+1,p+2,y_{p+2}} \, f^{y_{p+2},p+3,y_{p+3}} \, \cdots \,
  f^{y_{q-1},q,y_q} \right) \ \rho \left(  f^{q+1,q+2,z_{q+2}} \, f^{z_{q+2},q+3,z_{q+3}} \, \cdots \, f^{z_{n-1},n,z_n}   \right) .
}}
As in \nfinal, the numbers $p$ and $q$ are defined through the cardinality of the
permutations to be $p=|\sigma|+1$ and $q-p=|\pi|$.

\subsec{Duality between one-loop integrands and ${\cal M}_n^{F^4}$}
\subseclab\auchnoch

\noindent
This subsection is devoted to the close relationship between ${\cal M}_n^{F^4}$ and the one-loop
kinematic factor $K_n$. Our final expressions \fivefinal, \sixfinal, \sevenfinal\ and \nfinal\ for
$K_5,K_6,K_7$ and $K_n$ can be obtained from the corresponding ${\cal M}_n^{F^4}$ using a
well-defined one-to-one map between $d^{1pqr} (f^{bcd})^{n-4}$ colors factors and the
$(X_{rs})^{n-4}$ polynomials in the worldsheet integrand. The color basis choice of having leg one
attached to $d^{1 \ldots}$ corresponds to integrating by parts on the worldsheet such that only $X_{rs}$ with
$r,s\neq 1$ enter the minimal form of $K_n$.

Let us start with lower order examples for the $d^{1pqr} (f^{bcd})^{n-4} \leftrightarrow
(X_{rs})^{n-4}$ dictionary. First of all, $K_4 = {\cal A}^{F^4}(1,2,3,4)$ is related to ${\cal
M}_4^{F^4}$ via
\eqn\fourmap{
6 d^{1234} \ \ \longleftrightarrow \ \ 1 .
}
Comparing the representation \fivefinal\ for $K_5$ with \fivedress\ yields the five-points dictionary,
\eqn\fivemap{
6i f^{23a}d^{a145} \ \ \longleftrightarrow \ \ X_{23}.
}
The two six-point topologies $C_{1,234,5,6}, C_{1,23,45,6}$ in $K_6$ and ${\cal M}_6^{F^4}$ (given by \sixfinal\ and \sixdress, respectively) are accompanied by
\eqnn\sixmap
$$\eqalignno{
- \, 6 \, f^{23a}f^{a4b} d^{b156} \ \ &\longleftrightarrow \ \ X_{23} \, (X_{24} +
X_{34}) \cr
- \, 6 \, f^{23a}f^{45b} d^{ab16} \ \ &\longleftrightarrow \ \ X_{23} \, X_{45} \ ,&\sixmap
}$$
and the $C_{1,2345,6,7}, C_{1,234,56,7} $ and $C_{1,23,45,67}$ at seven-points are dressed by
\eqnn\sevenmap
$$\eqalignno{
- \, 6i \, f^{23a}f^{a4b} f^{b5c} d^{c167} \ \ &\longleftrightarrow \ \   X_{23} \,
(X_{24} + X_{34}) \, (X_{25}+X_{35}+X_{45}) \cr
- \, 6i \, f^{23a}f^{a4b} f^{56c} d^{bc17} \ \ &\longleftrightarrow \ \  X_{23} \,
(X_{24} + X_{34}) \, X_{56} \cr
- \, 6i \, f^{23a}f^{45b} f^{67c} d^{abc1} \ \ &\longleftrightarrow \ \  X_{23} \, X_{45}
\,X_{67} , &\sevenmap
}$$
see \sevenfinal\ for $K_7$ and \sevendress\ for ${\cal M}_7^{F^4}$, respectively.

Both sides of the mappings \fourmap\ to \sevenmap\ have the same symmetries in the labels $23\ldots n$ -- the left hand side because
of Jacobi identities, the right hand side due to algebraic identities such
as $X_{23}(X_{24}+X_{34})+{\rm cyc}(234)=0 \ \leftrightarrow \ f^{[23|a} f^{a|4]b}=0$ or
$$\eqalignno{
0 & = X_{23} \, (X_{24} + X_{34}) \, (X_{25}+X_{35}+X_{45}) \ - \ (4\leftrightarrow 5) \cr
& \quad + \ X_{45} \, (X_{42} + X_{52}) \, (X_{43}+X_{53}+X_{23}) \ - \ (2\leftrightarrow 3)
}$$
corresponding to $f^{23a} f^{a[4|b} f^{b|5]c} + f^{45a} f^{a[2|b} f^{b|3]c} =0$ (which
in turn reflects the ``third'' BRST symmetry $T_{23[45]} + T_{45[23]} = 0$ under the
map \treeten).

More generally, the three independent cubic subdiagrams contracted with the $x_p,y_{q},z_{r}$
indices of $d^{1x_py_{q}z_{r}}$ each correspond to a separate nested product of worldsheet
functions like $\prod_{k=3}^p\sum_{m=2}^{k-1} X_{mk}$:
$$
 f^{23x_3} \, f^{x_34x_4} \, \ldots \, 
f^{x_{p-1} p x_p} \ \ \longleftrightarrow \ \ X_{23} \, (X_{24}+X_{34}) \, \ldots \, (X_{2p} 
+ X_{3p}+\ldots + X_{p-1,p})\,.
$$
Combining the three subdiagrams with the central quartic vertex, we arrive at the following
dictionary between $d^{1pqr} (f^{bcd})^{n-4}$ color tensors and $(X_{rs})^{n-4}$ worldsheet
integrands:
\eqnn\nmap
$$\eqalignno{
  &6 i^n \, d^{1x_py_q z_n}  \, f^{23x_3} \, f^{x_34x_4} \, \ldots \, 
f^{x_{p-2},p-1,x_{p-1}} f^{x_{p-1} p x_p} \cr
& \ \ \times \ f^{p+1,p+2,y_{p+2}} \, f^{y_{p+2},p+3,y_{p+3}} \, \ldots \,
f^{y_{q-2},q-1,y_{q-1}} \, f^{y_{q-1},q,y_q} \cr
& \ \ \times \ f^{q+1,q+2,z_{q+2}} \, f^{z_{q+2},q+3,z_{q+3}} \, \ldots \,
f^{z_{n-2},n-1,z_{n-1}} \, f^{z_{n-1},n,z_n}  \cr
&\longleftrightarrow \ \  \left( \prod_{k=3}^p \sum_{m=2}^{k-1} X_{mk}  \right)  \,
\left( \prod_{k=p+2}^q \sum_{m=p+1}^{k-1} X_{mk}  \right) \,
\left( \prod_{k=q+2}^n \sum_{m=q+1}^{k-1} X_{mk}  \right) \ .
 &\nmap
}$$
Given the most general definition \nmap\ of the double-arrow notation, the final forms \nfinal\ and \nCf\ for $K_n$ and ${\cal M}_n^{F^4}$, respectively, are related by
\eqn\dict{
{\cal M}_n^{F^4} \ \ \longleftrightarrow \ \ K_n .
}
This map allows to construct the one-loop kinematic factor by knowledge of the corresponding color-dressed tree amplitude at order $\ap^2$.

\subsec{Proving total symmetry of $K_n$}
\subseclab\auchnochmal

\noindent
In this subsection, we use the ${\cal M}_n^{F^4} \leftrightarrow K_n$ duality \nmap\ to carry out
the outstanding proof that $K_n$ as given by \nfinal\ is completely symmetric in all labels
$(12\ldots n)$.

Representing $K_n$ and ${\cal M}_n^{F^4}$ in their minimal $S^{n-1}_3$ basis hides the total
permutation symmetry in $12\ldots n$. Leg number one is singled out in \nfinal\ and
\nCf\ on the level of both BRST invariants $C_{1,\ldots}$ and color factors $d^{1pqr}
(f^{bcd})^{n-4}$ and worldsheet functions $(X_{rs})^{n-4}$ since $r,s\neq 1$. Since the remaining
legs $23\ldots n$ enter on equal footing, it is sufficient to prove $1\leftrightarrow 2$ symmetry
of $K_n$ and ${\cal M}_n^{F^4}$. The explicit check would require several changes of basis -- firstly in
kinematic space from $C_{1,\ldots}$ to $C_{2,\ldots}$ using the identities in subsection \CsymSec,
secondly in color space $d^{1pqr} \mapsto d^{2pqr}$ and thirdly in the worldsheet integrand
$(X_{rs})^{n-4}$ from $r,s\neq 1$ to $r,s \neq 2$. We will instead apply an indirect argument.

The mapping \nmap\ between color factors and $(X_{rs})^{n-4}$ integrands respects not only the
standard Jacobi identities \Jacobi\ but also those relations which are required for the
aforementioned change of basis: The generalized Jacobi relations \genJacobi\ are dual to
integration by parts. The simplest non-trivial example can be found at five-points,
$$
f^{12a} d^{a345} + f^{13a} d^{a245} + f^{14a} d^{a235} + f^{15a} d^{a234} = 0 \ \ \longleftrightarrow \ \ \ X_{12} + X_{13} + X_{14} + X_{15} = 0
$$
where the validity of the $X_{1j}$ relation rests on integration against the Koba Nielsen factor, see subsection \IBP. At higher multiplicity, the form $\prod_{k=3}^p \sum_{m=2}^{k-1} X_{mk}$ of the worldsheet functions is sufficiently integration--by--parts--friendly such that they still obey four term identities of type \genJacobi, e.g.
$$
f^{12a} f^{a3b} d^{b456} =  f^{12a} (d^{a45b} f^{b36} + {\rm cyc}(456) ) \ \ \longleftrightarrow \ \ \ X_{12} ( X_{13} +  X_{23} ) = X_{12} ( X_{34} + X_{35} + X_{36})
$$
as well as
$$
f^{23a} f^{a1b} d^{b456} + {\rm sym}(1456) = 0  \ \ \longleftrightarrow \ \ \ X_{23}(X_{21}+ X_{24} + X_{25} + X_{26} +  2\leftrightarrow 3) = 0
$$
at six-points. Generalizations to higher multiplicity are straightforward.

Since the mapping \nmap\ preserves the generalized Jacobi relations \genJacobi, the hidden total
symmetry of ${\cal M}_n^{F^4}$ implies that of $K_n$. Our computation of ${\cal M}_n^{F^4}$
started with the manifestly $1\leftrightarrow 2$ symmetric expression \cdressed\ summing over all cyclically inequivalent permutations, so we can be sure that the representation \nCf\ is totally
symmetric. Our derivation of the final result \nfinal\ for $K_n$, on the other hand, started with the $V^1 \leftrightarrow U^2$ asymmetric prescription \onepresc\ and involved incomplete arguments about the absence of additional b-ghost contributions. It is quite assuring
to see that \nfinal\ must be totally symmetric as well -- if the b-ghost contributed to $K_n$
via OPE contractions, then this would probably modify its symmetry properties due to the asymmetric
response of $V^1$ and $U^{j \geq 2}$, suggesting their absence.

\subsec{Correspondence between color and kinematics in ${\cal M}_n^{F^4}$}

\noindent
It was argued in \BCJ\ that the symmetric role of kinematic numerators and color factors in SYM
amplitudes suggests to impose dual Jacobi identities in the kinematic sector. They have been
successfully applied to simplify the calculation of multiloop amplitudes in both SYM and gravity
\refs{\BernUE,\BernUF}. The BRST building blocks technique can be used to obtain local BCJ
numerators at tree-level for any number of external legs \BCJps\ through the low energy limit of
string amplitudes. Therefore, it seems worthwhile to search for possible BCJ generalizations at
the next order in the momentum expansion of the superstring.

So in this final subsection we show that the final form \nCf\ for the color-dressed ${\cal O}(\ap^2)$ amplitude ${\cal M}_n^{F^4}$ is symmetric under exchange of color and
kinematics. This observation has no direct relevance for one-loop amplitudes but it is an
interesting generalization of the color-kinematic-symmetric representation \BCJ
\eqn \YMsymm{
{\cal M}_n^{\rm YM} = \sum_{i}^{(2n-5)!!} {c_i \, n_i \over \prod^{n-3}_{\alpha_i} s_{\alpha_i}}
}
for color-dressed SYM tree amplitudes. The sum over $i$ encompasses all cubic diagrams with $n-3$
propagators $\prod_{\alpha_i} s_{\alpha_i}^{-1}$, and $c_i,n_i$ denote the associated color- and
kinematic structures. One rewarding property of \YMsymm\ is the fact that gravity tree amplitudes
can be immediately obtained by replacing color factors $c_i \mapsto \tilde n_i$ by another copy
$\tilde n_i$ of the kinematic numerators $n_i$, provided that the latter satisfy Jacobi identities
dual to the color factors $c_i$.

This encouraged us to build the ${\cal M}_n^{F^4}$ analogue
\Afsymm\ of \YMsymm, we regard it as the first step towards a double copy construction that could
ultimately yield a gravity analogue of ${\cal A}^{F^4}$ amplitudes. Instead of the cubic diagrams
in \YMsymm, the diagrams in ${\cal M}_n^{F^4}$ are built from one totally symmetry quartic vertex
and $n-4$ cubic vertices.

The expansion of ${\cal M}_n^{F^4}$ in terms of BRST invariants $C_{1,\ldots}$ takes a very
compact form, but since each $C_{1,\ldots}$ encompasses several kinematic poles (i.e.\ diagrams of
the form \Tbox), it is not immediately obvious from \nCf\ how the kinematic numerators associated
to these poles combine with color factors. In section \BRSTbb, we have constructed these
numerators in pure spinor superspace, they are quartic expressions $\langle T_{d_1 \ldots d_s}
T^i_{a_1\ldots a_p} T^j_{b_1\ldots b_q} \, T^k_{c_1\ldots c_r} \rangle$ in tree subdiagrams
$T_{\ldots}$ and $T^{i,j,k}_{\ldots}$ attached to a totally symmetric quartic vertex. As an artifact of inserting leg one
via unintegrated vertex operator $V^1$, each numerator obeys $1 \in \{ d_1,d_2 \ldots d_s\}$.

\bigskip
\moveright 0.7 cm
\vbox{\offinterlineskip
\halign{%
\strut \vrule\hfil # \hfil\vrule\hskip5pt
& \hfil # \hfil
& \hfil # \hfil
& \hfil # \hfil
& \hfil # \hfil
& \hfil # \hfil
& \hfil # \hfil\vrule \cr
\noalign{\hrule}
 multiplicity & 4 & 5 & 6 & 7 & 8 & $n$\cr
\noalign{\hrule}
$\ \ $diagrams per color-stripped ${\cal A}^{\rm YM}$ & 2 & 5 & 14 & 42 & 132 & $C(n-2)$ \cr
\noalign{\hrule}
$\ \ $diagrams per color-dressed ${\cal M }_n^{\rm YM}$ &3 &15 &105 &945 &10395 &$(2n-5)!!$\cr
\noalign{\hrule}
$\ \ $diagrams per color-stripped ${\cal A}^{F^4}$ &1 &5 &21 &84 &330 &${n-3\over 2} \, C(n-2)$\cr
\noalign{\hrule}
$\ \ $diagrams per color-dressed ${\cal M}_n^{F^4}$ &1 &10 &105 &1260 &17325 & ${n-3\over 3} \, (2n-5)!!$\cr
\noalign{\hrule}
}}
\medskip
{\leftskip=20pt\rightskip=20pt\noindent\ninepoint\baselineskip=10pt
{\bf Table 2.} Number of diagrams which compose the different types of amplitudes according
to their kinematic pole structure. Here, $C(k)$ denotes the $k^{\rm th}$ Catalan number $C(k)={ (2k)! \over k! (k+1)!}$.
\par }

\medskip
\noindent
The number of diagrams per color-dressed ${\cal M}_n^{F^4}$ is displayed in the last line of table 2\foot{In order to arrive at the diagrams per topology, note that there are $(2p-3)!!$ subdiagrams within all the
$T_{i_1 i_2 ... i_p}$ permutations (at fixed set $\{i_1 i_2 ... i_p\}$), corresponding to the
$(2n-5)!!$ cubic diagrams in an $n$-point color-dressed SYM tree amplitude. For
instance, there are three different diagrams
$${T_{123} \over s_{12} s_{123}} , \quad {T_{231} \over s_{23} s_{123}} , \quad {T_{312} \over s_{13} s_{123}} 
$$ corresponding to the $s$-, $t$- and $u$ channel in ${\cal M}_4^{\rm YM}$ and 15
different $T_{pqrs} / s^3$ subdiagrams.}. In order to resolve all of them, we start from the $\langle M_{d_1 \ldots d_s} M^i_{a_1\ldots a_p} M^j_{b_1\ldots b_q} \, M^k_{c_1\ldots c_r} \rangle$ constituents of $C_{1,\ldots}$ and expand the Berends--Giele currents in terms of BRST building blocks $\langle T_{d_1 \ldots d_s} T^i_{a_1\ldots a_p} T^j_{b_1\ldots b_q} \, T^k_{c_1\ldots c_r} \rangle$. Each individual kinematic diagram is associated with a separate color factor $d^{ijkl} (f^{bcd})^{n-4}$ which precisely matches its propagator structure. Of course, the color algebra makes use of the generalized Jacobi identities \Jacobi\ and \genJacobi, e.g. the five-point result \fivedress\ yields
\eqnn\fivedressedbcj
$$\eqalignno{
{1\over 6i}{\cal M}_5^{F^4} &=
        {1\over s_{23}} \langle V_{1}T^i_{23}T^j_{4}T_{5}^k \rangle f^{23a}d^{a145}
       + {1\over s_{24}} \langle V_{1}T^i_{24}T^j_{3}T_{5}^k \rangle f^{24a}d^{a135}\cr
&\quad + {1\over s_{25}} \langle V_{1}T^i_{25}T^j_{3}T_{4}^k \rangle f^{25a}d^{a134}
       + {1\over s_{34}} \langle V_{1}T^i_{34}T^j_{2}T_{5}^k \rangle f^{34a}d^{a125}\cr
&\quad + {1\over s_{35}} \langle V_{1}T^i_{35}T^j_{2}T_{4}^k \rangle f^{35a}d^{a124}
       + {1\over s_{45}} \langle V_{1}T^i_{45}T^j_{2}T_{3}^k \rangle f^{45a}d^{a123}\cr
&\quad + {1\over s_{12}} \langle T_{12}T^i_{3}T^j_{4}T_{5}^k \rangle f^{12a}d^{a345}
       + {1\over s_{13}} \langle T_{13}T^i_{2}T^j_{4}T_{5}^k \rangle f^{13a}d^{a245}\cr
&\quad + {1\over s_{14}} \langle T_{14}T^i_{2}T^j_{3}T_{5}^k \rangle f^{14a}d^{a235}
       + {1\over s_{15}} \langle T_{15}T^i_{2}T^j_{3}T_{4}^k \rangle f^{15a}d^{a234} . &\fivedressedbcj
}$$
Similarly at six- and seven-points, \sixdress\ and \sevendress\ become
\eqnn\sixdressedbcj
\eqnn\sevendressedbcj
$$\eqalignno{
- \,\frac{1}{6} \; {\cal M}_6^{F^4} = 45 \ &{\rm terms} \ 
\big[ \, f^{12a} \, f^{34b} \, d^{ab56} \, \frac{1}{s_{12} \, s_{34}}
\langle T_{12} \, T_{34}^i \, T_5^j \, T_6^k \rangle \big] \cr
+ \ 60 \ &{\rm terms} \ \big[ \, f^{12a} \, f^{a3b} \, d^{b456} \, \frac{1}{s_{12} \, s_{123}}
\langle T_{123} \, T_{4}^i \, T_5^j \, T_6^k \rangle \big] &\sixdressedbcj \cr
\frac{1}{6i} \; {\cal M}_7^{F^4}= 105 \ 
&{\rm terms} \ \big[ \, f^{12a} \, f^{34b} \, f^{56c} \, d^{abc7} \, \frac{1}{s_{12} \, s_{34} \,
s_{56}} \langle T_{12} \, T_{34}^i \, T_{56}^j \, T_7^k \rangle \big] \cr
+ \ 630 \ &{\rm terms} \ \big[ \, f^{12a} \, f^{a3b} \, f^{45c} \, d^{bc67} \, \frac{1}{s_{12} \,
s_{123} \, s_{45}} \langle T_{123} \, T_{45}^i \, T_6^j \, T_7^k \rangle \big] \cr
+ \ 420 \ &{\rm terms} \ \big[ \, f^{12a} \, f^{a3b} \, f^{b4c} \, d^{c567} \, \frac{1}{s_{12} \,
s_{123} \, s_{1234}} \langle T_{1234} \, T_{5}^i \, T_{6}^j \, T_7^k \rangle \big]  \cr
+ \ 105 \ &{\rm terms} \ \big[ \, f^{12a} \, f^{34b} \, f^{abc} \, d^{c567} \, \frac{1}{s_{12} \,
s_{34} \, s_{1234}} \, 2 \langle T_{12[34]} \, T_{5}^i \, T_{6}^j \, T_7^k \rangle \big]\,.&\sevendressedbcj
}$$
For each topology, we sum over all permutations that are inequivalent under the symmetries of $\langle T_{d_1 \ldots d_s} T^i_{a_1\ldots a_p} T^j_{b_1\ldots b_q} \, T^k_{c_1\ldots c_r} \rangle$, up to the aforementioned rule that $1 \in \{ d_1,d_2 \ldots d_s\}$ holds in each term. For instance, one of the suppressed terms in \sixdressedbcj\ reads $f^{23a} f^{45b}d^{ab61} \langle V_1
T_{23}^i T_{45}^j T_6^k\rangle / (s_{23}s_{45})$.

For higher multiplicity, this generalizes to
\eqn \Afsymm{
{\cal M}_n^{F^4} = \sum_{I}^{{1\over 3}(n-3)(2n-5)!!} {C_I \, N_I \over \prod^{n-4}_{\alpha_I} s_{\alpha_I}},
}
where the sum over $I$ encompasses all box diagrams with four tree subdiagrams at the corners,
$\prod_{\alpha_I} s_{\alpha_I}^{-1}$ denotes the associated $n-4$ propagators, and the numerator
contains a color- $C_I \leftrightarrow d^{ijkl} (f^{bcd})^{n-4}$ and a kinematic factor $N_I
\leftrightarrow \langle T_{\ldots} T_{\ldots}^i T_{\ldots}^j T_{\ldots}^k \rangle$.

Unfortunately, our superspace representation of these numerators $N_I$ does not yet lead to
kinematic Jacobi identities dual to the color relation $d^{a(ijk} f^{l)ab}=0$, e.g.
\eqn\RandomName{
\langle T_{12} T_3^i T_4^j T_5^k \rangle + \langle T_{13} T_2^i T_4^j T_5^k \rangle
+ \langle T_{14} T_2^i T_3^j T_5^k \rangle + \langle T_{15} T_2^i T_3^j T_4^k \rangle \neq 0 .
}
One could suspect that this is an artifact of the asymmetric role of label one
in $\langle T_{12} T_3^i T_4^j T_5^k \rangle$ and $\langle T_{1} T_{23}^i T_4^j T_5^k \rangle$.
It would be desirable to find an improved representation of the $N_I$ such that a strict duality holds
\eqn \newjacobi{
C_I + C_J + C_K + C_L = 0 \ \ \ \leftrightarrow \ \ \  N_I + N_J + N_K + N_L = 0 .
}
This is for instance achieved by the five-point box-numerators $\gamma_{ij}$ in \CarrascoMN. Finding such
a duality-satisfying representation for $n$-point kinematics and studying the significance of the gravity
amplitude obtained by replacing $C_I \mapsto \tilde N_I$ in \Afsymm\ is left for future work.

\newsec{Conclusions}
\seclab\concl

\noindent
In this article, we have derived BRST invariant worldsheet integrands $K_n$ for one-loop open superstring
amplitudes involving any number $n$ of massless gauge multiplets. Our main result \nfinal\ is
expressed in terms of kinematic building blocks $C_{1,\ldots}$ which are implicitly
given in terms of ${\cal O}(\ap^2)$ tree subamplitudes via \AFasC. Since we have used BRST
invariance in determining the associated worldsheet functions, our setup is by construction
blind to the hexagon anomaly \GSanom. A superspace treatment of anomalous amplitude
ingredients along the lines of \BerkovitsBK\ is left for future work.

Both the superspace kinematics $C_{1,\ldots}$ and the
associated worldsheet functions fall into a basis of dimension $S^{n-1}_3$, an unsigned Stirling
number of first kind. The same kind of symmetries also govern the color-dressed tree amplitude
${\cal M}_n^{F^4}$ at order $\ap^2$, so we point out a duality between its minimal form \nCf\ in a color basis and
the one-loop integrand $K_n$ given by \nfinal. The link is a one-to-one dictionary \nmap\ between color
factors $d^{ijkl} (f^{bcd})^{n-4}$ (encompassing one symmetrized four-trace and structure constants
otherwise) and worldsheet functions $X^{n-4}_{ij} \equiv (s_{ij} \eta_{ij})^{n-4}$ (built from $\eta_{ij}= \partial_i
\langle x(z_i,\bar z_i) x(z_j,\bar z_j) \rangle$) present in $K_n$.

A detailed analysis of the $S^{n-1}_3$ worldsheet integrals is left for future work. The only
comment we want to make at this point is that the integrand structure closely parallels the tree-level result from \refs{\MSSI,\MSSII}: Each $z_i \rightarrow z_j$ singularity in both the tree-level and the one-loop integrand is always accompanied by a corresponding Mandelstam numerator
$s_{ij}$, i.e.\ we have $s_{ij} \eta_{ij} = s_{ij} / z_{ij} + {\cal O}(z_{ij})$. This guarantees
that the integration does not introduce any poles in kinematic invariants, i.e. that the full
propagator structure due to open string exchange is captured by the $C_{1,\ldots}$. On the other hand, loop amplitudes
additionally involve non-analytic momentum dependencies, so the main challenge in further studying
the worldsheet integrals is to identify the polylogarithms that arise in both leading and
subleading orders in $\ap$.


\bigskip
\noindent
{\bf Acknowledgements:} We wish to thank Johannes Br\"odel for intensive discussions and
continuous inspiring email exchange. We are very grateful to Michael Green and Stephan Stieberger
for numerous insightful conversations and valuable advice. We are indebted to Johannes Br\"odel, Henrik Johansson, Stephan Stieberger and Stefan Theisen
for reading our draft and making helpful suggestions and to Bernhard Wurm for his creative ideas
about notation.
C.M. acknowledges support by the European Research Council Advanced Grant No. 247252 of Michael Green and would like to thank the Albert Einstein Institute in Potsdam for hospitality and financial support during completion of this work. O.S. is very grateful to Michael Green and his European Research Council Advanced Grant No. 247252 for
hospitality and financial support during preparation and completion of this work. Moreover, he would like to thank DAMTP for creating a stimulating atmosphere during the visits in Cambridge. 

\appendix{A}{On the uniqueness of the b-ghost zero mode contribution}
\applab\bghostapp

\noindent
The computation of higher-point amplitudes at one-loop might involve different
$d_\a$ zero-mode distributions  among the picture changing operators, the b-ghost
and the external vertices. In addition, the b-ghost might have OPE singularities with the
other operators, resulting in yet other types of contributions.

However, the following argument supports that the {\it zero-mode} b-ghost contribution at one-loop is unique
and given by $d^4\d'(N)$. In order to see this note that the zero-mode contribution of the picture changing operators
is fixed and given by $(d)^{10}(\l)^{10}\d^{10}(N)\d(J)(\t)^{11}\d^{11}(\l)$, which is responsible
among other things for absorbing all 11
bosonic zero-modes of $w_\a$ \MPS.
Now assume that the b-ghost zero-mode contribution
contains $(d)^{n} \d^{m}(N)$ and note that performing the zero-mode integral
$\int [{\cal D}N]\,d^{16}d (d)^{10+n}(\l)^{10}\d^{11}(\l)\d^{m}(N) \,\big[{\rm vertices}\big]$
has the net effect of replacing $(d)^{6-n} (N)^{m}$ zero-modes from the external vertices by
a function quadratic in $\l$,
\eqn\emprule{
(d)^{6-n} (N)^{m} \longrightarrow (\l)^2
}
since $[{\cal D}N]$ has ghost number $-8$.
From the expression \bghost\ for the b-ghost it follows that the possible
values are $n=0,1,2,4$ and $m=1,2,3$.
However $(d)^4 (N)^m$ and $(d)^5 (N)^m$ have no $(00002)$ component
for any value of $m$ \LiE\ and the zero-mode integral vanishes for $n=1,2$.
Therefore group theory alone does not exclude the possibility of the b-ghost contributing either
0 or 4 zero modes of $d_\a$ with varying number of derivatives of delta functions. So let us
analyze these possibilities in separate.

The possible zero-mode contribution from the b-ghost containing no $d_\a$ zero modes are given
by
\eqn\nozero{
N\Pi^2\d(N), \qquad N^2\Pi^2\d'(N),
}
but they both vanish due to oversaturated $N^{mn}$ zero modes using that $N\d(N)=0$. For the same
reason, any contribution $\sim J$ and $\sim J^2$ from the b-ghost (e.g. $(d)^4JN\d'''(N)$ or
$(d)^4(J)^2\d'''(N)$ at four $d_\alpha$ zero modes) is suppressed by the $\d(J) = 0$ from the
picture changing operator $Z_J$. That leaves the three contributions
\eqn\fourzero{
(d)^4\d'(N), \quad (d)^4 N\d''(N), \quad (d)^4(N)^2\d'''(N)
}
of uniform type under integration by parts.
Therefore the zero-mode contribution from the b-ghost is unique and given by $(d)^4\d'(N)$.
In this paper we studied the cohomology properties of precisely this class of terms in order
to anticipate its appearance in the final expression for the superspace kinematic factors.

When the b-ghost is allowed to contribute non-zero modes the number of possibilities increases, but
only those which also contain either $0$ or $4$ zero modes of $d_\a$ can have a nonzero impact on the amplitude.
As argued in \fiveptone, terms involving only one OPE contraction of the b-ghost
vanish because they are proportional to a derivative with respect to the position $z_0$ of
the b-ghost insertion. Since $z_0$ appears nowhere else in the correlation function, those terms
are total derivatives which integrate to zero due to the doubling trick. Having excluded single OPEs with the $b$ ghost, it follows that the
five-point amplitude gets no contribution at all from $b$ ghost OPEs \fiveptone, but
from six-points onwards these terms are not excluded. For example, the b-ghost term $(d)^4JN\d'''(N)$
can in principle have simultaneous OPEs involving $J$ and $N$ with different external vertices leading to factors
which are not manifestly total derivatives. This term requires two $d$'s and three $N$'s
from the integrated vertices which can be provided in case of six and more external states.

\appendix{B}{Symmetrized traces for six- and seven-point amplitudes}
\applab\ColorApp

\noindent
The six- and seven-point symmetrized traces can be computed using the 
{\tt color} package of {\tt FORM}. After rewriting the generated 
terms in the Kleiss--Kuijf basis of $f^{n-2}$ and in the ``Stirling'' basis of $d^{ijkl} (f^{bcd})^{n-4}$ one
gets for six-points
$$\eqalignno{
&{\rm Tr}(T^1 T^2 T^3 T^4 T^5 T^6) \ + \ {\rm Tr}(T^6 T^5 T^4 T^3 T^2 T^1) =  2 d^{123456} \cr
& +  {1 \over 5} f^{12 a}f^{a 3 b}f^{b 4 c}f^{c 56} 
 -  {1 \over 20} f^{12 a}f^{a 3 b}f^{b 5 c}f^{c 46} 
 -  {1 \over 20} f^{12 a}f^{a 4 b}f^{b 3 c}f^{c 56} 
 -  {1 \over 20} f^{12 a}f^{a 4 b}f^{b 5 c}f^{c 36} \cr &
 -  {1 \over 20} f^{12 a}f^{a 5 b}f^{b 3 c}f^{c 46} 
 +  {1 \over 30} f^{12 a}f^{a 5 b}f^{b 4 c}f^{c 36} 
 -  {1 \over 20} f^{13 a}f^{a 2 b}f^{b 4 c}f^{c 56} 
 +  {1 \over 30} f^{13 a}f^{a 2 b}f^{b 5 c}f^{c 46} \cr &
 -  {1 \over 20} f^{13 a}f^{a 4 b}f^{b 2 c}f^{c 56} 
 -  {1 \over 20} f^{13 a}f^{a 4 b}f^{b 5 c}f^{c 26} 
 -  {1 \over 20} f^{13 a}f^{a 5 b}f^{b 2 c}f^{c 46} 
 +  {1 \over 30} f^{13 a}f^{a 5 b}f^{b 4 c}f^{c 26} \cr &
 -  {1 \over 20} f^{14 a}f^{a 2 b}f^{b 3 c}f^{c 56} 
 +  {1 \over 30} f^{14 a}f^{a 2 b}f^{b 5 c}f^{c 36} 
 +  {1 \over 30} f^{14 a}f^{a 3 b}f^{b 2 c}f^{c 56} 
 +  {1 \over 30} f^{14 a}f^{a 3 b}f^{b 5 c}f^{c 26} \cr &
 -  {1 \over 20} f^{14 a}f^{a 5 b}f^{b 2 c}f^{c 36} 
 +  {1 \over 30} f^{14 a}f^{a 5 b}f^{b 3 c}f^{c 26} 
 -  {1 \over 20} f^{15 a}f^{a 2 b}f^{b 3 c}f^{c 46} 
 +  {1 \over 30} f^{15 a}f^{a 2 b}f^{b 4 c}f^{c 36} \cr &
 +  {1 \over 30} f^{15 a}f^{a 3 b}f^{b 2 c}f^{c 46}
 +  {1 \over 30} f^{15 a}f^{a 3 b}f^{b 4 c}f^{c 26} 
 +  {1 \over 30} f^{15 a}f^{a 4 b}f^{b 2 c}f^{c 36} 
 -  {1 \over 20} f^{15 a}f^{a 4 b}f^{b 3 c}f^{c 26} \cr &
 -  {1 \over 2} f^{23 a}d^{a 14 b}f^{b 56} 
 -  {1 \over 2} f^{23 a}d^{a 15 b}f^{b 46} 
 -  {1 \over 2} f^{23 a}d^{a 16 b}f^{b 45} 
 -  {1 \over 2} f^{24 a}d^{a 13 b}f^{b 56} 
 -  {1 \over 2} f^{24 a}d^{a 15 b}f^{b 36} \cr &
 -  {1 \over 2} f^{24 a}d^{a 16 b}f^{b 35} 
 -  {1 \over 2} f^{25 a}d^{a 13 b}f^{b 46} 
 -  {1 \over 2} f^{25 a}d^{a 14 b}f^{b 36} 
 -  {1 \over 2} f^{25 a}d^{a 16 b}f^{b 34} 
 -  {1 \over 2} f^{26 a}d^{a 13 b}f^{b 45} \cr &
 -  {1 \over 2} f^{26 a}d^{a 14 b}f^{b 35}
 -  {1 \over 2} f^{26 a}d^{a 15 b}f^{b 34} 
 -  {1 \over 2} f^{34 a}d^{a 12 b}f^{b 56} 
 -  {1 \over 2} f^{35 a}d^{a 12 b}f^{b 46} 
 -  {1 \over 2} f^{36 a}d^{a 12 b}f^{b 45} \cr &
 -  {1 \over 3} f^{23 a}f^{a 5 b}d^{b 146}
 -  {1 \over 3} f^{23 a}f^{a 6 b}d^{b 145} 
 -  {1 \over 3} f^{24 a}f^{a 3 b}d^{b 156} 
 -  {1 \over 3} f^{24 a}f^{a 5 b}d^{b 136} \cr &
 -  {1 \over 3} f^{24 a}f^{a 6 b}d^{b 135} 
 -  {1 \over 3} f^{26 a}f^{a 5 b}d^{b 134} 
 +  {2 \over 3} f^{34 a}f^{a 2 b}d^{b 156} 
 -  {1 \over 3} f^{34 a}f^{a 5 b}d^{b 126} \cr &
 -  {1 \over 3} f^{34 a}f^{a 6 b}d^{b 125} 
 +  {1 \over 3} f^{35 a}f^{a 2 b}d^{b 146} 
 +  {1 \over 3} f^{36 a}f^{a 2 b}d^{b 145} 
 -  {1 \over 3} f^{36 a}f^{a 5 b}d^{b 124} \cr &
 +  {1 \over 3} f^{45 a}f^{a 2 b}d^{b 136} 
 +  {1 \over 3} f^{45 a}f^{a 3 b}d^{b 126} 
 +  {1 \over 3} f^{46 a}f^{a 2 b}d^{b 135} 
 +  {1 \over 3} f^{46 a}f^{a 3 b}d^{b 125} \cr &
 -  {1 \over 3} f^{46 a}f^{a 5 b}d^{b 123} 
 +  {2 \over 3} f^{56 a}f^{a 2 b}d^{b 134} 
 +  {2 \over 3} f^{56 a}f^{a 3 b}d^{b 124} 
 +  {2 \over 3} f^{56 a}f^{a 4 b}d^{b 123}.
}$$
The seven-point expression is too big to be illuminating and was therefore omitted\foot{It
is commented out in the \TeX\ source file.}.
\appendix{C}{The higher-multiplicity BRST invariants}
\applab\Chigher

\noindent
In this appendix we list the explicit form of the BRST invariants which
appear in the eight-point amplitude.
$$\eqalignno{
C_{1,23456,7,8}& = \Big( M_1 \, M_{23456}^i  +  M_{612} \, M_{345}^i  +  M_{56123} \, M_4^i \ 
+ \ \big[ M_{12345} \, M_6^i +  M_{1234} \, M_{56}^i\cr
& +  M_{123} \, M_{456}^i  +  M_{12} \, M_{3456}^i  +  M_{6123} \, M_{45}^i  + M_{61234}
\, M_5^i  +  (2,3\leftrightarrow 6,5) \, \big] \Big) M_7^j M_8^k \cr
C_{1,2345,67,8}  &= \Big( M_1 \, M_{2345}^i \, M_{67}^j \ + \ M_{215} \, M_{34}^i \, M_{67}^j \
+ \ \big[ \, M_{71} \, M_{2345}^i \, M_6^j \ - \ (6 \leftrightarrow 7) \, \big] \cr
&\ + \  \big[ \, M_{12} \, M_{345}^i \ + \ M_{123} \, M_{45}^i \ + \ M_{1234} \, M_5^i\ + \
M_{4512}\, M_3^i \ - \ (2,3\leftrightarrow 5,4) \, \big] \, M_{67}^j \cr
&\ +  \ \big\{ \, \big[ \, M_{712} \, M_{345}^i \, M_6^j \ + \ M_{7123} \, M_{45}^i \, M_6^j \ + \
M_{71234} \, M_5^i \, M_6^j \ - \ M_{7125} \, M_{34}^i \, M_6^j \cr
&\ + \ M_{75123} \, M_4^i \, M_6^j \ + \ M_{57123} \, M_4^i \, M_6^j \ - \ (2,3\leftrightarrow
5,4) \, \big] \ - \ (6 \leftrightarrow 7) \, \big\} \Big) \, M_8^k \cr
C_{1,234,567,8} &= \Big( M_1 \, M_{234}^i \, M^j_{567} \ + \ \big[ \, M_{12} \, M_{34}^i \, M_{567}^j \ + \ M_{14} \, M_{32}^i \, M_{567}^j \ + \ M_{123} \, M_4^i \, M_{567}^j
\cr
& \ + \ M_{143} \, M_2^i \, M_{567}^j \ + \ M_{214} \, M_3^i \, M_{567}^j \ + \ (2,3,4
\leftrightarrow 5,6,7) \, \big] \ + \ \big[ \, M_{712} \, M_{34}^i \, M_{56}^j \cr
&\ + \ M_{7123} \, M_4^i \, M_{56}^j \ + \ M_{6712} \, M_{34}^i \, M_5^j \ + \ M_{6512} \, M_{34}^i \, M_7^j \ + \ M_{5124}  \, M_3^i \, M_{67}^j
\cr
&\ + \ M_{2157} \, M_{34}^i \, M_6^j \ + \ M_{67123} \, M_4^i \, M_5^j \ - \ M_{65124}  \, M_3^i
\, M_7^j \ - \ M_{32157}  \, M_4^i \, M_6^j \cr
& \ + \ M_{24157}  \, M_3^i \, M_6^j \ + \ (2\leftrightarrow 4) \ + \ (5 \leftrightarrow 7)\,
\big]  \Big)  \, M_8^k \cr
C_{1,234,56,78} &= \Big( M_1 \, M_{234}^i \ + \ M_{214} \, M_3^i  \ + \ \big[ \, M_{12} \,
M_{34}^i \ + \ M_{123} \, M_{4} \ + \ (2 \leftrightarrow 4) \, \big] \, \Big) \, M_{56}^j \,
M_{78}^k \cr
& \ + \ \Big(  \big[ \, M_{15}\, M_{234}^i \, M_{6}^j \ - \ M_{16}\, M_{234}^i \, M_{5}^j \,
\big] \,M_{78}^k \ + \ \big[ \,M_{216} \, M_{34}^i \,M_5^j \ + \ M_{6123} \, M_4^i \, M_5^j \cr
& \ + \ M_{5124} \, M_3^i \, M_6^j  \ + \ (2 \leftrightarrow 4) \ - \ (5 \leftrightarrow 6) \,
\big] \, M_{78}^k \ + \ (5,6 \leftrightarrow 7,8)\, \Big) \cr
& \ + \ \big[ \,M_{617} \, M_{234}^i \, M_5^j \, M_8^k \ - \ (5 \leftrightarrow 6) \ - \ (7 \leftrightarrow 8) \, \big] \ + \ \big[ \, ( M_{7126} +M_{7162})\, M_{34}^i \, M_5^j \, M_8^k
\cr
& \ + \ ( M_{75123}+M_{57123}) \, M_4^i \, M_6^j \, M_8^k \ - \ ( M_{75124}+M_{57124}) \, M_3^i \,
M_6^j \, M_8^k \cr
& \ + \ (2 \leftrightarrow 4)  \ - \ (5 \leftrightarrow 6) \ - \ (7 \leftrightarrow 8)\,\big] 
}$$

\listrefs

\bye

%% file: harvmacM
%
%
%
\def\unredoffs{} \def\redoffs{\voffset=-.31truein\hoffset=-.48truein}
\def\speclscape{}
%
%
%
%
%
\newbox\leftpage \newdimen\fullhsize \newdimen\hstitle \newdimen\hsbody
\tolerance=1000\hfuzz=2pt
\catcode`\@=11 
\ifx\hyperdef\UNd@FiNeD\def\hyperdef#1#2#3#4{#4}\def\hyperref#1#2#3#4{#4}\fi
\def\bigans{b }
\def\answ{b }
%
\ifx\answ\bigans\message{(This will come out unreduced.}
\magnification=1200\unredoffs\baselineskip=16pt plus 2pt minus 1pt
\hsbody=\hsize \hstitle=\hsize 

\else\message{(This will be reduced.} \let\l@r=L
\magnification=1000\baselineskip=16pt plus 2pt minus 1pt \vsize=7truein
\redoffs \hstitle=8truein\hsbody=4.75truein\fullhsize=10truein\hsize=\hsbody
\output={\ifnum\pageno=0 
  \shipout\vbox{\speclscape{\hsize\fullhsize\makeheadline}
    \hbox to \fullhsize{\hfill\pagebody\hfill}}\advancepageno
  \else
  \almostshipout{\leftline{\vbox{\pagebody\makefootline}}}\advancepageno
  \fi}
\def\almostshipout#1{\if L\l@r \count1=1 \message{[\the\count0.\the\count1]}
      \global\setbox\leftpage=#1 \global\let\l@r=R
 \else \count1=2
  \shipout\vbox{\speclscape{\hsize\fullhsize\makeheadline}
      \hbox to\fullhsize{\box\leftpage\hfil#1}}  \global\let\l@r=L\fi}
\fi
%
\newcount\yearltd\yearltd=\year\advance\yearltd by -2000

\def\Title#1#2{\nopagenumbers\abstractfont\hsize=\hstitle\rightline{#1}%
\vskip 1in\centerline{\titlefont #2}\abstractfont\vskip .5in\pageno=0}
\def\Date#1{\vfill\leftline{#1}\tenpoint\supereject\global\hsize=\hsbody%
\footline={\hss\tenrm\hyperdef\hypernoname{page}\folio\folio\hss}}%
%

\def\draftmode{\message{ DRAFTMODE }\def\draftdate{{\rm preliminary draft:
\number\month/\number\day/\number\yearltd\ \ \hourmin}}%
\headline={\hfil\draftdate}\writelabels\baselineskip=20pt plus 2pt minus 2pt
 {\count255=\time\divide\count255 by 60 \xdef\hourmin{\number\count255}
  \multiply\count255 by-60\advance\count255 by\time
  \xdef\hourmin{\hourmin:\ifnum\count255<10 0\fi\the\count255}}}
\def\nolabels{\def\wrlabeL##1{}\def\eqlabeL##1{}\def\reflabeL##1{}}
\def\writelabels{\def\wrlabeL##1{\leavevmode\vadjust{\rlap{\smash%
{\line{{\escapechar=` \hfill\rlap{\sevenrm\hskip.03in\string##1}}}}}}}%
\def\eqlabeL##1{{\escapechar-1\rlap{\sevenrm\hskip.05in\string##1}}}%
\def\reflabeL##1{\noexpand\llap{\noexpand\sevenrm\string\string\string##1}}}
\nolabels
%
\global\newcount\secno \global\secno=0
\global\newcount\meqno \global\meqno=1
\def\s@csym{}
\def\newsec#1{\global\advance\secno by1%
{\toks0{#1}\message{(\the\secno. \the\toks0)}}%
\global\subsecno=0\eqnres@t\let\s@csym\secsym\xdef\secn@m{\the\secno}\noindent
{\bf\hyperdef\hypernoname{section}{\the\secno}{\the\secno.} #1}%
\writetoca{{\string\hyperref{}{section}{\the\secno}{\vskip2pt \bf \the\secno\quad}} {\bf #1}}%
\par\nobreak\medskip\nobreak}
\def\eqnres@t{\xdef\secsym{\the\secno.}\global\meqno=1\bigbreak\bigskip}
\def\sequentialequations{\def\eqnres@t{\bigbreak}}\xdef\secsym{}
\global\newcount\subsecno \global\subsecno=0
\def\subsec#1{\global\advance\subsecno by1%
{\toks0{#1}\message{(\s@csym\the\subsecno. \the\toks0)}}%
\ifnum\lastpenalty>9000\else\bigbreak\fi
\noindent{\it\hyperdef\hypernoname{subsection}{\secn@m.\the\subsecno}%
{\secn@m.\the\subsecno.} #1}\writetoca{\string\hskip1.45cm
{\string\hyperref{}{subsection}{\secn@m.\the\subsecno}{\secn@m.\the\subsecno.}}
{#1}}\par\nobreak\medskip\nobreak}
\def\appendix#1#2{\global\meqno=1\global\subsecno=0\xdef\secsym{\hbox{#1.}}%
\bigbreak\bigskip\noindent{\bf Appendix \hyperdef\hypernoname{appendix}{#1}%
{#1.} #2}{\toks0{(#1. #2)}\message{\the\toks0}}%
\xdef\s@csym{#1.}\xdef\secn@m{#1}%
\writetoca{{\string\hyperref{}{appendix}{#1}{\vskip2pt \bf {#1}\quad}} {\bf #2}}%
\par\nobreak\medskip\nobreak}
%
%
\def\checkm@de#1#2{\ifmmode{\def\f@rst##1{##1}\hyperdef\hypernoname{equation}%
{#1}{#2}}\else\hyperref{}{equation}{#1}{#2}\fi}
\def\eqnn#1{\DefWarn#1\xdef #1{(\noexpand\relax\noexpand\checkm@de%
{\s@csym\the\meqno}{\secsym\the\meqno})}%
\wrlabeL#1\writedef{#1\leftbracket#1}\global\advance\meqno by1}
\def\f@rst#1{\c@t#1a\em@ark}\def\c@t#1#2\em@ark{#1}
\def\eqna#1{\DefWarn#1\wrlabeL{#1$\{\}$}%
\xdef #1##1{(\noexpand\relax\noexpand\checkm@de%
{\s@csym\the\meqno\noexpand\f@rst{##1}1}{\hbox{$\secsym\the\meqno##1$}})}
\writedef{#1\numbersign1\leftbracket#1{\numbersign1}}\global\advance\meqno by1}
\def\eqn#1#2{\DefWarn#1%
\xdef #1{(\noexpand\hyperref{}{equation}{\s@csym\the\meqno}%
{\secsym\the\meqno})}$$#2\eqno(\hyperdef\hypernoname{equation}%
{\s@csym\the\meqno}{\secsym\the\meqno})\eqlabeL#1$$%
\writedef{#1\leftbracket#1}\global\advance\meqno by1}
\def\xeqn{\expandafter\xe@n}\def\xe@n(#1){#1}
\def\xeqna#1{\expandafter\xe@n#1}
\def\eqns#1{(\e@ns #1{\hbox{}})}
\def\e@ns#1{\ifx\UNd@FiNeD#1\message{eqnlabel \string#1 is undefined.}%
\xdef#1{(?.?)}\fi{\let\hyperref=\relax\xdef\next{#1}}%
\ifx\next\em@rk\def\next{}\else%
\ifx\next#1\xeqn#1\else\def\n@xt{#1}\ifx\n@xt\next#1\else\xeqna#1\fi
\fi\let\next=\e@ns\fi\next}

\def\DefWarn#1{\ifx\UNd@FiNeD#1\else
\immediate\write16{*** WARNING: the label \string#1 is already defined ***}\fi}
%
\newskip\footskip\footskip14pt plus 1pt minus 1pt 
\def\footnotefont{\ninepoint}\def\f@t#1{\footnotefont #1\@foot}
\def\f@@t{\baselineskip\footskip\bgroup\footnotefont\aftergroup\@foot\let\next}
\setbox\strutbox=\hbox{\vrule height9.5pt depth4.5pt width0pt}
\global\newcount\ftno \global\ftno=0
\def\foot{\global\advance\ftno by1\def\foot@rg{\hyperref{}{footnote}%
{\the\ftno}{\the\ftno}\xdef\foot@rg{\noexpand\hyperdef\noexpand\hypernoname%
{footnote}{\the\ftno}{\the\ftno}}}\footnote{$^{\foot@rg}$}}
%
\newwrite\ftfile
\def\footend{\def\foot{\global\advance\ftno by1\chardef\wfile=\ftfile
\hyperref{}{footnote}{\the\ftno}{$^{\the\ftno}$}%
\ifnum\ftno=1\immediate\openout\ftfile=\jobname.fts\fi%
\immediate\write\ftfile{\noexpand\smallskip%
\noexpand\item{\noexpand\hyperdef\noexpand\hypernoname{footnote}
{\the\ftno}{f\the\ftno}:\ }\pctsign}\findarg}%
\def\footatend{\vfill\eject\immediate\closeout\ftfile{\parindent=20pt
\centerline{\bf Footnotes}\nobreak\bigskip\input \jobname.fts }}}
\def\footatend{}
%
%
\global\newcount\refno \global\refno=1
\newwrite\rfile
\def\ref{[\hyperref{}{reference}{\the\refno}{\the\refno}]\nref}
\def\nref#1{\DefWarn#1%
\xdef#1{[\noexpand\hyperref{}{reference}{\the\refno}{\the\refno}]}%
\writedef{#1\leftbracket#1}%
\ifnum\refno=1\immediate\openout\rfile=\jobname.refs\fi
\chardef\wfile=\rfile\immediate\write\rfile{\noexpand\item{[\noexpand\hyperdef%
\noexpand\hypernoname{reference}{\the\refno}{\the\refno}]\ }%
\reflabeL{#1\hskip.31in}\pctsign}\global\advance\refno by1\findarg}
\def\findarg#1#{\begingroup\obeylines\newlinechar=`\^^M\pass@rg}
{\obeylines\gdef\pass@rg#1{\writ@line\relax #1^^M\hbox{}^^M}%
\gdef\writ@line#1^^M{\expandafter\toks0\expandafter{\striprel@x #1}%
\edef\next{\the\toks0}\ifx\next\em@rk\let\next=\endgroup\else\ifx\next\empty%
\else\immediate\write\wfile{\the\toks0}\fi\let\next=\writ@line\fi\next\relax}}
\def\striprel@x#1{} \def\em@rk{\hbox{}}
\def\lref{\begingroup\obeylines\lr@f}
\def\lr@f#1#2{\DefWarn#1\gdef#1{\let#1=\UNd@FiNeD\ref#1{#2}}\endgroup\unskip}
\def\semi{;\hfil\break}
\def\addref#1{\immediate\write\rfile{\noexpand\item{}#1}} 
\def\listrefs{\footatend\vfill\supereject\immediate\closeout\rfile\writestoppt
\baselineskip=\footskip\centerline{{\bf References}}\bigskip{\parindent=20pt%
\frenchspacing\escapechar=` \input \jobname.refs\vfill\eject}\nonfrenchspacing}
\def\startrefs#1{\immediate\openout\rfile=\jobname.refs\refno=#1}
\def\xref{\expandafter\xr@f}\def\xr@f[#1]{#1}
\def\refs#1{\count255=1[\r@fs #1{\hbox{}}]}
\def\r@fs#1{\ifx\UNd@FiNeD#1\message{reflabel \string#1 is undefined.}%
\nref#1{need to supply reference \string#1.}\fi%
\vphantom{\hphantom{#1}}{\let\hyperref=\relax\xdef\next{#1}}%
\ifx\next\em@rk\def\next{}%
\else\ifx\next#1\ifodd\count255\relax\xref#1\count255=0\fi%
\else#1\count255=1\fi\let\next=\r@fs\fi\next}
%

%
\newwrite\ffile\global\newcount\figno \global\figno=1
\def\fig{fig.~\hyperref{}{figure}{\the\figno}{\the\figno}\nfig}
\def\nfig#1{\DefWarn#1%
\xdef#1{fig.~\noexpand\hyperref{}{figure}{\the\figno}{\the\figno}}%
\writedef{#1\leftbracket fig.\noexpand~\xfig#1}%
\ifnum\figno=1\immediate\openout\ffile=\jobname.figs\fi\chardef\wfile=\ffile%
{\let\hyperref=\relax
\immediate\write\ffile{\noexpand\medskip\noexpand\item{Fig.\ %
\noexpand\hyperdef\noexpand\hypernoname{figure}{\the\figno}{\the\figno}. }
\reflabeL{#1\hskip.55in}\pctsign}}\global\advance\figno by1\findarg}
\def\listfigs{\vfill\eject\immediate\closeout\ffile{\parindent40pt
\baselineskip14pt\centerline{{\bf Figure Captions}}\nobreak\medskip
\escapechar=` \input \jobname.figs\vfill\eject}}
\def\xfig{\expandafter\xf@g}\def\xf@g fig.\penalty\@M\ {}
\def\figs#1{figs.~\f@gs #1{\hbox{}}}
\def\f@gs#1{{\let\hyperref=\relax\xdef\next{#1}}\ifx\next\em@rk\def\next{}\else
\ifx\next#1\xfig #1\else#1\fi\let\next=\f@gs\fi\next}
\def\figin{\epsfcheck\figin}\def\figins{\epsfcheck\figins}
\def\epsfcheck{\ifx\epsfbox\UNd@FiNeD
\message{(NO epsf.tex, FIGURES WILL BE IGNORED)}
\gdef\figin##1{\vskip2in}\gdef\figins##1{\hskip.5in}
\else\message{(FIGURES WILL BE INCLUDED)}%
\gdef\figin##1{##1}\gdef\figins##1{##1}\fi}
\def\DefWarn#1{}
\def\figinsert{\goodbreak\midinsert}
\def\ifig#1#2#3{\DefWarn#1\xdef#1{fig.~\noexpand\hyperref{}{figure}%
{\the\figno}{\the\figno}}\writedef{#1\leftbracket fig.\noexpand~\xfig#1}%
\figinsert\figin{\centerline{#3}}\medskip\centerline{\vbox{\baselineskip12pt
\advance\hsize by -1truein\noindent\wrlabeL{#1=#1}\footnotefont%
{\bf Fig.~\hyperdef\hypernoname{figure}{\the\figno}{\the\figno}:} #2}}
\bigskip\endinsert\global\advance\figno by1}
\newwrite\lfile
{\escapechar-1\xdef\pctsign{\string\%}\xdef\leftbracket{\string\{}
\xdef\rightbracket{\string\}}\xdef\numbersign{\string\#}}
\def\writedefs{\immediate\openout\lfile=labeldefs.tmp \def\writedef##1{%
{\let\hyperref=\relax\let\hyperdef=\relax\let\hypernoname=\relax
 \immediate\write\lfile{\string\def\string##1\rightbracket}}}}%
\def\writestop{\def\writestoppt{\immediate\write\lfile{\string\pageno
 \the\pageno\string\startrefs\leftbracket\the\refno\rightbracket
 \string\def\string\secsym\leftbracket\secsym\rightbracket
 \string\secno\the\secno\string\meqno\the\meqno}\immediate\closeout\lfile}}
\def\writestoppt{}\def\writedef#1{}
\def\seclab#1{\DefWarn#1%
\xdef #1{\noexpand\hyperref{}{section}{\the\secno}{\the\secno}}%
\writedef{#1\leftbracket#1}\wrlabeL{#1=#1}}
\def\subseclab#1{\DefWarn#1%
\xdef #1{\noexpand\hyperref{}{subsection}{\secn@m.\the\subsecno}%
{\secn@m.\the\subsecno}}\writedef{#1\leftbracket#1}\wrlabeL{#1=#1}}
\def\applab#1{\DefWarn#1%
\xdef #1{\noexpand\hyperref{}{appendix}{\secn@m}{\secn@m}}%
\writedef{#1\leftbracket#1}\wrlabeL{#1=#1}}
\newwrite\tfile \def\writetoca#1{}
\def\leaderfill{\leaders\hbox to 1em{\hss.\hss}\hfill}
\def\writetoc{\immediate\openout\tfile=\jobname.toc
   \def\writetoca##1{{\edef\next{\write\tfile{\noindent ##1
   \string\leaderfill  {\string\hyperref{}{page}{\noexpand\number\pageno}%
                       {\noexpand\number\pageno}} \par}}\next}}}
\newread\ch@ckfile
\def\listtoc{\immediate\closeout\tfile\immediate\openin\ch@ckfile=\jobname.toc
\ifeof\ch@ckfile\message{no file \jobname.toc, no table of contents this pass}%
\else\closein\ch@ckfile\centerline{\bf Contents}\nobreak\medskip%
{\baselineskip=16pt\footnotefont\parskip=0pt\catcode`\@=11\input\jobname.toc
\catcode`\@=12\bigbreak\bigskip}\fi}
\catcode`\@=12 
%
\edef\tfontsize{\ifx\answ\bigans scaled\magstep3\else scaled\magstep4\fi}
\font\titlerm=cmr10 \tfontsize \font\titlerms=cmr7 \tfontsize
\font\titlermss=cmr5 \tfontsize \font\titlei=cmmi10 \tfontsize
\font\titleis=cmmi7 \tfontsize \font\titleiss=cmmi5 \tfontsize
\font\titlesy=cmsy10 \tfontsize \font\titlesys=cmsy7 \tfontsize
\font\titlesyss=cmsy5 \tfontsize \font\titleit=cmti10 \tfontsize
\skewchar\titlei='177 \skewchar\titleis='177 \skewchar\titleiss='177
\skewchar\titlesy='60 \skewchar\titlesys='60 \skewchar\titlesyss='60
\def\titlefont{\def\rm{\fam0\titlerm}
\textfont0=\titlerm \scriptfont0=\titlerms \scriptscriptfont0=\titlermss
\textfont1=\titlei \scriptfont1=\titleis \scriptscriptfont1=\titleiss
\textfont2=\titlesy \scriptfont2=\titlesys \scriptscriptfont2=\titlesyss
\textfont\itfam=\titleit \def\it{\fam\itfam\titleit}\rm}
 \ifx\answ\bigans\else scaled\magstep1\fi
\ifx\answ\bigans\def\abstractfont{\tenpoint}\else
\font\absit=cmti10 scaled \magstep1
\font\abssl=cmsl10 scaled \magstep1
\font\absrm=cmr10 scaled\magstep1 \font\absrms=cmr7 scaled\magstep1
\font\absrmss=cmr5 scaled\magstep1 \font\absi=cmmi10 scaled\magstep1
\font\absis=cmmi7 scaled\magstep1 \font\absiss=cmmi5 scaled\magstep1
\font\abssy=cmsy10 scaled\magstep1 \font\abssys=cmsy7 scaled\magstep1
\font\abssyss=cmsy5 scaled\magstep1 \font\absbf=cmbx10 scaled\magstep1
\skewchar\absi='177 \skewchar\absis='177 \skewchar\absiss='177
\skewchar\abssy='60 \skewchar\abssys='60 \skewchar\abssyss='60
\def\abstractfont{\def\rm{\fam0\absrm}
\textfont0=\absrm \scriptfont0=\absrms \scriptscriptfont0=\absrmss
\textfont1=\absi \scriptfont1=\absis \scriptscriptfont1=\absiss
\textfont2=\abssy \scriptfont2=\abssys \scriptscriptfont2=\abssyss
\textfont\itfam=\absit \def\it{\fam\itfam\absit}\def\footnotefont{\tenpoint}%
\textfont\slfam=\abssl \def\sl{\fam\slfam\abssl}%
\textfont\bffam=\absbf \def\bf{\fam\bffam\absbf}\rm}\fi
\def\tenpoint{\def\rm{\fam0\tenrm}
\textfont0=\tenrm \scriptfont0=\sevenrm \scriptscriptfont0=\fiverm
\textfont1=\teni  \scriptfont1=\seveni  \scriptscriptfont1=\fivei
\textfont2=\tensy \scriptfont2=\sevensy \scriptscriptfont2=\fivesy
\textfont\itfam=\tenit \def\it{\fam\itfam\tenit}\def\footnotefont{\ninepoint}%
\textfont\bffam=\tenbf \def\bf{\fam\bffam\tenbf}\def\sl{\fam\slfam\tensl}\rm}
\font\ninerm=cmr9 \font\sixrm=cmr6 \font\ninei=cmmi9 \font\sixi=cmmi6
\font\ninesy=cmsy9 \font\sixsy=cmsy6 \font\ninebf=cmbx9
\font\nineit=cmti9 \font\ninesl=cmsl9 \skewchar\ninei='177
\skewchar\sixi='177 \skewchar\ninesy='60 \skewchar\sixsy='60
\def\ninepoint{\def\rm{\fam0\ninerm}
\textfont0=\ninerm \scriptfont0=\sixrm \scriptscriptfont0=\fiverm
\textfont1=\ninei \scriptfont1=\sixi \scriptscriptfont1=\fivei
\textfont2=\ninesy \scriptfont2=\sixsy \scriptscriptfont2=\fivesy
\textfont\itfam=\ninei \def\it{\fam\itfam\nineit}\def\sl{\fam\slfam\ninesl}%
\textfont\bffam=\ninebf \def\bf{\fam\bffam\ninebf}\rm}
%
%

\hyphenation{anom-aly anom-alies coun-ter-term coun-ter-terms}
\def\inv{^{\raise.15ex\hbox{${\scriptscriptstyle -}$}\kern-.05em 1}}

\def\Dsl{\,\raise.15ex\hbox{/}\mkern-13.5mu D} 
\def\dsl{\raise.15ex\hbox{/}\kern-.57em\partial}

\def\lspace{\ifx\answ\bigans{}\else\qquad\fi}
\def\lbspace{\ifx\answ\bigans{}\else\hskip-.2in\fi} 
\def\boxeqn#1{\vcenter{\vbox{\hrule\hbox{\vrule\kern3pt\vbox{\kern3pt
	\hbox{${\displaystyle #1}$}\kern3pt}\kern3pt\vrule}\hrule}}}
\def\mbox#1#2{\vcenter{\hrule \hbox{\vrule height#2in
		\kern#1in \vrule} \hrule}}  
%

\def\darr#1{\raise1.5ex\hbox{$\leftrightarrow$}\mkern-16.5mu #1}

\def\half{{\textstyle{1\over2}}} 
\def\roughly#1{\raise.3ex\hbox{$#1$\kern-.75em\lower1ex\hbox{$\sim$}}}

\global\newcount\subsubsecno \global\subsubsecno=0
\def\subsubsec#1{\global\advance\subsubsecno by1%
{\toks0{#1}\message{(\the\secno\the\subsecno\the\subsubsecno. \the\toks0)}}%
\ifnum\lastpenalty>9000\else\bigbreak\fi
\noindent{\it\hyperdef\hypernoname{subsubsection}{\the\secno.\the\subsecno\the\subsubsecno}%
{\the\secno.\the\subsecno.\the\subsubsecno.} #1}
\par\nobreak\medskip\nobreak}